\documentclass[twocolumn]{aastex631}
\usepackage{amsmath}
\usepackage{upgreek}
\usepackage{CJK}
\usepackage{graphicx}
\usepackage{subfigure}
\usepackage{enumerate}
\usepackage{multirow}
\usepackage{threeparttable}
\usepackage{booktabs}
\usepackage{hyperref}
\usepackage{natbib}
\usepackage{makecell}

\submitjournal{AAS Journals}

\shorttitle{Mass Deviation of S-type Planets}
\shortauthors{Huang X.M. et al.}
\shortauthors{}

\graphicspath{{./}{figs/}}

\begin{document}


\title{Dynamical Mass Demography of Single S-type Planets Shaped by Stellar Companions}

\correspondingauthor{Jianghui Ji}

\author{Xiumin Huang}
\affiliation{CAS Key Laboratory of Planetary Sciences, Purple Mountain Observatory, Chinese Academy of Sciences, Nanjing 210023, China}
\affiliation{Division of Space Research and Planetary Sciences, Physics Institute, University of Bern, Gesellschaftsstrasse 6, 3012 Bern, Switzerland}
\email{huangxm@pmo.ac.cn}

\author{Jianghui Ji}
\affiliation{CAS Key Laboratory of Planetary Sciences, Purple Mountain Observatory, Chinese Academy of Sciences, Nanjing 210023, China}
\affiliation{School of Astronomy and Space Science, University of Science and Technology of China, Hefei 230026, China}
\affiliation{CAS Center for Excellence in Comparative Planetology, Hefei 230026, China}
\email{jijh@pmo.ac.cn}

\begin{abstract}
  More than 700 S-type planets have been identified from Gaia and TESS observations, including over 150 transiting targets, enabling the precise dynamical orbits and planetary mass characterization in binary environments. Here we present the first systematic N-body dynamical mass determination for 40 single S-type (Satellite-type orbit) planets in close binaries using an improved Markov Chain Monte Carlo (MCMC) framework with Gaussian priors to model stellar companion perturbations. {For binary systems with $a_\mathrm{B}<100$ au, companion-induced perturbations can bias the RV semi-amplitude and consequently the inferred planetary masses. The impact is strongest for short-period sub-Jovian planets, whose masses decrease by up to 20\% when transitioning from Keplerian to N-body modeling, while massive Jovian planets with $M_{\mathrm{Jup}} < M_{\mathrm{p}} < 5$ $M_{\mathrm{Jup}}$ show only a 7.2\% reduction.} Applying the NAFF stability criterion to TOI-4633, $i_{\mathrm{mut}} = 70^{\circ}$ is the upper limit for the fast rejection of unstable orbital solutions. Our results reveal that intermediate-mass planets exhibit both positive or negative biases in the inferred masses and influence the observed population of the sub-Jovian desert.

\end{abstract}


\section{Introduction}\label{sec:1}

\begin{figure}
\includegraphics[width=\columnwidth,height=7cm]{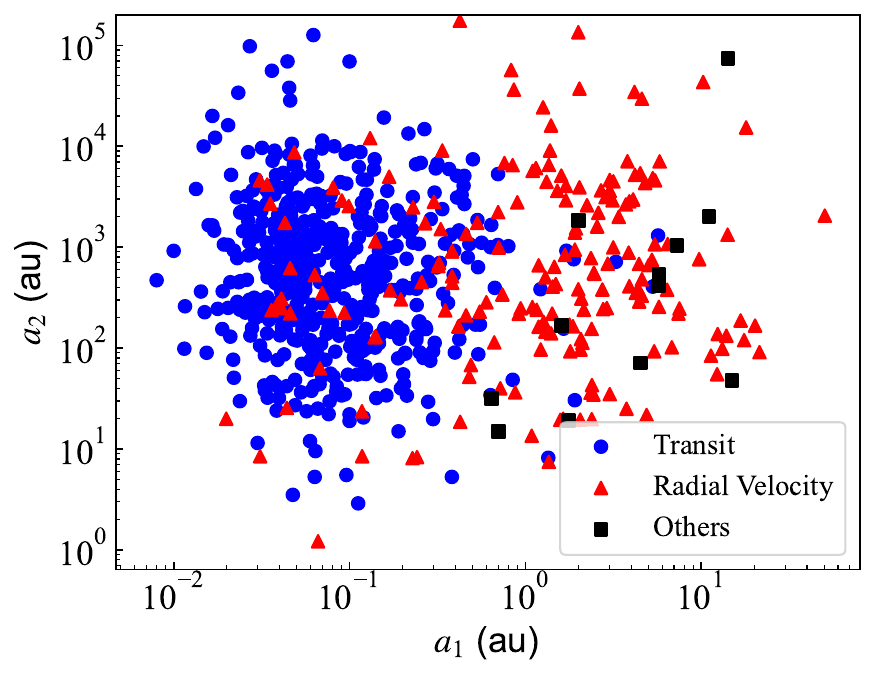}
    \caption{Distribution of orbital semi-major axes for S-type planets and binary systems. Here, $a_{\mathrm{1}}$ denotes the semi-major axis of the S-type planet’s orbit, and $a_{\mathrm{2}}$ represents the semi-major axis of the binary orbit. More than 720 S-type planets have been identified to date, with transiting planets comprising over half of the sample and predominantly residing at $a_{\mathrm{1}} < 1$ au \citep{Huang2022}.}
    \label{fig:transit_distribution}
\end{figure}

The census of exoplanets in binary systems has expanded rapidly, with more than 90\% of these planets being S-type planets. S-type planets are circumstellar planets that orbit one component of a binary system in satellite-like orbits, whereas the other class, known as P-type planets, orbit both stars in planetary-type or circumbinary orbits. The confirmed S-type planets increased from $\sim$100 to more than 700 known planets over the past three years\footnote{\url{https://exoplanet.eu/planets_binary/}}. Among these newly identified systems, 497 planets are found in relatively close-in transiting orbits. In particular, approximately 90 of these new S-type planets were detected by the Transiting Exoplanet Survey Satellite (TESS) \citep{ricker2015}. The distribution of this growing population is presented in Figure \ref{fig:transit_distribution}.

Even though the revealing of binary systems is accomplished through multi-source data including Gaia Data Releases, new transit observations, high-accuracy RV detection, and innovative methods for deriving stellar parameters, the transit method remains an effective tool for the first step of discovering and characterizing exoplanets. \citet{Crida2018} provided the most accurate physical parameters of the constraints of mass, radius, density, and internal composition to date for 55 Cnc e. According to the observations of host stars of TESS exoplanets with stellar companions at 1 to 1000 au and implications for small planet detection \citep{Lester2021}, the large pixel scale of 21 arcsec of the TESS telescope leads to light blending between the companion star and the host star, leading to the averaged 6.1\% systematic deviation of the radius of exoplanets. At the same time, the companion star could inhibit planet formation or affect its orbital evolution \citep{Gong2017,Gong2018}.

\citet{Steffen2016} discussed the sensitivity bias of Transit Timing Variation (TTV) and radial velocity (RV) mass measurements in the mass-radius distribution of exoplanets. For planets of the same radius, RV is more sensitive to short-period, high-mass planets, whereas TTV is more sensitive to planets with large radii and long periods. RV is needed to detect high-inclination systems or non-resonant planets. The combination of both methods could provide comprehensive information about the mass of the planet and the trajectory of planet formation \citep{barragan2019}.

The new generation of spectrographs, including ESPRESSO (Echelle SPectrograph for Rocky Exoplanets and Stable Spectroscopic Observations) \citep{Pepe2010} and HARPS (High-Accuracy Radial velocity Planet Searcher) \citep{Wilken2010, Wilken2012, Cosentino2012}, can achieve RV precision of $0.10 \sim 0.30$ m/s. Such high RV measurement precision presents exciting opportunities for the determination of the mass of transiting rocky planets.

Traditional RV analysis neglects interplanetary interactions, leading to uncertainties in the measurement of orbital parameters. For a hierarchical system with a massive secondary companion, significant mutual interaction between the planet and the companion is supposed to be considered under the Jacobi framework to present the real dynamics \citep{Lee2003}. This is especially true in multi-planet systems, where the mass of the planet can only be constrained by $M_{\mathrm{p}} \sin i$. \citet{Pal2010} proposed an algorithm that incorporates interplanetary perturbation effects into the RV model, thereby improving the accuracy of the measurement of planetary masses and orbital parameters. Considering the near circular orbits of multiple transiting planets, the Lagrange orbital elements $k =\sqrt  e \cos \omega$ and $h = \sqrt e \sin \omega$ were used to replace the argument of perihelion $\omega$ and the eccentricity to avoid numerical instabilities in near circular orbits in \citet{Pal2010}. The mean longitude $\lambda$ is used instead of the transit time of the periastron $\tau_p$ to simplify the description of the phase.

Previous studies have demonstrated that the joint analysis of radial velocity (RV) and transit data can effectively constrain the orbital architectures and dynamical evolution of exoplanet systems. PlanetPack \citep{Baluev2013,Baluev2018} focuses on the exploration of planetary planets and dynamic analysis of RV data. \citet{Lopez2019} conducted a detailed study of the properties of six exoplanets in the K2-138 system, which has the longest known resonance chain. \citet{dawson2021} utilized the N-body integration of REBOUND \citep{Rein2012} and PyTTV algorithms, combined with stellar parameters, to jointly fit the TTV and RV data, and studied the pairing of planetary resonance in the TOI-216 system.


Classical pure-Keplerian orbital fitting approaches suffer from several modeling biases, as they neglect planet-planet dynamical interactions and constrain orbital parameters at a reference epoch rather than the instantaneous physical state of the system. Dynamical variations induced in radial-velocity signals provide a more realistic framework for accurately determining planetary masses. Previous studies have demonstrated that dynamical interactions can play an important role in the interpretation of planetary observations \citep{Laughlin2001, Gozdziewski2006, Gozdziewski2008, Rivera2010}. \citet{Laughlin2001} first showed that Keplerian models become inadequate for resonant systems such as GJ 876, where mutual planetary perturbations produce detectable signatures in the RV data. \citet{Rivera2010} further developed a self-consistent dynamical RV modeling approach by simultaneously fitting planetary masses, orbital elements, resonant angles, and long-term stability, which has since served as a benchmark for dynamical analyses of multi-planet systems.


Beyond resonant orbital configurations, long-term dynamical stability analyses have also been widely employed to constrain planetary system parameters. \citet{Gozdziewski2006} and \citet{Gozdziewski2008} combined Newtonian orbital fitting with stability-constrained optimization by incorporating the Mean Exponential Growth factor of Nearby Orbits (MEGNO) as an additional constraint. \citet{Cloutier2017} reported the RV detection of a non-transiting planet in GJ 1132 and generated planetary orbital solutions by considering the Lagrange stability criterion. \citet{Marshall2020} performed stability analyses for three exoplanet systems and verified the reliability of the angular momentum deficit (AMD) criterion through N-body simulations, providing guidance for future observations. \citet{Trifonov2021} developed the Exo-Striker framework, which enables joint fitting of RV and transit observations and incorporates dynamical stability analyses for multi-planet systems. To overcome the mass–inclination degeneracy inherent in RV measurements, dynamical signatures associated with orbital precession driven by the von Zeipel-Lidov-Kozai (vZLK) mechanism \citep{Zeipel1910,Kozai1962,Lidov1962,Lee2003,Naoz2013} have also been utilized to constrain planetary masses \citep{Rosenthal2019,Jensen2022,Huang2025,Qin2025}.  \citet{Qin2025} derived the true mass of the super-Jupiter HD 41004 Ab, finding $M_{\mathrm{p}} > 3 M_{\mathrm{J}}$ for a high-inclination configuration. Long-term RV monitoring can constrain mutual inclinations and dynamical architectures, providing critical information on planetary masses, internal compositions, and formation pathways. These studies highlight the importance of incorporating dynamical constraints to achieve more accurate planetary mass measurements beyond conventional Keplerian RV analyses.

Except for conducting the pure-RV fitting, the synergy of RV with high precision astrometry measurements \citep{Ji2022,Ji2024,Huang2025} is sufficient to detect weak signals from orbital dynamical evolution, such as the release of Gaia DR4, which will provide the possibility of detecting potential distant companions in the planetary system.

\begin{figure*}
\begin{center}
\includegraphics[width=1.5\columnwidth,height=10cm]{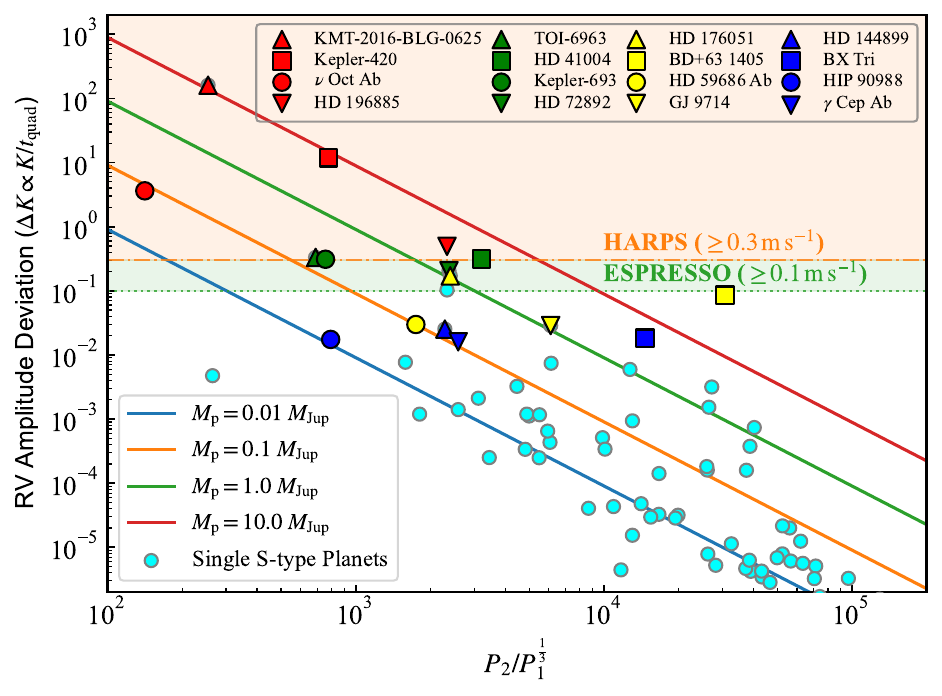}
    \caption{The relationship between $\Delta K$ for single S-type planets and the period ratio function $P_2/P_1^{\frac{1}{3}}$. $\Delta K$ of systems above the green dashed line could exceed the ESPRESSO observation limit. Solid lines correspond to four different sets of continuous initial conditions with $M_{\mathrm{p}}$ = [0.01, 0.1, 1.0, 10.0] $M_{\mathrm{Jup}}$, $M_{\mathrm{A}}$ = 1.0 $M_{\odot}$, $M_{\mathrm{B}}$ = 0.5 $M_{\odot}$, $e_1 = 0.3$, $e_2 = 0.5$ and the mutual inclination $i_{\mathrm{mut}}$ = 60$^{\circ}$. The RV observation baseline $T_{\mathrm{obs}}$ is set to be 30 yrs. Among total 95 confirmed single S-type systems with $a_2 < 100$ au, the primary targets with potentially detected $\Delta K$ are marked with triangles, squares and circles. The magnitude of $\Delta K$ for each system decreases sequentially from top to bottom and from left to right as shown in the upper right legend.}
    \label{fig:dK_distribution}
\end{center}
\end{figure*}

Here, we present a systematic N-body reanalysis of RV observations for 40 confirmed single-planet S-type systems to derive their dynamical planetary masses. By comparing self-consistent N-body and conventional Keplerian models, we quantify how planet–planet dynamical effects influence the inferred masses and bulk densities of planets. Combining RV and transit constraints allows us to break the minimum mass ($M_{\mathrm{p}}\sin i$) degeneracy and recover true planetary masses, providing new constraints on the formation, evolution, and atmospheric properties of planets in binary systems.


The paper is organized as follows. Section \ref{sec:2.1} quantifies the observability of dynamical variations in the RV semi-amplitude $\Delta K$ as a function of planetary mass, binary mass, and the orbital period ratio between the inner planetary orbit and the outer binary orbit. Section \ref{sec:2.2} presents a systematic N-body dynamical analysis of RV observations, from which we derive the dynamical masses of 40 S-type planets, either as $M_{\mathrm{p}}\sin i$ or the true mass $M_{\mathrm{p}}$ when transit constraints are available. Section \ref{sec:2.3} presents the demographic analysis of N-body planetary masses for the S-type planet population.


Section \ref{sec:3} explores the dynamical stability of the TOI-4633 system in highly inclined configurations using a NAFF-based Bayesian indicator. Section \ref{sec:4} summarizes our main findings, and Section \ref{sec:5} discusses future directions, including additional stability diagnostics and the application of N-body modeling to larger astrometric planet samples. The catalog of derived dynamical planetary masses is presented in Appendix \ref{sec:appendixB}.

\section{RV N-body Fitting}\label{sec:2}

\subsection{Observability of RV Amplitude Deviation}\label{sec:2.1}

\begin{table*}[htbp]
\centering
\renewcommand{\arraystretch}{1.2}
	\caption{RV N-body fitting results of S-type planets and the stellar companion in selected four close binaries}
	\label{tab:TOI-6963}
\begin{tabular*}{\textwidth}{@{\extracolsep{\fill}}lcccc@{}}
\toprule
\multirow{2}*{Parameters} & \multicolumn{2}{c}{TOI - 6963 \citep{Barber2024}}& \multicolumn{2}{c}{HD 72892 \citep{Jenkins2017}}\\
\cline{2-3}  \cline{4-5}
 & Kepler Model & N-body Model & Kepler Model & N-body Model \\
 \hline
$K_1(\mathrm{~m} \mathrm{~s}^{-1})$ &
$42.6803^{+	9.8612}_{-9.5927}$ &  $42.5420^{+9.7402	}_{-9.3864}$&
$347.5861^{+5.7754}_{-57.2275}$ &$353.2604^{+3.1651}_{-3.1452}$\\

$P_1 $ (days) &
$8.86^{+0.46}_{-0.52}$ & $8.84^{+0.45}_{-0.51}$&
$39.47^{+0.01}_{-0.42}$ & $39.47^{+3.01E-03}_{-2.95E-03}$ \\

$a_1 $ (au) &
$0.0744^{+0.0025}_{-0.0029}$ &$ 0.0743^{+0.0025}_{-0.0029}$ &
$0.2287^{+3.10E-5}_{-0.0017}$ & $0.2284^{+3.7648E-04}_{-3.4657E-04}$\\

$e_1$ &
$0.0460^{+0.1967}_{	-0.0460}$ & $0.1126^{+0.0943}_{-0.0945}$&
$0.1826^{+0.0087}_{-0.0782}$ &$0.1821\pm	0.0032$\\

$\omega_1$ (deg) &
$37.92^{+123.95	}_{-144.37}$ & $45.26^{+23.97}_{-24.04}$&
$-16.11^{+1.27}_{-36.64}$ & $-15.52^{+0.58}_{-0.59}$ \\

$T_{\mathrm{p,1}}$ (JD) &
$2457055.00^{+4.51}_{-4.26}$ & $2457054.27^{+5.28}_{-3.83}$&
$2455701.47^{+4.55}_{-0.37}$ & $2455701.48\pm0.08$\\

$m_1 \mathrm{sin}i_1$ ($M_{\oplus}$) &
$106.07^{+26.72}_{-25.48}$ & $101.90^{+25.41}_{-24.05}$&
$1691.21^{+	28.19}_{-283.42	}$ & $1719.43^{+15.45}_{-15.35}$\\

$\Delta \mathrm{BIC}$ & \multicolumn{2}{c}{3.783} & \multicolumn{2}{c}{1.510} \\
\toprule
\multirow{2}*{Parameters}& \multicolumn{2}{c}{HIP 90988 \citep{Jones2021}}& \multicolumn{2}{c}{HD 59686 \citep{Ortiz2016}}\\
\cline{2-3}  \cline{4-5}
&Kepler Model & N-body Model & Kepler Model & N-body Model \\
\hline
$K_1(\mathrm{~m} \mathrm{~s}^{-1})$ &
$43.8749^{+	5.5707}_{-0.2795}$ & $44.2435^{+9.0644}_{	-0.6138	}$&
$136.1324^{+1.8098 }_{-4.1017}$ &
$136.7892^{+0.5141	}_{	-0.7788}$ \\

$P_1 $ (days) &
$454.17	^{+13.31}_{-0.23}$ &
$455.29^{+12.56}_{-1.33}$&
$299.45^{+17.54	}_{-0.16}$ &
$299.23^{+0.08}_{-0.06}$\\

$a_1 $ (au) &
$1.2626	^{+0.0246}_{0.0004}$&
$1.2646^{+0.0232}_{-0.0025}$&
$1.0861^{+0.0421}_{-0.0016}$ &
$1.0856\pm0.0002$\\

$e_1$ &
$0.0410^{+0.0670}_{-0.0301}$&
$0.0732^{+0.0620}_{-0.0391}$&
$0.0569^{+0.0346}_{-0.0131}$ & $0.0595^{+0.0059}_{-0.0077}$\\

$\omega_1$ (deg) &
$181.85^{+40.59}_{-14.60}$&
$185.13	^{+	18.02}_{-22.62}$&
$131.4590^{+91.95}_{-16.55}$ &
$132.2567^{+2.1934}_{-0.8446}$\\

$T_{\mathrm{p,1}}$ (JD) &
$2455309.20^{+10.73}_{-0.04	}$&
$2455309.70^{+4.81}_{-0.47}$&
$2451536.33^{+84.83}_{-63.35}$ &
$2451540.89^{+1.95}_{-0.31}$\\

$m_1 \mathrm{sin}i_1$ ($M_{\oplus}$) &
$617.2915^{+85.1050	}_{-4.0369	}$&
$611.5410^{+132.00}_{-9.0731}$&
$2181.0532^{+39.1905}_{-77.7553}$&
$2185.5162^{+8.4187	}_{	-12.6071}$\\

$\Delta \mathrm{BIC}$ & \multicolumn{2}{c}{-7.095} & \multicolumn{2}{c}{9.962} \\
 \hline
\end{tabular*}
\end{table*}

Based on long-baseline RV observations extended to more than 30 years and the utilization of the latest generation of spectrographs, including ESPRESSO and HARPS, with precision of up to 0.50 m/s, {we have an unprecedented opportunity to characterize exoplanet masses and orbital architectures by exploiting their mutual gravitational interactions}.

Previous studies have visualized the theoretical RV signal in the N-body framework and indicated a drift in the RV semi-amplitude $K$ \citep{Huang2025,Qin2025}. High-precision RV data with an accuracy of 1 m/s can detect such dynamical effects. Here we theoretically import the von-Zeipel-Kozai-Lidov timescale \citep{Zeipel1910,Kozai1962,Lidov1962,Naoz2013,Naoz2016} to calculate $\Delta K$. Here $\Delta K$ is the maximum variation of the semi-amplitude of the RV signal drift caused by the companion star during the observation time span $T_{\text{obs}}$, which should be included in the formula of $\Delta K$ to assess the detectability of the binaries' perturbation effect.

According to the equations of motion under the quadrupole perturbation in \citet{Naoz2013}, the orbital precession timescale under the quadrupole level effect is:
\begin{equation}
\label{equ:timescale}
\begin{aligned}
t_{\mathrm{quad}} & \sim \frac{16}{15} \frac{a_2^3\left(1-e_2^2\right)^{3 / 2} \sqrt{M_\mathrm{A}+M_\mathrm{p}}}{a_1^{3 / 2} M_\mathrm{B} k}\\
& =\frac{16}{30 \pi} \frac{M_\mathrm{A}+M_\mathrm{B}+M_\mathrm{p}}{M_\mathrm{B}} \frac{P_2^2}{P_1}\left(1-e_2^2\right)^{3 / 2}
\end{aligned}
\end{equation}
where $M_\mathrm{A}$ is the mass of the primary star, $M_\mathrm{B}$ is the stellar mass of the secondary star, and $M_\mathrm{p}$ is the planetary mass, $k$ equals the square root of the gravitational constant. The indices 1 and 2 represent the planetary orbit and the binary orbit, respectively.

The variation of $e_1$ and $i_1$ is expressed in $t_{\mathrm{quad}}$:

\begin{equation}
\label{equ:de_dt}
\begin{aligned}
C_e &= 2 \sqrt{1-e_1^2} e_1 \text{sin}^2i_{\mathrm{mut}} \text{sin}(2w_1)\\
C_i & = \frac{-e_1^2 \text{sin}(2i_{\mathrm{mut}}) \text{sin}(2w_1)}{\sqrt{1-e_1^2}}
\end{aligned}
\end{equation}

\begin{equation}
\label{equ:di_dt}
\begin{aligned}
\frac{de_1}{dt} &= \frac{C_e}{t_{\mathrm{quad}}}\\
\frac{di_\mathrm{mut}}{dt} &= \frac{C_i}{t_{\mathrm{quad}}},
\end{aligned}
\end{equation}
when $i_2$ = 0 in the invariable plane frame, $i_1 = i_{\mathrm{mut}}$.

Finally, we compute $\Delta K$ with the substitution of $de_1/dt$ and $di_1/dt$ in Equation \ref{equ:di_dt}:

\begin{equation}
\label{equ:deltaK}
\begin{aligned}
 K &= \left(\frac{2 \pi G}{P_1}\right)^{1 / 3} \frac{M_{\mathrm{p}} \sin i_1}{(M_{\mathrm{A}}+M_{\mathrm{p}})^{2 / 3}} \frac{1}{\sqrt{1-e_1^2}},\\
\Delta K &= (\frac{\partial K}{\partial e_1} \frac{de_1}{dt} + \frac{\partial K}{\partial i_1} \frac{di_1}{dt})T_{\text{obs}}\\
& = \left(\frac{e_1}{1-e_1^2}C_e + \text{cot}i_{\mathrm{mut}}C_i\right)\frac{K}{t_{\mathrm{quad}}}T_{\text{obs}}
\end{aligned}
\end{equation}
$G$ is the gravitational constant and $\frac{K}{t_{\mathrm{quad}}} \propto (\frac{P_2}{P_1^{1/3}})^{-2}$.

The maximum $\Delta K$ for single S-type planets with $a_2 < 100$ au are plotted in Figure \ref{fig:dK_distribution}. The solid lines correspond to four different sets of continuous initial conditions with $M_{\mathrm{p}}$ = [0.01, 0.1, 1.0, 10.0] $M_{\mathrm{Jup}}$, $M_{\mathrm{A}}$ = 1.0 $M_{\odot}$, $M_{\mathrm{B}}$ = 0.5 $M_{\odot}$, $e_1 = 0.3$, $e_2 = 0.5$ and the mutual inclination $i_{\mathrm{mut}}$ = 60$^{\circ}$. The RV observation baseline $T_{\mathrm{obs}}$ is set as 30 years. For binaries in blue dots where $e_1$, $e_2$, $i_{\mathrm{mut}}$ are not all known, we made the same assumptions as the solid lines. Cyan dots above the green dashed line have the potential to exceed the ESPRESSO observation limit ($\geq$ 0.1 m/s). Thus, the non-interactive Keplerian orbital model is no longer applicable when considering the precession of planetary orbits in these hierarchical systems.

The N-body RV fitting of six wider close-binary systems: GJ 86, $\tau$ Bootis, GJ 3021, HD 196885, HD 41004 and HD 164509 have been conducted in \citet{Huang2025,Qin2025}, and the minimum mass deviation between the Keplerian and N-body frames could reach 0.2 $M_{\mathrm{Jup}}$. However, before the release of Gaia Data Release 4 \citep{gaia2016,gaia2018,gaia2021}, the true masses of these planets could not be determined, and the astrometry data would also incorporate the effects of N-body perturbations, resulting in deviations in both planetary orbital inclinations and precise mass determinations.

Given the emerging group of transiting S-type planets, this section is dedicated to revealing the dynamical masses of planets in a wider range of binaries, including true planetary masses and minimum planetary masses. To use the simplest hierarchical-triple-body model as the first step, only single-planet systems are resolved in Section \ref{sec:2.2} and Section \ref{sec:2.3}.

\subsection{RV N-body Fitting Results}\label{sec:2.2}

The first group consists of 29 single S-type planets in close-binaries with $a_2 < 100 $ au. Considering the selection effect of transit observations on binaries orbits, 95\% of single transiting S-type planets are located in binaries with $a_2 > 100 $ au. To use a sample as comprehensive as possible to describe the mass-radius relationship of planets under the dynamical model, we also imported 11 planets with precisely knowable planetary radius but with $ 100 < a_2 < 500 $ au. The remaining samples with $a_2 > 500$ au were used solely as background control samples and dynamical retrieval for these planets is outside the scope of this article. The RV data and uncertainties used in this work are all from the original papers listed in the references.

From the angle of fitting models, if the RV observation baseline is $T_{\mathrm{obs}} < P_2/10$, we use the single-planet RV model, if $T_{\mathrm{obs}} < P_2/4$, we fit the RV signal from the binaries as a linear trend factor $\dot{V}$. For $T_{\mathrm{obs}} > P_2/4$, we could make an attempt to fit the complete two-body orbits.

In terms of methodology, the N-body radial Velocity fitting procedures are the same as \cite{Huang2025}. We modified the Markov Chain Monte Carlo (MCMC) sampler $emcee$ \citep{Foreman-Mackey2013} with the incorporation of the REBOUND \citep{Rein2012} N-body package using the IAS15 \citep{Rein2015} integrator. At every single RV observation epoch, we calculated the theoretical RV signal induced by the planet and the outer companion star B. For simplicity, we selected orbital configurations at the first observational epoch as the model parameters. The orbital eccentricities of most selected transiting planets are close to 0, to avoid invalid solutions for $\omega$ caused by an eccentricity approaching zero, we choose basic fitting parameters such as $K_1$, $P_1$, $\sqrt{e}_1\mathrm{cos}\omega_1$, $\sqrt{e}_1\mathrm{sin}\omega_1$, $\tau_{\mathrm{p,1}}$, and the RV offset $v_{\mathrm{off}}$. The prior guess values of these fitting parameters are given by the Keplerian fitting results from the original detection literature, which were also set as the initial conditions of the N-body integrator.


For binary systems with largely unconstrained orbital parameters, we assume $i_{\mathrm{B}} = 45^{\circ}$ and $e_{\mathrm{B}} = 0.3$. The adopted inclination $i_{\mathrm{B}} = 45^{\circ}$ corresponds to the lower threshold for activating secular resonances; a larger inclination would naturally lead to stronger dynamical perturbations and larger deviations in $K$. The binary eccentricity $e_{\mathrm{B}}$ = 0.3 is adopted based on the mean orbital eccentricity of currently detected S-type binary systems.

Given that N-body fitting consumes tens of times more computational time than Kepler fitting, to efficiently conduct the dynamical fitting programs, we employed a unified Gaussian prior model for the fitting parameters. The mean of the Gaussian prior function was derived from the observational data itself. To clarify the difference between the Keplerian model and the N-body model, we define the planetary mass deviation as $\Delta M_{\mathrm{p}} = M_{\mathrm{p,N-body}}-M_{\mathrm{p,Kepler}}$ hereafter.

Given primary targets listed in Figure \ref{fig:dK_distribution} and excluding HD 196885, HD 41004, $\gamma$ Cep, which already have solutions to the dynamical planetary masses in the previous work. The RV fitting results for four relatively stable systems where the perturbation effect is potentially observed are presented in Table \ref{tab:TOI-6963}, including TOI-6963, HD 72892, HIP 90988, HD 59686. The Keplerian and N-body results here are both from this work fitted by the Gaussian prior Keplerian and N-body model. The original detection papers are given by references, fitted RV data and uncertainties used are all from the original papers as well. HD 59686 is a typical case to indicate the difference between the fitting results of the Keplerian model and the N-body model given sufficient long baseline for retrieving the orbit of binary. To compare the performance of the Keplerian model and the N-body model, we used the Bayesian Information Criterion (BIC) \citep{Schwarz1978,Kass1995,Fulton2017},

\begin{equation}
\label{equ:de_dt}
\begin{aligned}
\mathrm{BIC} &=\chi^2+k \ln (N) \\
\Delta \mathrm{BIC} &= \chi^2_{\mathrm{Kep}} - \chi^2_{\mathrm{N-body}}
\end{aligned}
\end{equation}
where $\chi^2$ is the standard weighted residual sum of squares, $k$ is the number of model parameters, and $N$ is the total number of observation data points. Following the conservative criteria proposed by \citet{Kass1995}, a threshold of $\Delta$BIC $\geq 10$ was adopted as decisive evidence to justify the inclusion of dynamical effects. For HD 59686, the two-Keplerian model yields $\chi^2_{\mathrm{Kep}}$ = 21.9, while the planet-companion N-body model reduces the value to $\chi^2_{\mathrm{N-body}}$ = 11.9, resulting in $\Delta$ BIC $\approx 10$. This provides moderate evidence that the RV data favor the dynamically interacting model. TOI-6963 and HD 72892 show weaker signatures of dynamical perturbations, with 1$<$$\Delta$BIC$<10$, suggesting that N-body effects may be present but are not strongly required by the current data. HIP 90988 has $\Delta$BIC of -7.095; however, both models yield poor fits $\chi^2$ over 200, indicating that additional complexities beyond the adopted models may be needed.


For statistical retrieval purposes, we have compiled the Keplerian masses and N-body masses of 40 systems into a catalog of single S-type planets, as shown in the Appendix \ref{sec:appendixB}. Due to the instability of the systems with the set $i_{\mathrm{B}} = 45^{\circ}$ and the insufficient accuracy of the radial velocity data, we were unable to retrieve all 95 cases with our general batch fitting program at the same time.

\subsection{Demography of Planetary Mass}\label{sec:2.3}

\begin{figure*}
\begin{center}
       \subfigure[]{\includegraphics[width=\columnwidth,height=7cm]{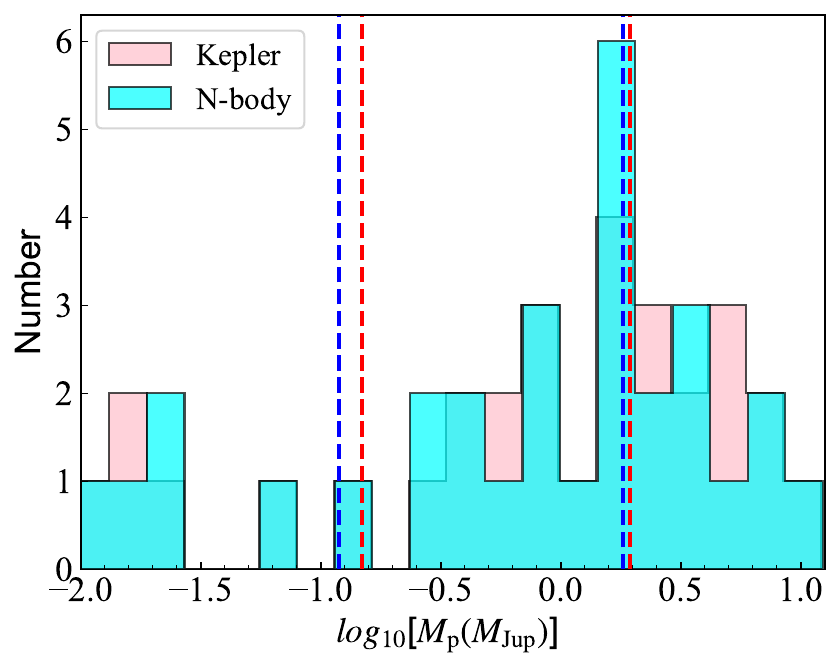}}
       \subfigure[]{\includegraphics[width=\columnwidth,height=7cm]{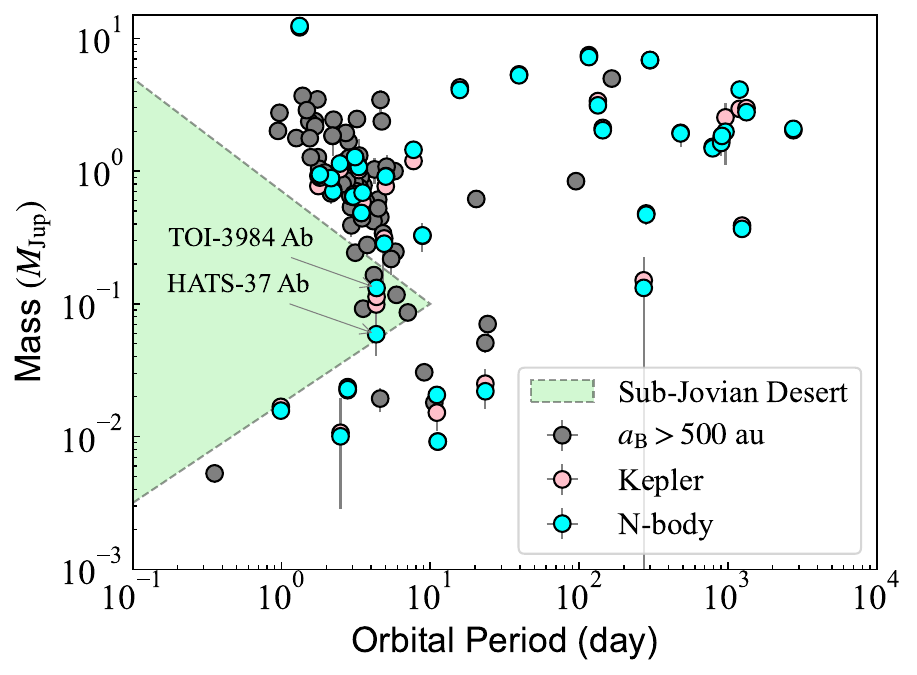}}
    \caption{(a) The histogram plot of the first 30 target S-type planets with $a_2 < 100 $ au in Kepler and N-body model. We utilized the Kernel Density Estimation (KDE) to estimate the regional smoothed peak value of planetary masses. {Left two lines represent the smoothed peak values of $M_{\mathrm{p,Kepler}} = 0.15$ $M_{\mathrm{Jup}}$ (red) and $M_{\mathrm{p,N-body}} = 0.12$ $M_{\mathrm{Jup}}$ (blue) for intermediate-mass planets with $M_{\mathrm{p}} < 1$ $M_{\mathrm{Jup}}$, the decreasing rate is 20\%. The right blue and red dashed lines represent the smoothed regional peak values of $M_{\mathrm{p}}$ for Jovian planets with 1 $M_{\mathrm{Jup}} < M_{\mathrm{p}} < 5$ $M_{\mathrm{Jup}}$, $M_{\mathrm{p,Kepler}} = 1.94$ $M_{\mathrm{Jup}}$ (red) shifts slightly to $M_{\mathrm{p,N-body}} = 1.80$ $M_{\mathrm{Jup}}$ (blue), with a decline rate of 7.2\%. } (b) The mass--orbital period distribution of 29 S-type planets in close-in binaries together with 11 transiting planets in wider binaries. Pink dots for the Keplerian masses and cyan dots for the N-body masses, gray dots in the background refer to the remaining single S-type planets.}
    \label{fig:kepler_density}
    \end{center}
\end{figure*}

In this section, we rebuilt the mass demography of selected single S-type planets in the N-body frame. The main conclusions will be introduced as the relative mass difference between two models and the bidirectional shift trend of short-period intermediate-mass planets.

\begin{figure}
\begin{center}
    \includegraphics[width=\columnwidth,height=7.5 cm]{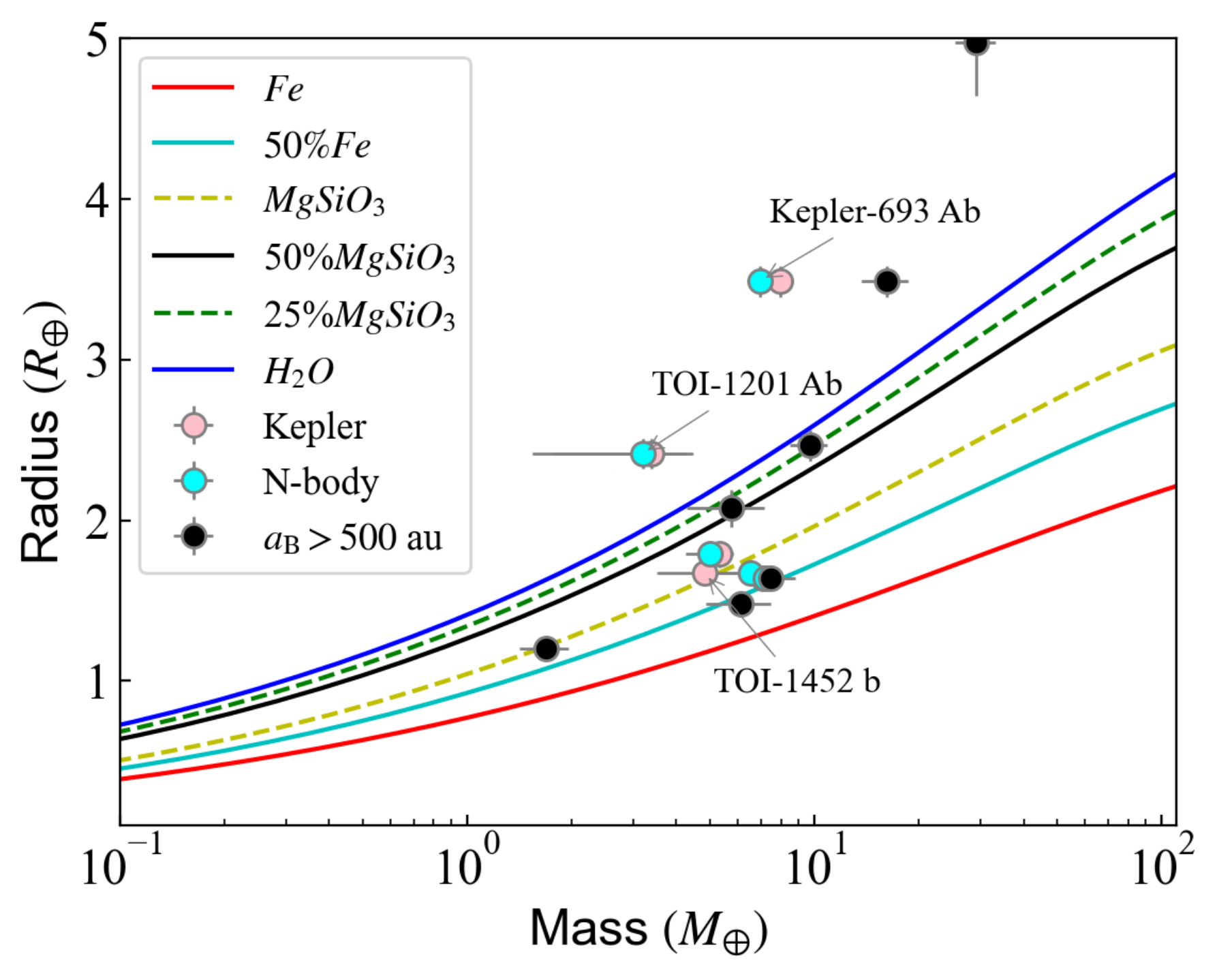}
    \caption{Mass-radius diagram for transiting sub-Jovian S-type planets with masses $< 100$ $M_{\oplus}$. We retrieval planetary masses of single transiting S-type planets in pinks with the N-body model, cyan dots are the dynamical mass for those with same radius. Comparing with the Keplerian model, our N-body fitting results slightly shift TOI-1452 b to right with higher density, but shift TOI-1201 Ab and Kepler-693 Ab to left with lower density. Density curves in different colors are adopted from \citet{Zeng2016}.}
    \label{fig:terr_density}
    \end{center}
\end{figure}

With the dynamical planetary masses derived from Section \ref{sec:2}, Figure \ref{fig:kepler_density} shows the mass distribution of target S-type planets in the Keplerian and N-body models, including the histogram plot in the Keplerian and N-body models in Figure \ref{fig:kepler_density} (a) and the mass--orbital period distribution of 40 targets in Figure \ref{fig:kepler_density} (b).


\citet{Matsakos2016} combined the transit and radial velocity data and found a sub-Jovian desert \citep{Owen2018,Matsakos2016} in short-period orbits. Distributions of the the planetary mass and orbital period in Figure \ref{fig:kepler_density} revealed a consistent feature of close-in sub-Jovian desert as \citet{Matsakos2016}, but for single S-type planets.
In Figure \ref{fig:kepler_density} (a), we utilized the Kernel Density Estimation (KDE) to estimate the regional smoothed peak value of planetary masses. {For intermediate-mass planets with $M_{\mathrm{p}} < 1 $ $M_{\mathrm{Jup}}$, the difference between the peak value of Keplerian and N-body mass is $\Delta M_{\mathrm{p}} = -0.03$ $M_{\mathrm{Jup}}$ and the decreasing rate is 20\%. For Jovian planets with $M_{\mathrm{Jup}} < M_{\mathrm{p}} < 5$ $M_{\mathrm{Jup}}$, $M_{\mathrm{p,Kepler}} $ shifts slightly to $M_{\mathrm{p,N-body}}$ with $\Delta M_{\mathrm{p}} = - 0.13$ $M_{\mathrm{Jup}}$ with a decreasing rate of 7.2\%.}

{Our results in Figure \ref{fig:kepler_density}(b) indicate that short-period sub-Jovian planets with $P_1 \sim 10$ days are more susceptible to gravitational perturbations from their stellar companions. In the Keplerian model, the perturbation-induced deviations in the RV signal are treated as the constant parameters ($K_1$, $e_1$), whereas the N-body model accounts for these perturbations more self-consistently. The dynamical fitted $K_1$ and $e_1$ are adjusted frequently in real time to simulate both the constant part at the first epoch and the remaining oscillating part during orbital dynamics evolution, especially leading to both positive and negative shifts in the inferred planetary masses for intermediate-mass planets. In contrast, for more massive Jovian planets, the RV signal is dominated by the planet itself, and the perturbation from the stellar companion represents a relatively small contribution, resulting in only a minor shift between the Keplerian and N-body solutions.}

According to the distribution of pink dots in the Keplerian model of Figure \ref{fig:kepler_density} (b), TOI-3984 Ab and HATS-37 Ab are located in the middle of sub-Jovian desert, while TOI-3984 Ab moves to the upper limit and HATS-37 Ab moves to the lower limit of the region in the N-body model. {Therefore, once the binary orbits are accurately characterized in the future, these two cases of observed sub-Jovian planets could potentially be shifted out of the desert region in the N-body framework due to the change in the inferred planetary masses.}

Deviations of the masses of giants are induced by distances, eccentricities, mutual inclinations of binary orbits, and the stellar mass of companions. Considering the differences in the sensitivity of various systems to binary dynamics, only the regional difference of planetary mass are given for Jovian planets and sub-Jovian planets. {We conclude that giant planets with $\Delta M_{\mathrm{p}} < 0$ may have had their masses overestimated in previous Keplerian analyses, because part of the inferred planetary mass could be attributed to dynamical effects arising from orbital interactions. For transiting planets with measured radii, an overestimated mass would imply lower bulk densities than previously inferred.}

\begin{figure*}
\begin{center}
       \subfigure[]{\includegraphics[width=\columnwidth,height=7cm]{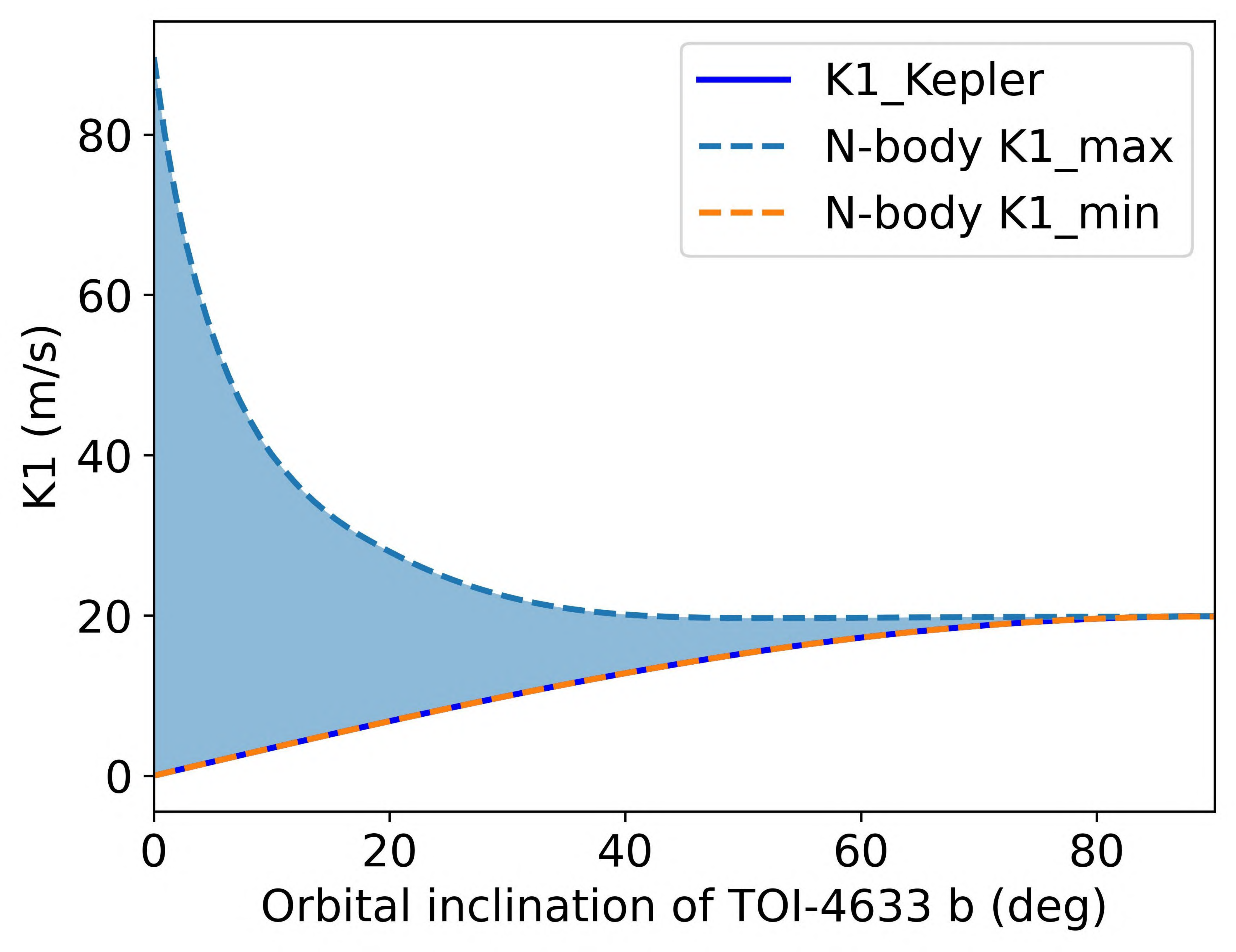}}
         \subfigure[]{\includegraphics[width=\columnwidth,height=7cm]{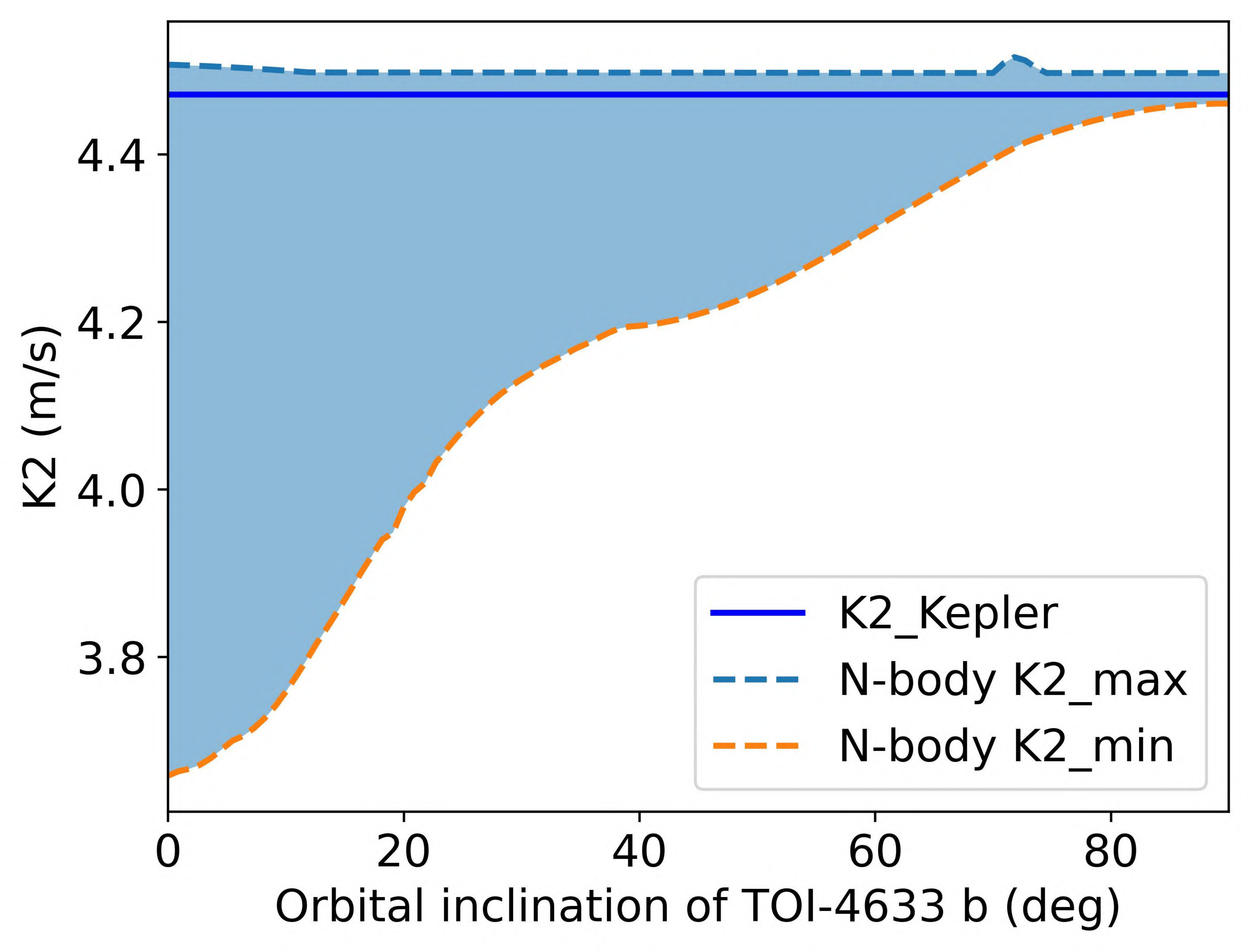}}
    \caption{The RV semi-amplitude K  induced by planets TOI-4633 b and TOI-4633 c under the perturbation of a stellar companion. Selected as the most extreme case of orbital structures, in panel (a), the maximum deviation between the Keplerian model and the N-body model $\Delta K_b= 90$ m/s when  $i_{\mathrm{b}} = 0^{\circ}$, in panel (b), the maximum deviation between the Keplerian model and the N-body model $\Delta K_c= 0.75$ m/s when $i_{\mathrm{b}} = 0^{\circ}$. }
    \label{fig:TOI_4633_K}
\end{center}
\end{figure*}

\citet{Gaidos2024} constructed the mass-radius diagram of 15 small planets with $R_{\mathrm{p}} < 8$ $R_{\oplus}$ with $a_2 < 500$ au, which is indistinguishable from that of planets orbiting single stars.



To clearly visualize the effect of density shift in the N-body frame, we plot mass-radius distributions for sub-Jovian planets in Figure \ref{fig:terr_density}. Compared with the Keplerian solutions, the N-body model slightly shifts TOI-1452 b to the right with higher density, but shifts TOI-1201 Ab and Kepler-693 Ab to the lower side of density. Density curves in different colors are adopted from \citet{Zeng2016}.

\section{Stability Constrains on mutual inclinations}\label{sec:3}

\subsection{The NAFF Indicator}\label{subsec:3.1}
\begin{figure*}
\begin{center}
\includegraphics[width=1.5\columnwidth,height=10cm]{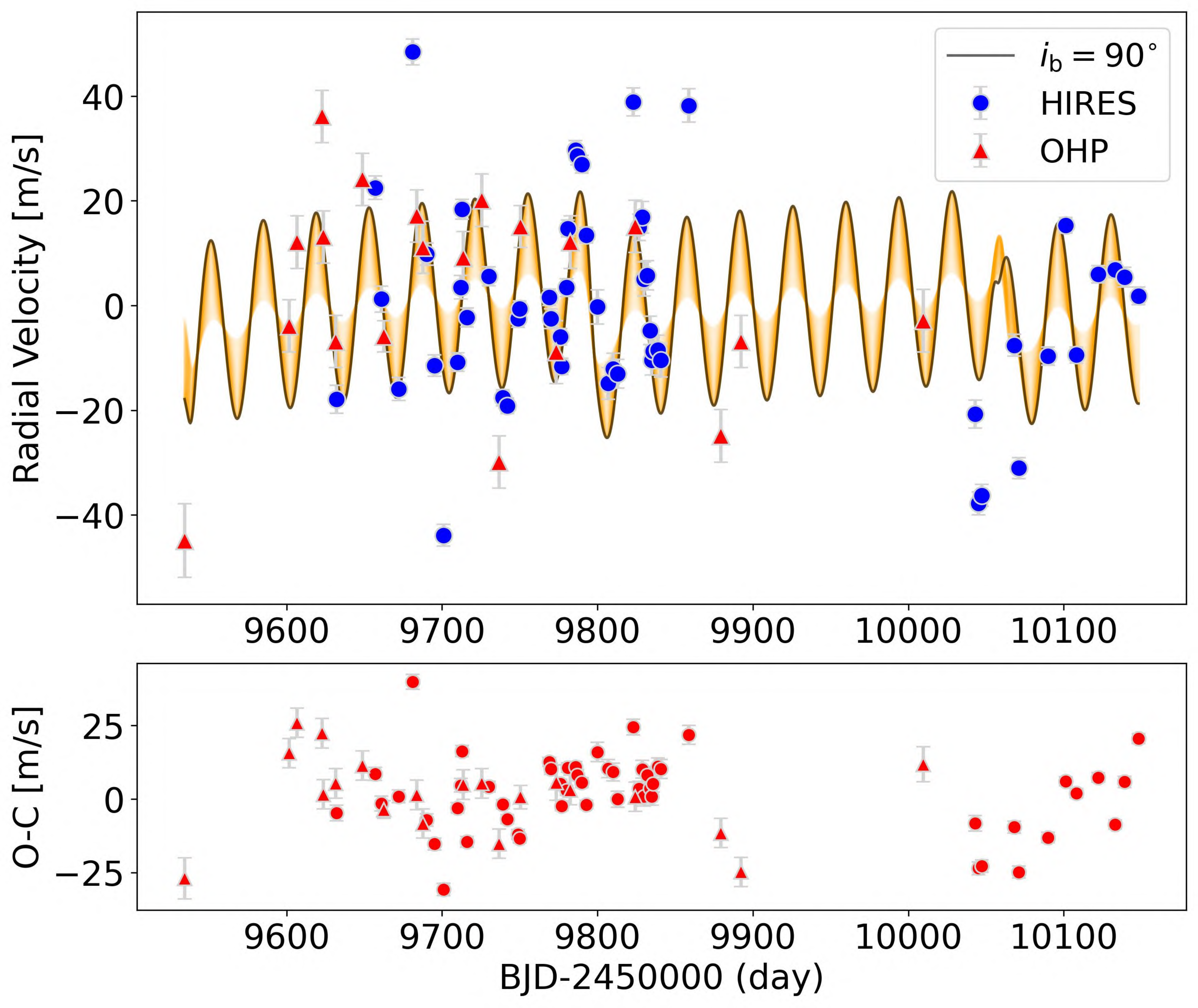}
    \caption{N-body RV fitting results of TOI-4633 with two planets: $M_{\mathrm{b}} = 98.86$ $M_{\oplus}$, $M_{\mathrm{c}} = 41.95$ $M_{\oplus}$. Red triangles and blue dots are the observed RV data from OHP and HIRES. The black solid line is the best-fit curve of the system with $i_b = 90^{\circ}$, other cases with $i_b$ uniformly distributes in ($10^{\circ} - 90^{\circ}$) are plotted in orange for comparison.}
    \label{fig:TOI_4633_fit}
\end{center}
\end{figure*}

According to \citet{Zhang2026}, more than 12 transiting S-type planets around target stars in the TOIs catalog were detected to have an outer companion of a planet-mass or brown dwarf. \citet{Kane2014} conducted a dynamic analysis on four multi-planet systems with extreme orbital eccentricity diversity ($e_{\mathrm{in}} > 0.5$ while $e_{\mathrm{out}} \sim 0$ ), revealing the stability constraints that medium-mass planets with $M_{\mathrm{p}} > 0.2 -0.5$  $M_{\mathrm{J}}$ and $ P < 1000$ days are less possible to embed in the system. Furthermore, the eccentric oscillations and precession periods can last for hundreds of years, which is feasible to capture with the current observational precision of ESPRESSO.

The Numerical Analysis of Fundamental Frequencies (NAFF) indicator \citep{Laskar1990,Laskar1993,Correia2005} was first introduced by \citet{Laskar1988} to be applied in the investigation of the secular evolution of the solar system via a semi-analytical approach. The NAFF indicator is built on the basis of frequency analysis. It was used to estimate the main frequencies in a high precision dynamical system. A more rigorous process of NAFF calibration was proposed by \citet{Couetdic2010}. Combining Bayesian statistics with the fast chaos index NAFF significantly improves the accuracy of the planetary orbital parameters and mass constraints \citep{stalport2022}.

We adopted the formula for the calculation of the NAFF Bayesian criterion \citep{stalport2022}, where the mean motion drift of every planetary orbit is calculated. After normalization of  the drift between planetary orbits, the NAFF stability indicator is expressed as

\begin{equation}
    \mathrm{NAFF}=\max _j\left[\log _{10} \frac{\left|n_{j, 2}-n_{j, 1}\right|}{n_{j, 0}}\right],
\end{equation}
where $n_{j,1}$ and $n_{j,2}$ are the estimated mean motions during the first and second half of the integration time for every planet $j$ in the system, and $n_{j,0}$ is the initial Keplerian mean motion of planet $j$.

NAFF indicates the mean motion drift of every planetary orbit, the value marks the start of the weakly chaotic distribution. According to \citet{stalport2022}, the NAFF diffusion index does not have a universal critical value that separates stable and unstable systems. Instead, the stability threshold is determined empirically from the distribution of the diffusion index for the ensemble of sampled orbital solutions. The histogram of the diffusion index first exhibits a dominant peak at small frequency variations, corresponding to quasi-periodic orbits whose frequency changes are primarily due to numerical errors and finite integration time. At larger diffusion indices, the distribution is associated with orbits undergoing significant frequency diffusion, indicative of resonant or chaotic evolution. Therefore the stability threshold is located at the first local minimum separating these two populations. This choice minimizes the overlap between regular and chaotic trajectories and has been widely adopted in frequency-map analysis as an empirical criterion for classifying orbital stability.

The efficient NAFF Bayesian criterion is combined with the MCMC RV fitting results in this section to provide novel evidence on the stability of the specific S-type planetary system TOI-4633. For single-planet binary systems, the stable posterior distributions of eccentricity and orbital inclinations are constrained by the mutual inclination between planet b and star B. The prior condition could be selected as the uniform distribution of mutual inclination between planet and stellar companion $i_{\mathrm{mut}}=10^{\circ} - 90^{\circ}$. In Section \ref{subsec:3.2}, we concern the orbital inclination dispersion of two planets in the binary system TOI-4633. The prior condition was selected as the uniform distribution of $i_{\mathrm{b}}=10^{\circ} - 90^{\circ}$, in the presence of star B.

\subsection{Stable Criteria for Mutual Inclination in TOI-4633}\label{subsec:3.2}

\begin{table*}
\centering
	\caption{Binary system parameters of TOI-4633}
	\label{tab:TOI-4633}
	\begin{tabular}{lcccc}
		\hline\hline
        \multicolumn{5}{c}{Binary parameters \citep{Bailer2018,Eisner2024}}\\
        \hline
        Distance (pc) & \multicolumn{4}{r}{$95.20 \pm 0.24$} \\
		$M_{\mathrm{A}}\left(M_{\odot}\right)$ & \multicolumn{4}{r}{$1.10 \pm 0.06$}  \\
        $M_{\mathrm{B}}\left(M_{\odot}\right)$ & \multicolumn{4}{r}{$1.05 \pm 0.06$}  \\
        Semi-major axis $a_{\mathrm{B}}$ (au) & \multicolumn{4}{r}{$48.6_{-3.5}^{+4.4}$}\\
        Eccentricity $e_{\mathrm{B}}$ & \multicolumn{4}{r}{$0.91_{-0.03}^{+0.03}$}\\
        Inclination $i_{\mathrm{B}}$ (deg) & \multicolumn{4}{r}{$90.1_{-0.4}^{+0.4}$}\\
        $\omega_{\mathrm{B}}$ (deg) & \multicolumn{4}{r}{$110.5_{-2.1}^{+2.1}$}\\
        $\Omega_{\mathrm{B}}$ (deg) & \multicolumn{4}{r}{$123.5_{-2.9}^{+3.3}$}\\
        Orbital Period $P_{\mathrm{B}}$(years) & \multicolumn{4}{r}{$231_{-24}^{+32}$}\\
        \hline
         Planetary parameters& \multicolumn{2}{c}{TOI-4633 b} & \multicolumn{2}{c}{TOI-4633 c}\\
         &\citet{Eisner2024} & N-body (this work) & \citet{Eisner2024} & N-body (this work)\\
         \hline
        Orbital period P($\mathrm{days}$) & $34.15 \pm 0.15$ & $34.0364^{+0.2185}_{-0.1238}$ &$271.9445_{-0.0040}^{+0.0039}$ & $271.6947^{+0.3818}_{-0.9001}$ \\
        Doppler semi-amplitude $K$ (ms$^{-1}$)& $19.97_{-2.30}^{+2.29}$ & $18.37^{+0.73}_{-0.80}$ & $4.58_{-2.29}^{+2.56}$ & $7.02^{+8.87}_{-1.44}$ \\
        Time of periastron $\tau_{p}$ (JD)& $2459796.64_{-1.19}^{+1.10}$ & $2459543.1119_{-7.0921}^{+2.1199}$ & -- & $2459537.5480_{-4.3127}^{+0.7351}$\\
        Semi-major axis $a$ (au) & -- & $0.2122^{+0.0009}_{-0.0005}$ & $0.847 \pm 0.061$ & $0.8475^{+0.0008}_{-0.0019}$ \\
        Eccentricity $e$ & $0.096_{-0.065}^{+0.102}$ & $0.0004^{+0.0019}_{-0.0003}$ & $0.117_{-0.085}^{+0.186}$ & $0.6920^{+0.0715}_{-0.0864}$\\
        Argument of periastron $\omega$ (deg) & $-43.9_{-72.8}^{+104.8}$ & $275.3734	^{+63.7950}_{-42.3352}$  & $-21_{-108}^{+131}$ & $124.3109^{+2.7395}_	{-3.7384}$\\
        $M_{\mathrm{p}}$sin$i$ $(M_{\oplus})$ & $106.8_{-12.8}^{+13.0}$ & $114.27		^{+4.20}_{-4.42}$ &  $47.8_{-23.8}^{+27.6}$& $41.95^{+94.27}_		{-3.20}$\\
         Inclination $i$ (deg) & -- & -- & $89.888_{-0.064}^{+0.069}$ & -- \\
        \hline
        \multicolumn{5}{c}{Other parameters} \\
        \hline
        OHP/SOPHIE RV-offset ( $\mathrm{m} \mathrm{s}^{-1}$ ) & \multicolumn{2}{c}{$2.3_{-2.0}^{+1.9}$} & \multicolumn{2}{c}{$-2.50_{-3.04}^{+2.26}$}\\
        Keck/HIRES RV-offset ( $\mathrm{m} \mathrm{s}^{-1}$ ) & \multicolumn{2}{c}{$-0.7 \pm 3.4$} & \multicolumn{2}{c}{$-0.22_{-0.74}^{+0.80}$}\\
		\hline
	\end{tabular}
\end{table*}

TOI-4633 is one of the S-type systems with the most compact orbit of binaries. TOI-4633 A is a bright G-type star (Vmag = 9.0) located at a distance of $95.2 \pm 0.24$ pc with stellar mass of $1.1 \pm 0.06$ $M_{\odot}$. The properties of the binary system are listed in Table \ref{tab:TOI-4633}.  Continual observations of TESS conducted by the Planet Hunters TESS citizen science project validate one Neptune-size planet TOI-4633 c \citep{Eisner2024}, and the total available transit data confirm that the transiting object (TOI-4633 c) has an orbital period of $\sim$ 271.94 days.

Combination of the follow-up spectroscopic data and new high-resolution spectral imaging data revealed a bound companion star TOI-4633 B, and the brightness ratio of the two stars equals 0.7. As constrained by Bayesian parameter estimation, the semi-major axis of the binary orbit is $a_{B} = 48.6_{-3.5}^{+4.4}$ au, the eccentricity is $e_{\mathrm{B}} = 0.91\pm{0.03}$, and the binary orbital inclination is determined as $i_{\mathrm{B}} = 90.1^{\circ}\pm0.4$. The orbital period of star B is derived to be $231_{-24}^{+32}$ yr.

According to \citet{Eisner2024}, TOI-4633 hosts another Jupiter-sized planet candidate TOI-4633 b. The accurate eccentricity and the inclination of planet b is still to be decided due to the gravitational interaction between planets and from the companion star B. In Figure \ref{fig:TOI_4633_K}, we calculate the semi-amplitude of the radial velocity induced by both TOI-4633 b and TOI-4633 c under the perturbation of a stellar companion. The maximum deviation between the Keplerian model and the N-body model $\Delta K_b= 90$ m/s and $\Delta K_c = 0.75$ m/s when $i_{\mathrm{b}} = 90^{\circ}$.

\begin{figure*}
\begin{center}
\includegraphics[width=2\columnwidth,height=12cm]{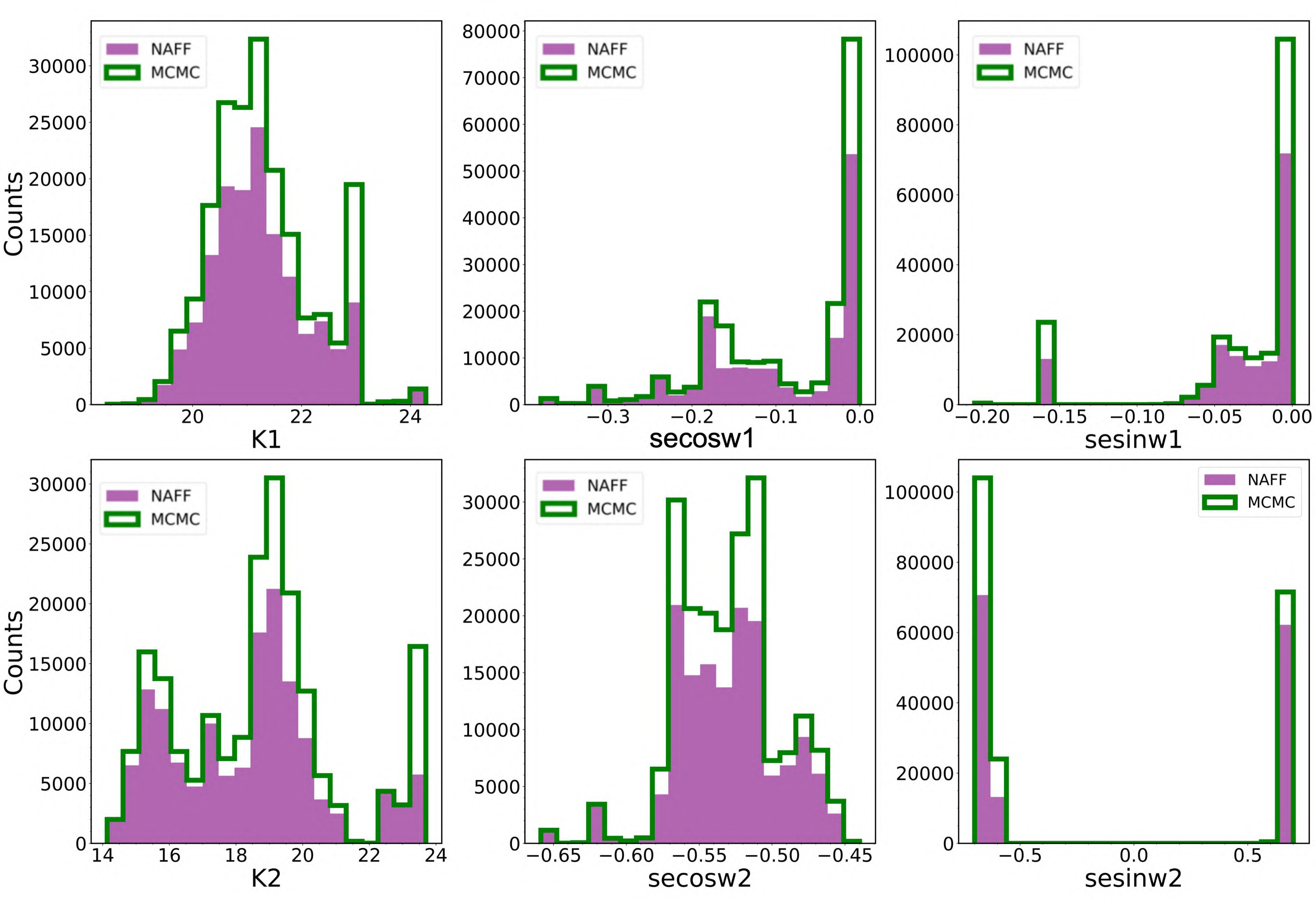}
    \caption{Contradistinction before and after the NAFF stability analysis on the MCMC fitting results. Green hollow histograms: the posterior distribution of MCMC RV fitting parameters before stability analysis. Purple filled histograms: the posterior distribution of RV fitting parameters for both planets in TOI-4633 system with the condition of $i_{\mathrm{mut}} = 20^{\circ}$ and the constrains of NAFF stability threshold shrink the range of both ($K_1$, $e_1$) and ($K_2$, $e_2$). Note: secos$\omega$1 = $\sqrt{e_1}$cos$\omega_1$, sesin$\omega$1 = $\sqrt{e_1}$sin$\omega_1$, secos$\omega$2 = $\sqrt{e_2}$cos$\omega_2$, sesin$\omega$2 = $\sqrt{e_2}$sin$\omega_2$.}
\label{fig:TOI4633_NAFF_stab_posterior}
\end{center}
\end{figure*}

The results of the N-body radial velocity fitting of TOI-4633 with two planets are implemented in Figure \ref{fig:TOI_4633_fit}, which annotates the fitted planetary masses $M_{\mathrm{p}} = 98.86$ $M_{\oplus}$, $m_c = 41.95$ $M_\oplus$. The red triangles and blue dots are the RV data observed from OHP and HIRES. The solid black line is the best-fit curve of the system with $i_b = 90^{\circ}$, other cases with $i_b$ uniformly distributed in ($0^{\circ} - 90^{\circ}$) are colored orange for comparison.

The specific results of the NAFF stability indicator in the TOI-4633 system are listed in the Appendix \ref{sec:appendixA}. For each case of the mutual orbital inclination between planet b and planet c in $i_{\mathrm{mut}} = 10^{\circ} - 90^{\circ}$, the distribution of NAFF and the chaos threshold are presented in Figure \ref{fig:TOI4633_NAFF} in the Appendix \ref{sec:appendixA}. When $i_{\mathrm{mut}} = 10^{\circ}$ and $20^{\circ}$, the NAFF stability indicator is close to -2.97, while for other cases with $i_{\mathrm{mut}} \in [30^{\circ}, 90^{\circ}]$, NAFF = -2.96.

We use the indicator of NAFF = -2.97 for $i_{\mathrm{mut}} = 10^{\circ}$ and $20^{\circ}$, then set the last $2\times10^6$ MCMC samples of orbital configurations as initial conditions for orbital integration. After integrating the binary planet systems for $10^3$ times of the inner planet's orbital period, if NAFF $< -2.97$, {orbits are relative more regular with a smaller frequency diffusion index and weaker chaos}. The simulation results are plotted in Figure \ref{fig:TOI4633_NAFF_stab_posterior}, which indicates the consistent posterior distributions before and after the stability analysis with $i_{\mathrm{mut}} = 20^{\circ}$, except that $K_2$ being more convergent and moves slightly to the left.

When the mutual inclination between the planetary orbits is equal to $30^{\circ}-90^{\circ}$, for one simulation with an integration time of $10^3$ times the orbital period of the inner planet, we use the NAFF threshold of -2.96. The orbits were stable under the perturbation of NAFF $< -2.96$. We plot the posterior distributions of RV fitting parameters for both planets as well. Comparing the distributions before and after NAFF filtering, the main differences are observed on the orbital eccentricities and planetary masses, which is consistent with the conclusions about the HD 37124 system in \citet{stalport2022}. Thus, we plot the results of $K_1$, $\sqrt{e_1}$cos$\omega_1$, $\sqrt{e_1}$sin$\omega_1$, $K_2$, $\sqrt{e_2}$cos$\omega_2$, $\sqrt{e_2}$sin$\omega_2$ here. More cases with mutual inclinations of $70^{\circ}$ and $90^{\circ}$ are shown in the Appendix \ref{sec:appendixA}. When $i_{\mathrm{mut}} < 70^{\circ}$, the posterior filtered distributions of all fitting parameters are consistent with the case of $i_{\mathrm{mut}} = 20^{\circ}$ in Figure \ref{fig:TOI4633_NAFF_stab_posterior}, when $i_{\mathrm{mut}} = 90^{\circ}$, the posterior distribution of both ($K_1$, $e_1$) and ($K_2$, $e_2$) is significantly reduced under the constraint of the NAFF criterion. Thus, we conclude that the mutual inclination between TOI-4633 b and c is likely less than $70^{\circ}$ inferred from the NAFF stability indicator.

NAFF indicator is particularly efficient for screening the chaos appearing before instability in $10^6$ posterior orbital solutions generated by our MCMC sampling. Our main conclusion is that TOI-4633 system is exhibiting chaotic diffusion first $10^3$ orbits when the mutual planetary inclination $i_{\mathrm{mut}}  > 70^{\circ}$ based on the N-body fitting model. Thus $i_{\mathrm{mut}} = 70^{\circ}$ is the upper limit for fast rejection of unstable orbital solutions.

\section{Conclusions}\label{sec:4}

Recent observational advances from TESS and Gaia have significantly improved our understanding of exoplanetary systems, especially through the rapid growth in the detection and characterization of binary and multi-body systems in recent years. Using the latest catalog of transiting single S-type planets, we performed N-body dynamical analyzes to determine planetary masses for single-planet transiting S-type systems and investigated the long-term stability of the two-planet system TOI-4633.

Section \ref{sec:2.1} first derived the theoretical relationship between $\Delta K$ and the period ratio function $P_2/P_1^{\frac{1}{3}}$ for single S-type planets with $a_\mathrm{2} < 100$ au. $\Delta K$ in systems including KMT-2016-BLG-0625, Kepler-420, $\nu$ Oct, HD 196885, TOI-6963, HD 41004, Kepler-693 and HD 72892 are probably detectable in RV observations.

In Section \ref{sec:2.2}, we develop a Bayesian N-body dynamical fitting framework by combining MCMC sampling with direct N-body integrations to constrain the dynamical masses of 40 selected single-planet systems in close-in binary architectures. Compared with conventional Keplerian models, the inclusion of dynamical perturbations reduces uncertainties of the posterior model parameters. Using the Bayesian Information Criterion (BIC), we identify the necessity to include the dynamical RV model in HD 59686. The best-fitting RV solutions and the complete catalog of derived planetary parameters are listed in Table \ref{tab:catalog} in Appendix \ref{sec:appendixB}.

{In Section \ref{sec:2.3}, we apply the dynamical fitting results to revise the demographics of single S-type planets. For sub-Jovian planets ($M_{\mathrm{p}} < 1~M_{\mathrm{Jup}}$) with orbital periods of around 10 days, the planetary masses inferred from Keplerian models decrease by approximately 20\% when adopting the N-body model. In comparison, the relative mass difference between the two models is only 7.2\% for Jovian planets. Therefore, short-period sub-Jovian planets with $P_1 \sim 10$ days are more susceptible to changes in the inferred orbital parameters when gravitational perturbations from the stellar companion are considered.}  TOI-3984 Ab and HATS-37 Ab are located in the sub-Jovian desert, TOI-3984 Ab's mass shifts toward the upper boundary and HATS-37 Ab shifts to the lower end. This displacement may suggest a relatively more inflated structure of HATS-37 Ab than previously thought.

{As discussed above, short-period sub-Jovian planets show the largest deviations between Keplerian and N-body solutions, suggesting that their inferred masses are particularly sensitive to dynamical interactions and can be significantly improved through N-body modeling. Nevertheless, the necessity of N-body fitting depends on the planetary regime. For low-mass rocky planets or distant massive gas giants, the perturbation induced by the stellar companion is expected to be relatively weak, and Keplerian modeling may remain sufficient. Therefore, the aim of this work is not to establish a universally superior model for all exoplanets, but rather to identify the planetary populations for which gravitational perturbations introduce significant biases in orbital and mass determinations and where N-body modeling becomes essential.}

In Section \ref{sec:3}, we extend the dynamical framework to investigate the stability of the two-planet system TOI-4633, providing a methodology that can be further applied to multi-planet systems in binary stars. The NAFF-based stability criterion effectively constrains the range of stable orbital configurations obtained from the MCMC fitting under different mutual-inclination scenarios. Our results suggest that the mutual inclination of the TOI-4633 system is likely lower than $70^{\circ}$. However, accurate orbital inclinations require future confirmation through high-precision astrometric observations.
$i_{\mathrm{mut}} = 70^{\circ}$ is the upper limit for fast rejection of unstable orbital solutions.

\section{Discussion} \label{sec:5}

In this work, we show that the inferred dynamical masses of individual transiting S-type planets can be significantly affected by the properties of their stellar companions, including companion mass, orbital separation, and eccentricity. {Although the current statistical mass -- radius analysis reveals only weak individual-level differences in planetary masses,} future surveys with larger samples and improved dynamical constraints will be essential for identifying systematic biases in planetary mass determinations. {Furthermore, the sub-Jovian desert is not entirely devoid of planets, and the revised masses and orbital parameters obtained from N-body modeling may reveal additional planets occupying this sparsely populated region. Therefore, constructing a more complete sample of short-period exoplanets near the sub-Jovian desert will be essential for future dynamical retrieval studies.}

The effects of uncertainties in stellar masses and planetary eccentricities on the inferred planetary masses are beyond the scope of this work. For clarification, we take HD 72892 in Table \ref{tab:TOI-6963} as an example. Assuming the current 5\% uncertainty in the stellar mass, the corresponding uncertainty in $\Delta K$ is 0.0102 m/s. With the initial conditions of $K_1=347.5861$ m/s, $M_{\mathrm{A}}=1.02M_{\odot}$, $P_1=39.47$ days, $e_1=0.1826$, and $i_1=\pi/2$, the inferred planetary mass $M_{\mathrm{p}}$ changes by 3.29\% according to Equation \ref{equ:deltaK}. This demonstrates that the uncertainty in the stellar mass can introduce a substantially larger effect on the inferred planetary mass than the direct variation in $\Delta K$ itself. If the stellar mass is fixed, an eccentricity uncertainty of $\sigma_{e_1}=0.1$ induces an uncertainty in $\Delta K$ of 0.0494 m/s. With the initial conditions of $K_1=347.5861$ m/s, $M_{\mathrm{A}}=1.02M_{\odot}$, $P_1=39.47$ days, $e_1=0.1826$, and $i_1=\pi/2$, the inferred planetary mass changes by 3.73\% according to Equation \ref{equ:deltaK}. This estimate indicates that uncertainties in planetary eccentricity can also have a significant impact on the inferred planetary mass.

Furthermore,  we consider other facts that could induce the planetary mass deviation from RV measurements,  including the potential distant planet and the stellar activities. The Lomb-scargle (LS) periodogram analysis of fitted systems shows little prefer of the second planet. Inferring from the bisector span (BIS) in the original RV observation paper,  the differential effect likely does not arise from the stellar activity, which has been subtracted during the detrending process of the raw RV data. HIP 94235 is one of the systems with strong pollutions from stellar activity, which is moving in the AB Doradus group \citep{zuckerman2004}. {The planetary mass limit of $M_{\mathrm{p}} < 390~M_{\oplus}$ obtained by \citet{zhou2022} is a $3\sigma$ upper bound from the original analysis, corresponding to an uncertainty of $1\sigma = 111~M_{\oplus}$}. In Table \ref{tab:catalog}, we also give the Keplerian and N-body planetary mass with $3\sigma$ upper limit derived in this work, without subtracting the effect of stellar activity. However the actual planetary mass of HIP 94235 is estimated as $0.0352 \pm 0.0044$ $M_{\mathrm{Jup}}$ \citep{zhou2022} by further observations of transit.

Future long-term RV monitoring with extended baselines and improved precision will allow tighter constraints on the dynamical mass deviations identified in this work. The forthcoming Gaia DR4 release is expected to substantially increase the number of known exoplanet systems in binary environments and provide direct mass constraints for a larger population of planets. Combining these observations with advanced N-body fitting frameworks, such as $orvara$ \citep{Brandt2021}, Exo-striker \citep{Trifonov2019}, or Nii-body \citep{Jia2026}, will enable the construction of a comprehensive dynamical mass catalog, offering valuable constraints on planetary formation and internal structure theories in close-binary systems. Furthermore, for multi-planet systems, the joint analysis of high-precision RV and TTV observations will provide a powerful approach to revealing additional planetary companions through their mutual gravitational interactions.

\section*{Acknowledgements}
We sincerely thank the Scientific Editor and the anonymous reviewers for their constructive comments and insightful suggestions, which have significantly improved the quality and clarity of this manuscript. This work is financially supported by the National Natural Science Foundation of China (grant Nos.12533011,12033010,12473076), the CAS Scholarship, the Strategic Priority Research Program on Space Science of the Chinese Academy of Sciences (Grant No. XDA 15020800), the Natural Science Foundation of Jiangsu Province (Grant No. BK20221563), the Foreign Expert Project (Grant No. S20240145), the Foundation of Minor Planets of the Purple Mountain Observatory and the Jiangsu Funding Program for Excellent Postdoctoral Talent. This work has been carried out within the framework of the NCCR PlanetS supported by the Swiss National Science Foundation under grant 51NF40\_205606.

\section*{Data Availability}
The radial velocity data analyzed in this work are adopted from the original papers listed in Table \ref{tab:catalog} and are publicly available from the Strasbourg astronomical Data Center via \url{https://cds.unistra.fr}. The derived orbital parameters will be archived and made available online.

\software{emcee \citep{Foreman-Mackey2013},
          REBOUND \citep{Rein2012},
          matplotlib \citep{Hunter2007},
          Scipy \citep{Virtanen2020},
          Astropy \citep{astropy2013}}
\bibliography{ms}{}
\bibliographystyle{aasjournal}

\appendix\label{sec:Appendix}

\section{Computation of NAFF Chaos Threshold}\label{sec:appendixA}

\begin{figure*}[h]
\includegraphics[width=\columnwidth,height=15cm]{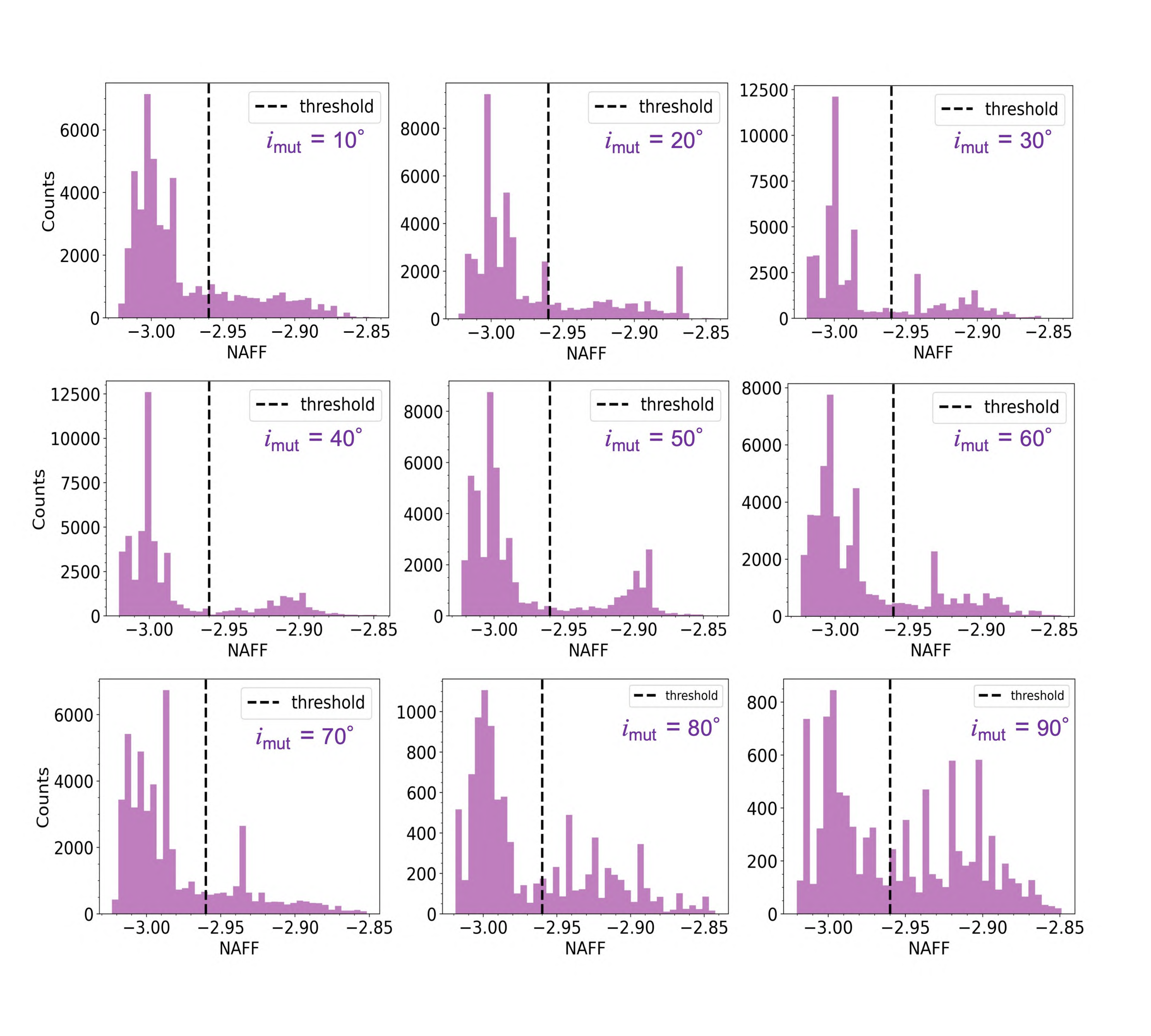}
    \caption{Distribution of the NAFF chaotic threshold for TOI-4633 planetary system with different cases of $i_{\mathrm{mut}}$ between planet b and planet c. The histograms are drawn by integrating $5 \times 10^5$ MCMC samples from N-body RV fitting for one thousand orbits of the inner planet b. NAFF threshold is the value above which the variations of the estimated mean motions are considered too large for the system to be stable on long timescales, when $i_{\mathrm{mut}}$ > 80$^{\circ}$, the emerging narrow peaks are not Gaussian noise, instead they could be discrete clusters of NAFF frequencies due to the extreme chaotic orbital parameter spaces. The fact that this is strongest for $i_{\mathrm{mut}}$ = 90$^{\circ}$ also suggests the dynamics are producing several preferred secular frequencies rather than one smooth distribution.}
    \label{fig:TOI4633_NAFF}
\end{figure*}
\clearpage

\begin{figure*}
\begin{center}
\includegraphics[width=\columnwidth,height=12cm]{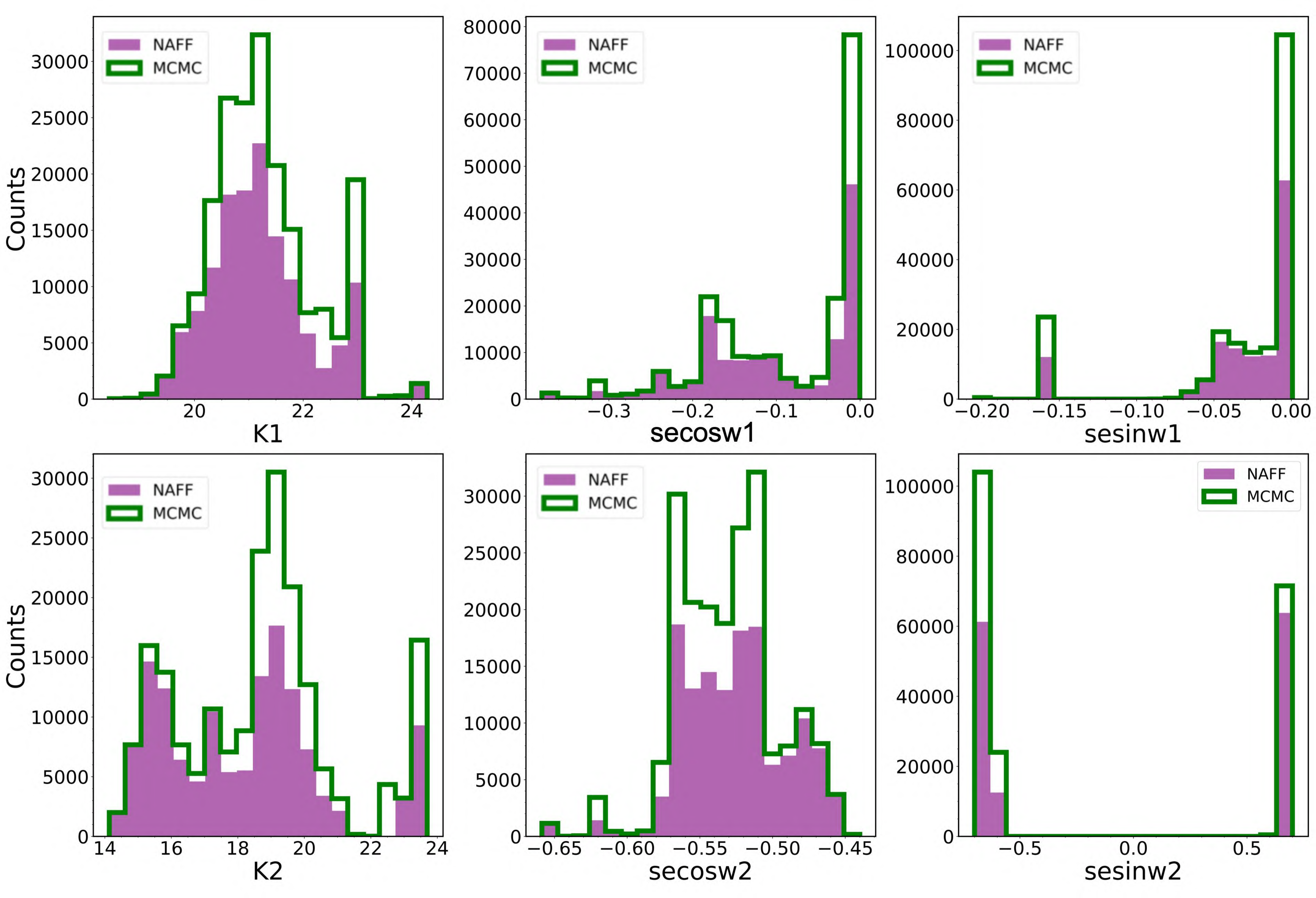}
    \caption{Contradistinction before and after the NAFF stability analysis on the MCMC fitting results. Green hollow histograms: the posterior distribution of MCMC RV fitting parameters before stability analysis. Purple filled histograms: the posterior distribution of RV fitting parameters for both planets in TOI-4633 system with the condition of $i_{\mathrm{mut}} = 70^{\circ}$ and the constrains of NAFF stability threshold shrink the range of both ($K_1$, $e_1$) and ($K_2$, $e_2$). Note: secosw1 = $\sqrt{e_1}$cos$\omega_1$, sesinw1 = $\sqrt{e_1}$sin$\omega_1$, secosw2 = $\sqrt{e_2}$cos$\omega_2$, sesinw2 = $\sqrt{e_2}$sin$\omega_2$.}
\label{fig:TOI4633_NAFF_stab_posterior_70}
\end{center}
\end{figure*}
\clearpage

\begin{figure*}
\begin{center}
\includegraphics[width=\columnwidth,height=12cm]{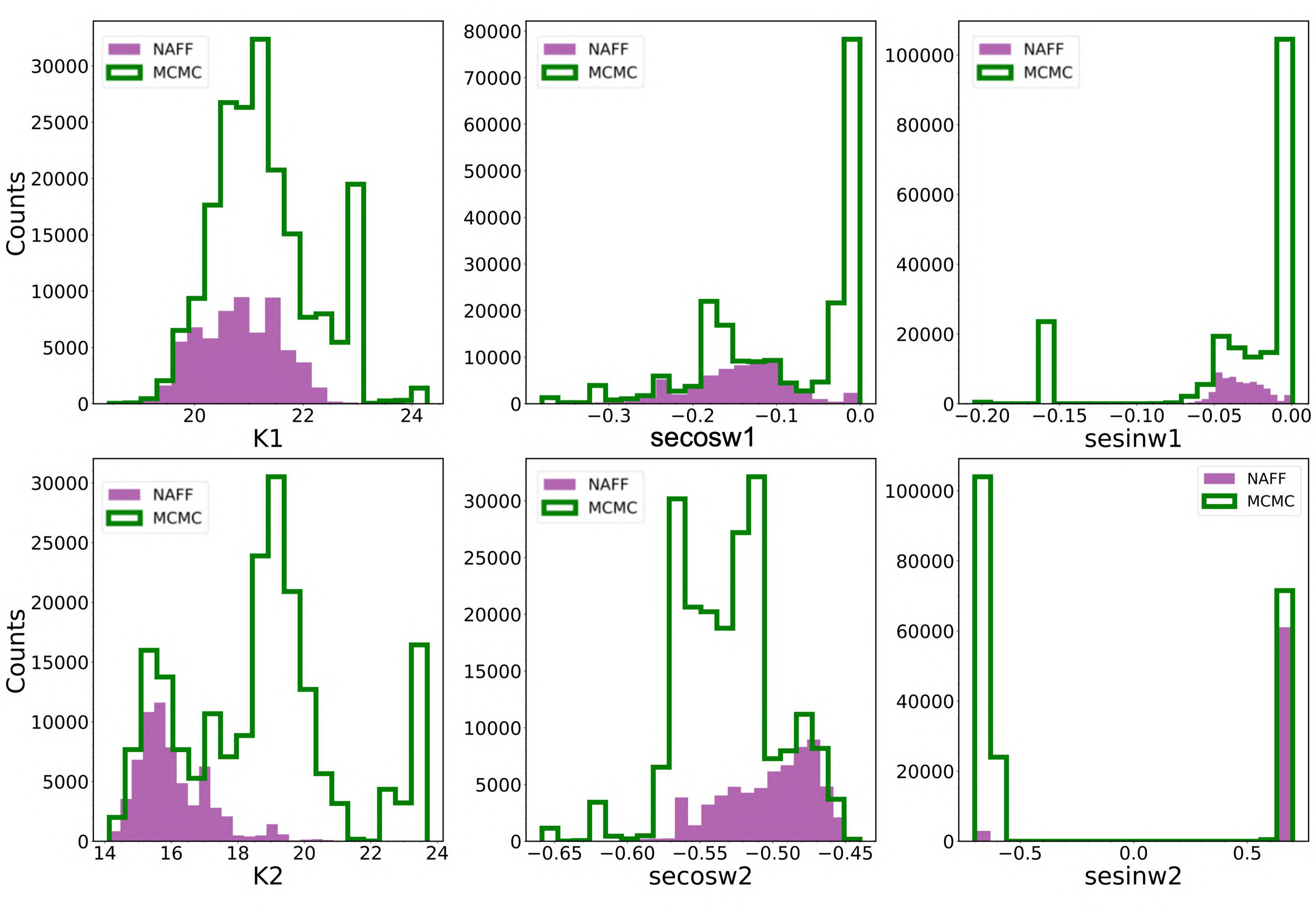}
    \caption{Contradistinction before and after the NAFF stability analysis on the MCMC fitting results. Green hollow histograms: the posterior distribution of MCMC RV fitting parameters before stability analysis. Purple filled histograms: the posterior distribution of RV fitting parameters for both planets in TOI-4633 system with the condition of $i_{\mathrm{mut}} = 90^{\circ}$. The NAFF indicator favors the parameters to the lower edge of both ($K_1$, $e_1$) and ($K_2$, $e_2$). The integration samples are the same as those in Figure \ref{fig:TOI4633_NAFF_stab_posterior} Note: secosw1 = $\sqrt{e_1}$cos$\omega_1$, sesinw1 = $\sqrt{e_1}$sin$\omega_1$, secosw2 = $\sqrt{e_2}$cos$\omega_2$, sesinw2 = $\sqrt{e_2}$sin$\omega_2$.}
\label{fig:TOI4633_NAFF_stab_posterior_90}
\end{center}
\end{figure*}

\section{RV N-body Fitting of Single S-type Planets}\label{sec:appendixB}

\subsection{RV N-body Fitting Curves and Residuals}

\begin{figure*}
\begin{center}
       \subfigure[]{\includegraphics[width=0.32\columnwidth,height=6cm]{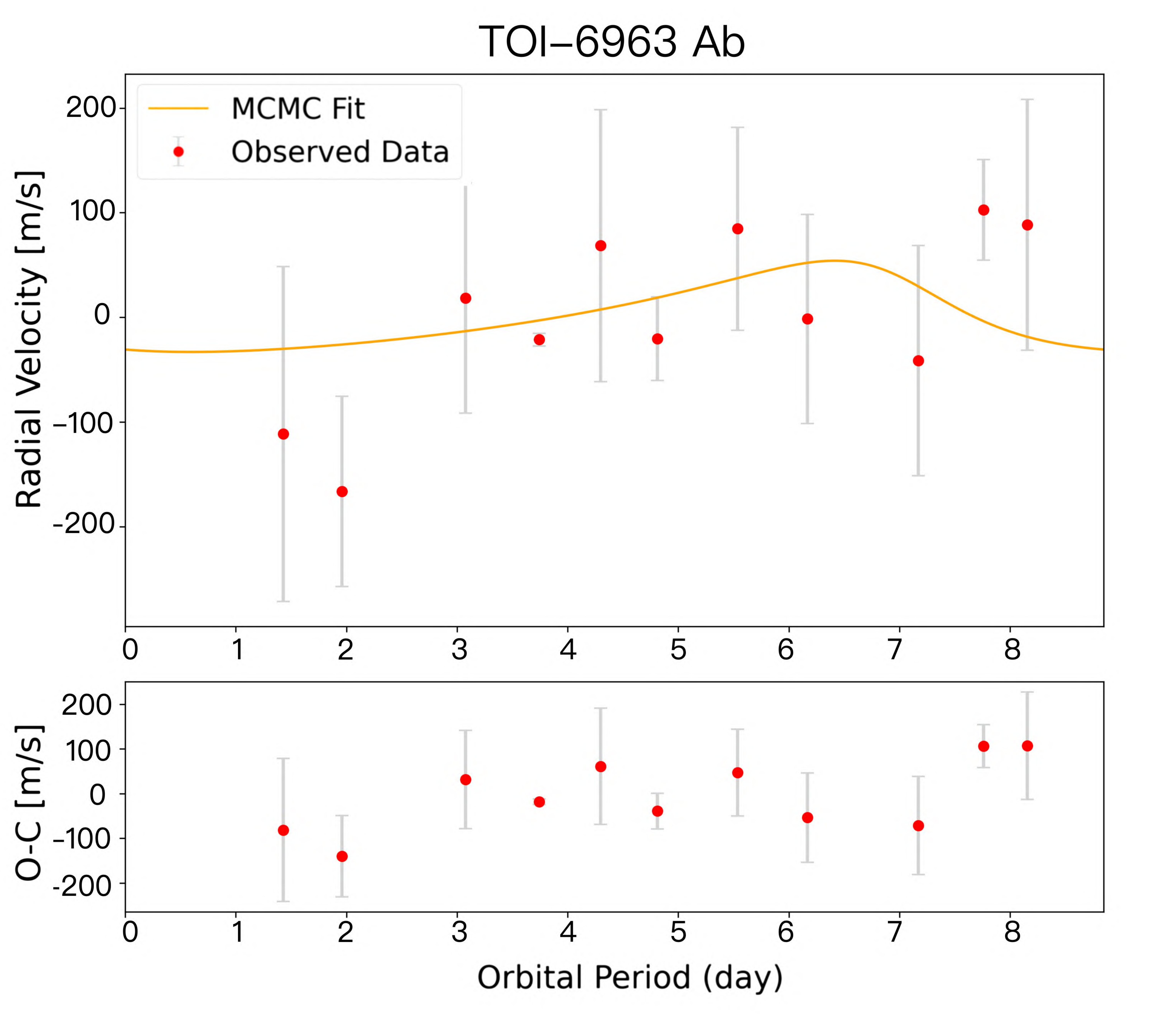}}
         \subfigure[]{\includegraphics[width=0.32\columnwidth,height=6cm]{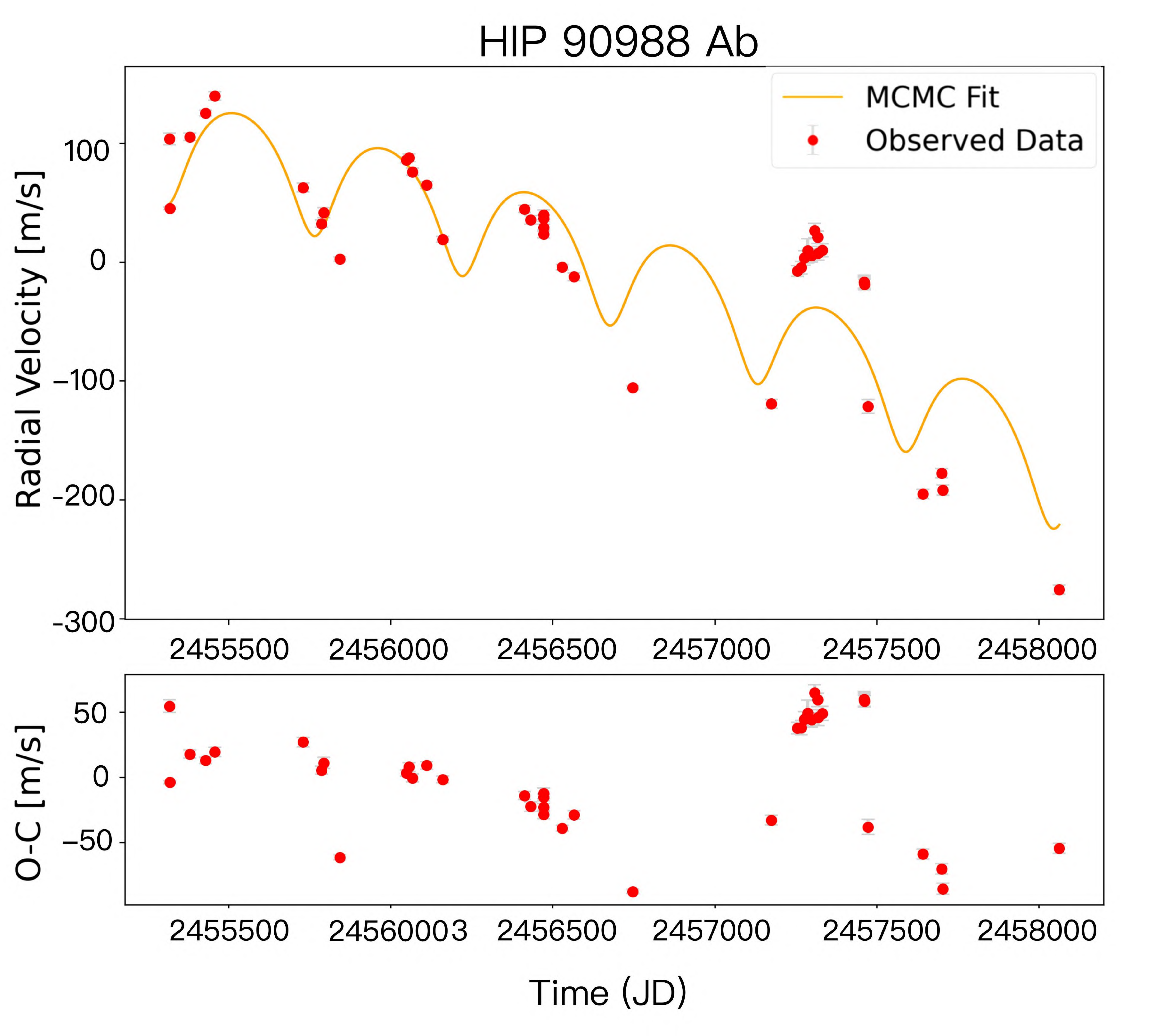}}
         \subfigure[]{\includegraphics[width=0.32\columnwidth,height=6cm]{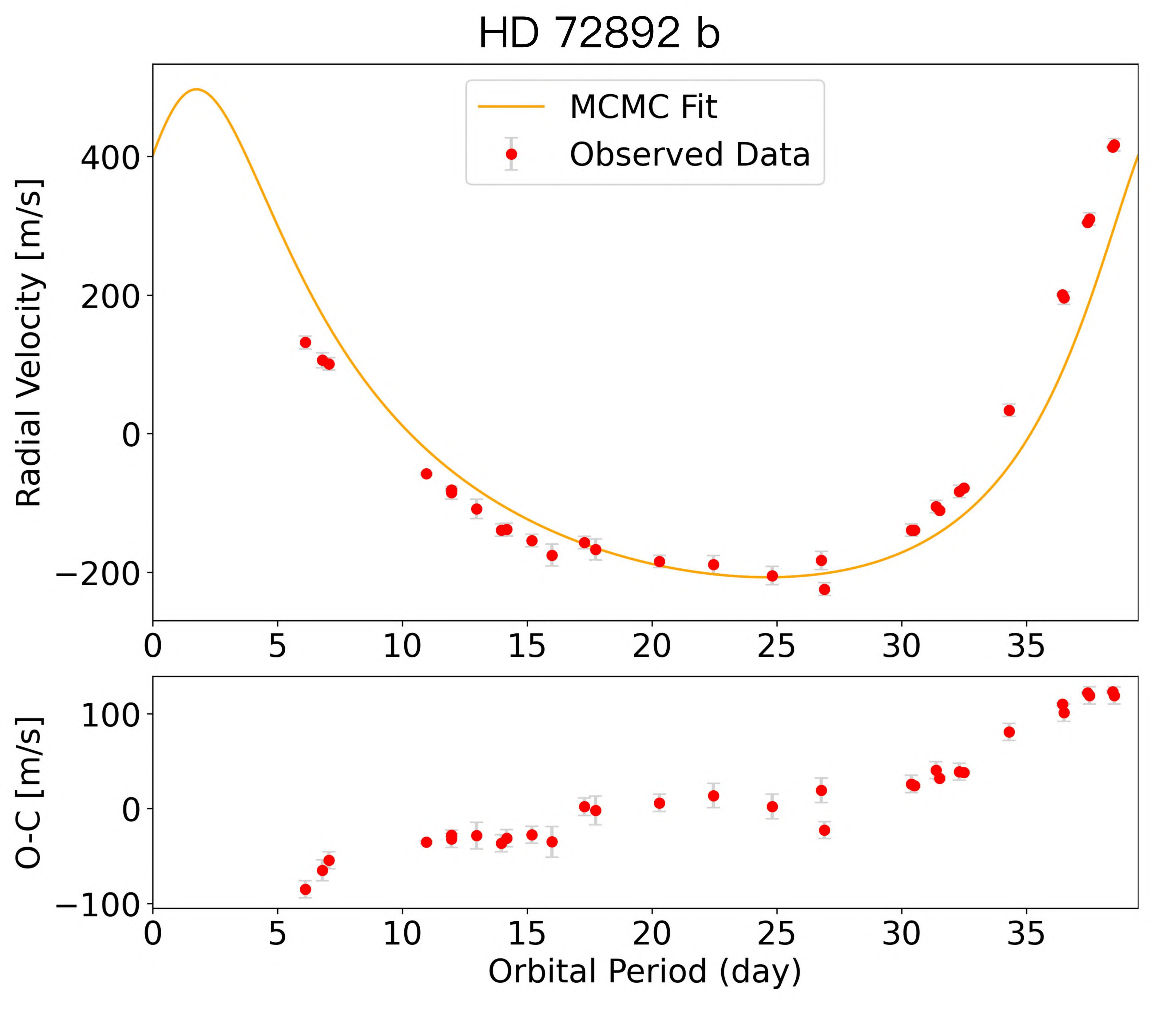}}\\
         \subfigure[]{\includegraphics[width=0.32\columnwidth,height=6cm]{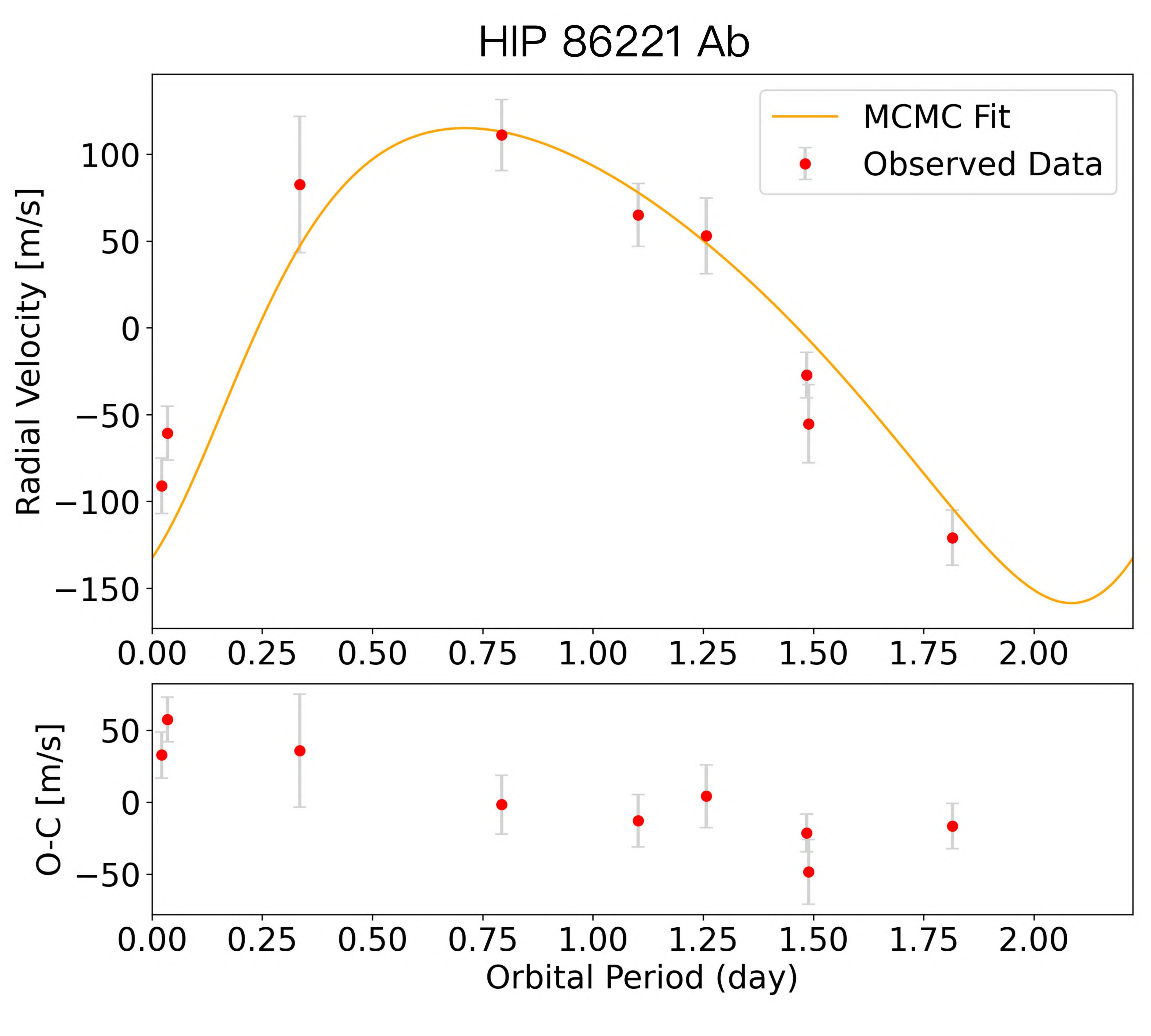}}
         \subfigure[]{\includegraphics[width=0.32\columnwidth,height=6cm]{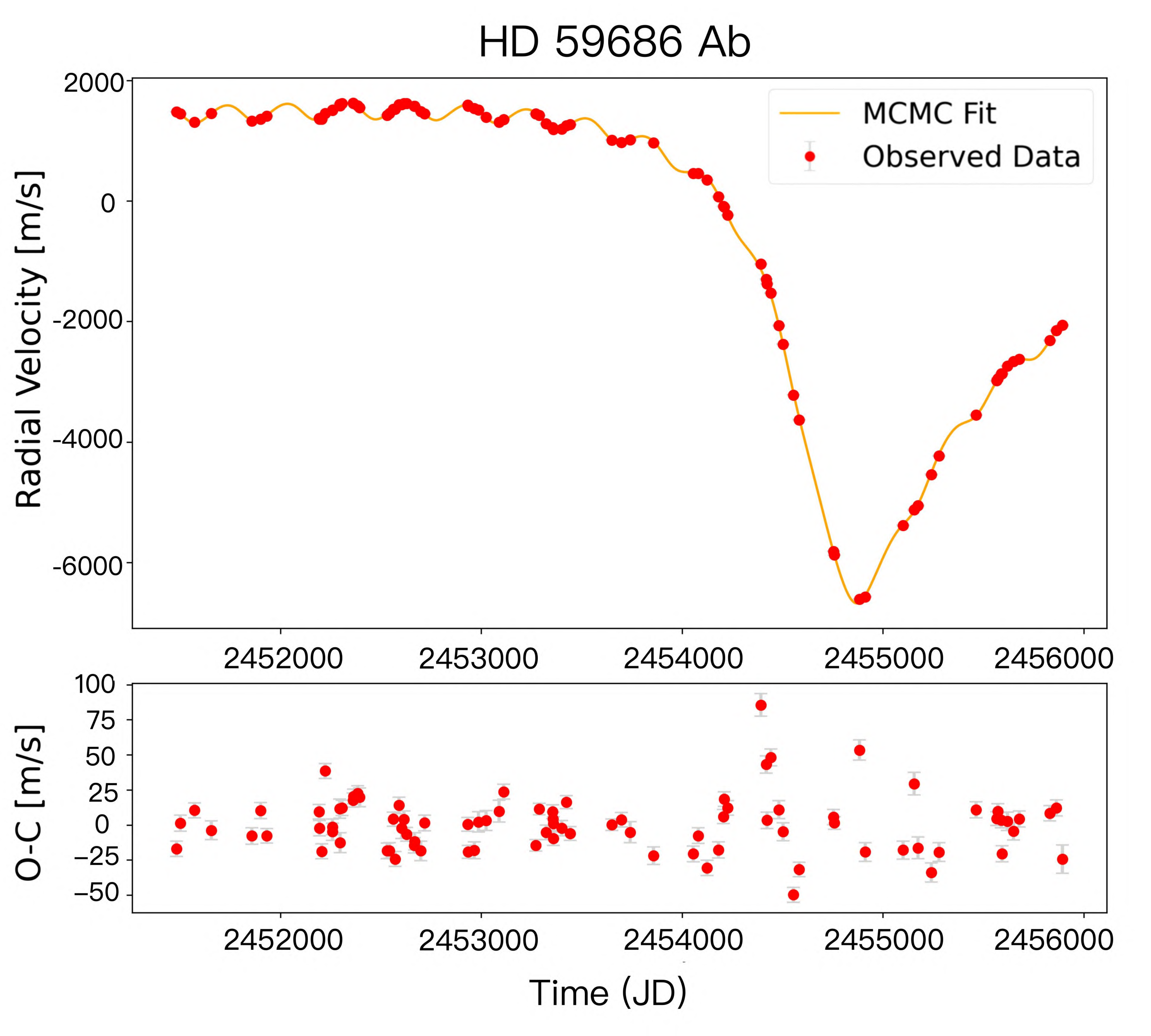}}
         \subfigure[]{\includegraphics[width=0.32\columnwidth,height=6cm]{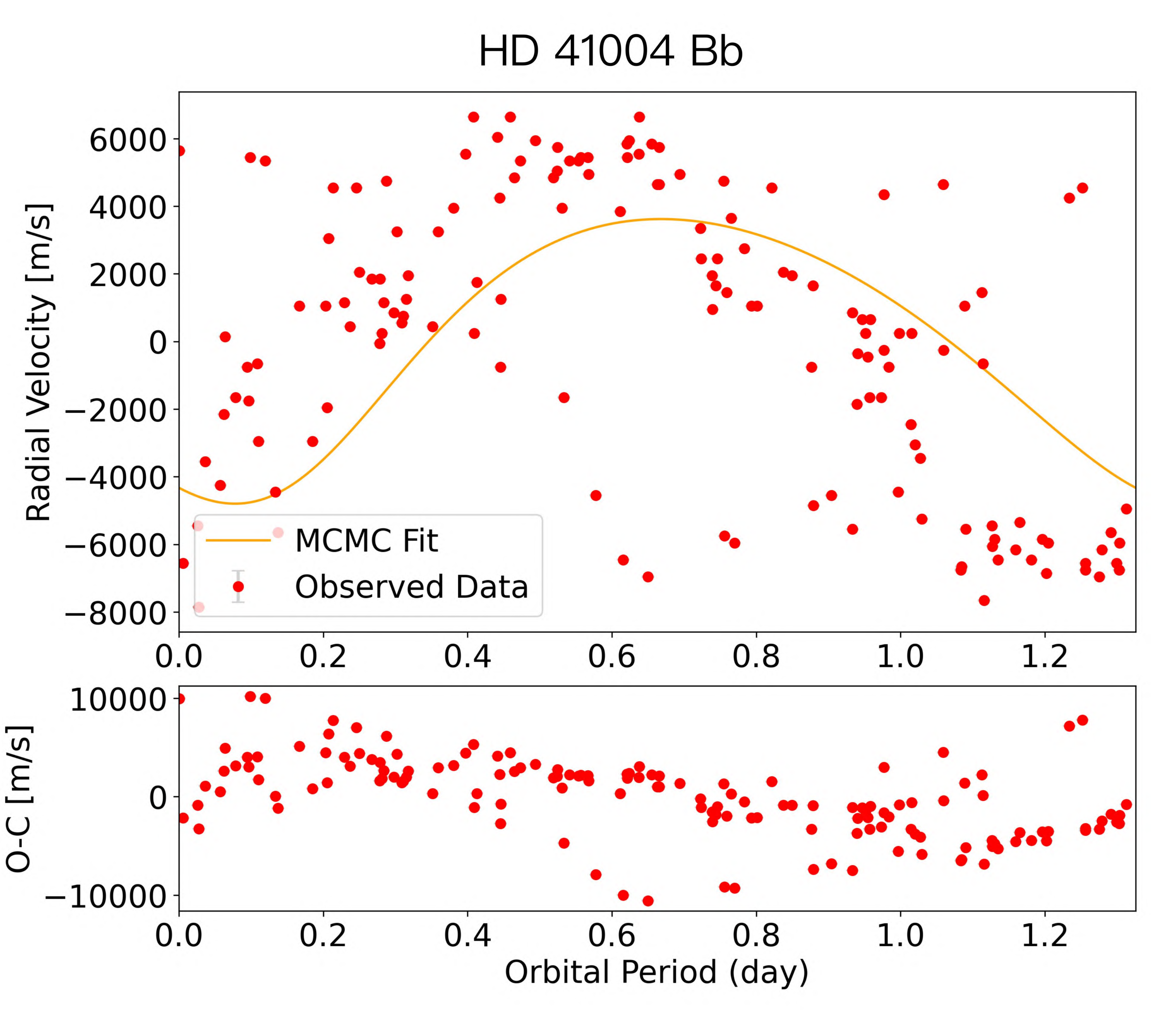}}\\
         \subfigure[]{\includegraphics[width=0.32\columnwidth,height=6cm]{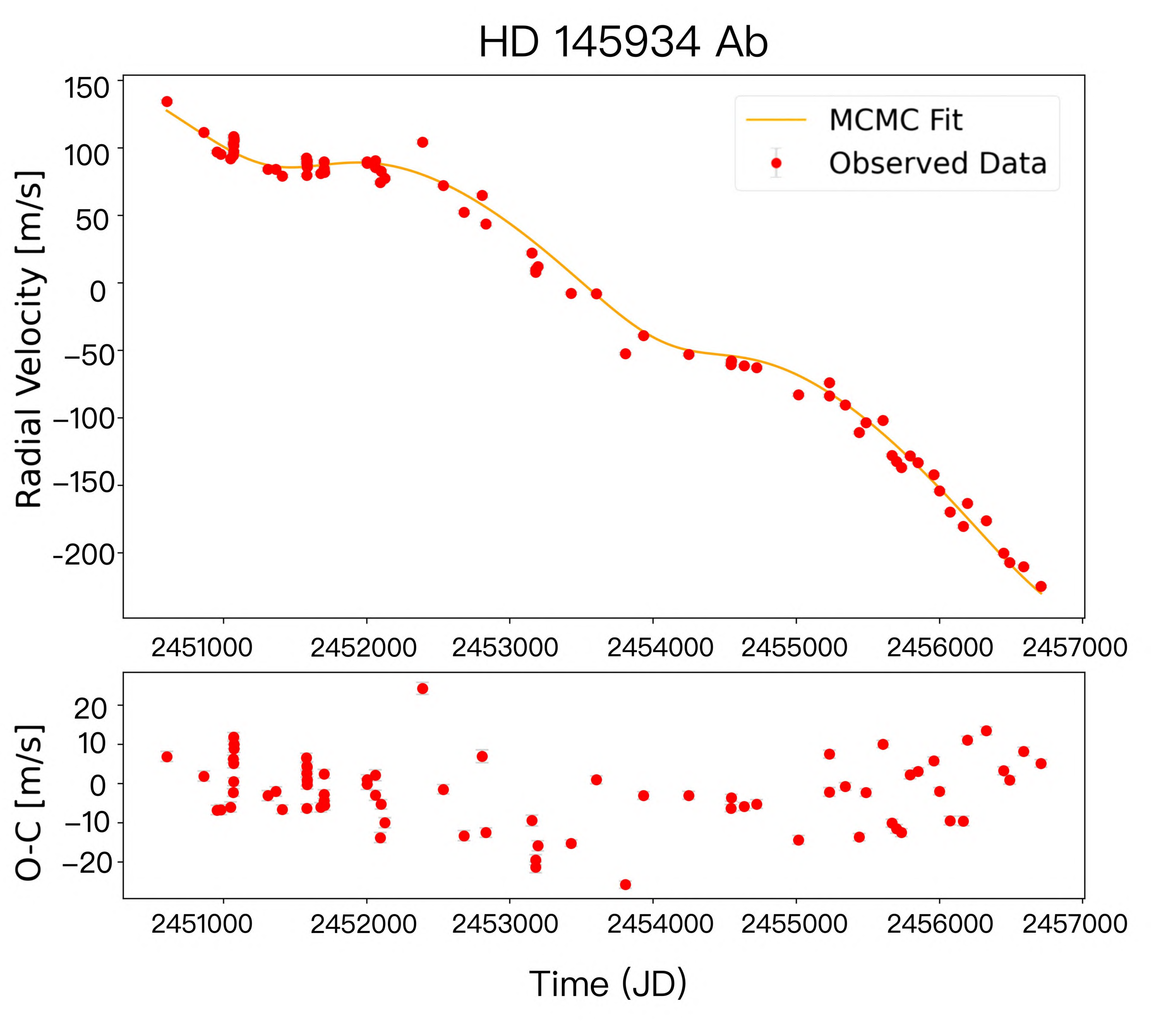}}
         \subfigure[]{\includegraphics[width=0.32\columnwidth,height=6cm]{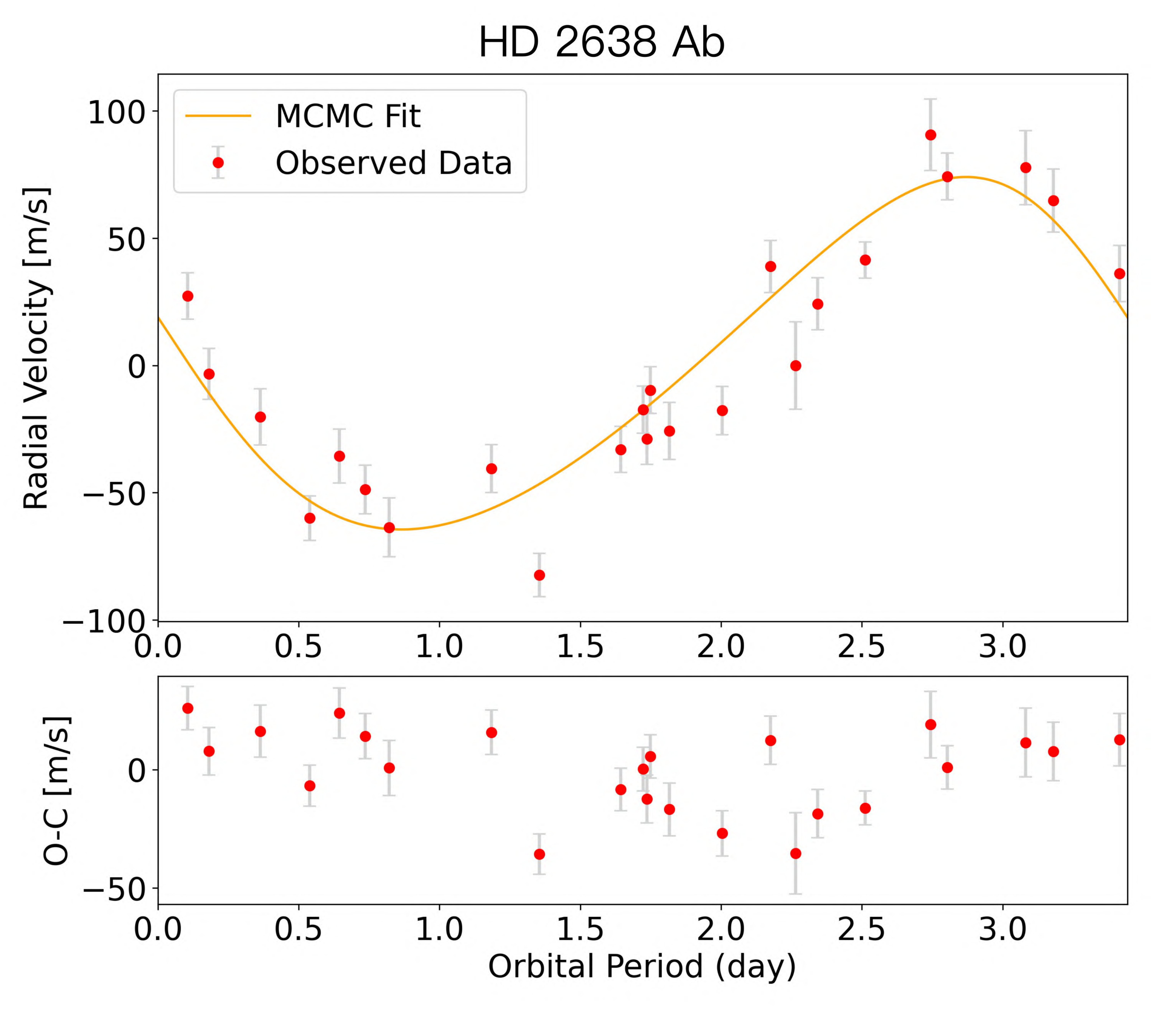}}
         \subfigure[]{\includegraphics[width=0.32\columnwidth,height=6cm]{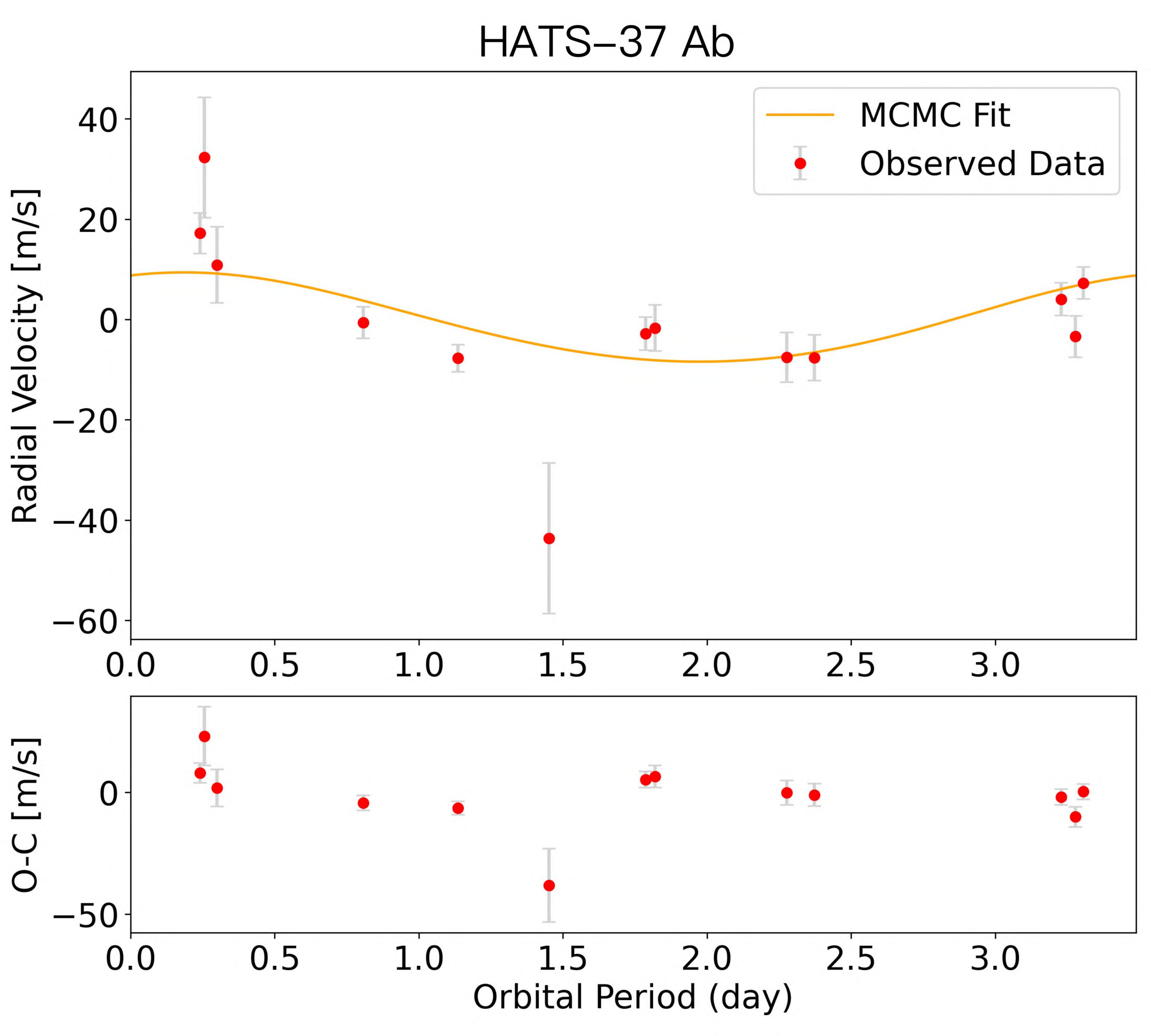}}\\
    \caption{N-body RV fitting curves of nine close-binaries with $a_{\mathrm{B}} < 30$ au, including TOI-6963 Ab, HIP 90988 Ab, HD 72892 b, HIP 86221 Ab, HD 59686 Ab, HD 41004 Bb, HD 145934 Ab, HD 2638 Ab and HATS-37 Ab. We derived the dynamical minimum masses of these planets with the mass deviations $|\Delta M_{\mathrm{p}}|$ = 4.1626 $M_{\oplus}$, 5.7505 $M_{\oplus}$, 28.2157 $M_{\oplus}$, 0.1869 $M_{\oplus}$, 4.4630 $M_{\oplus}$, 88.5395 $M_{\oplus}$, 15.2994 $M_{\oplus}$, 0.2561 $M_{\oplus}$ and 0.3961 $M_{\oplus}$ respectively.}
    \label{fig:rv_30au}
\end{center}
\end{figure*}

\begin{figure*}
\begin{center}
       \subfigure[]{\includegraphics[width=0.32\columnwidth,height=6cm]{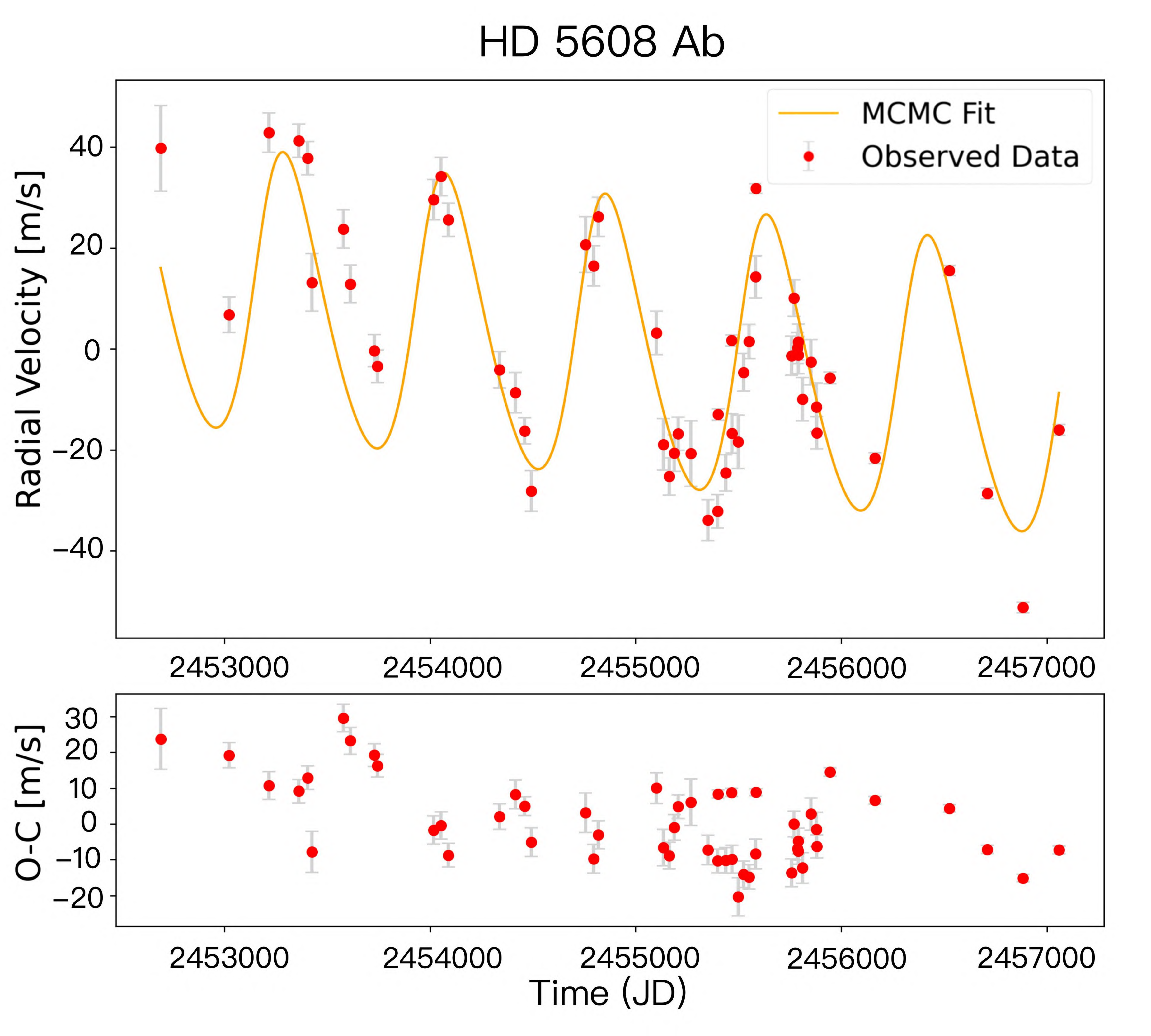}}
         \subfigure[]{\includegraphics[width=0.32\columnwidth,height=6cm]{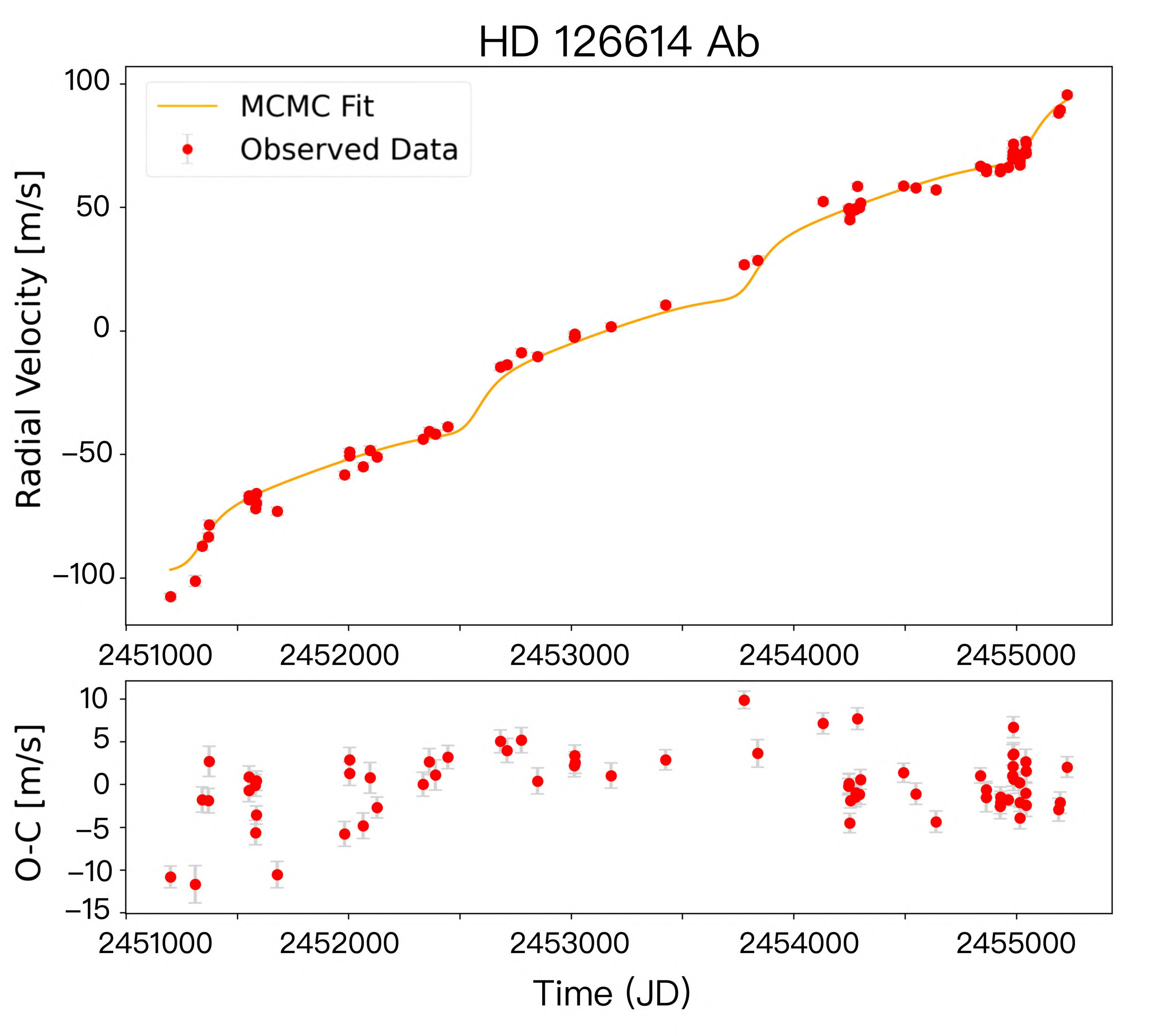}}
         \subfigure[]{\includegraphics[width=0.32\columnwidth,height=6cm]{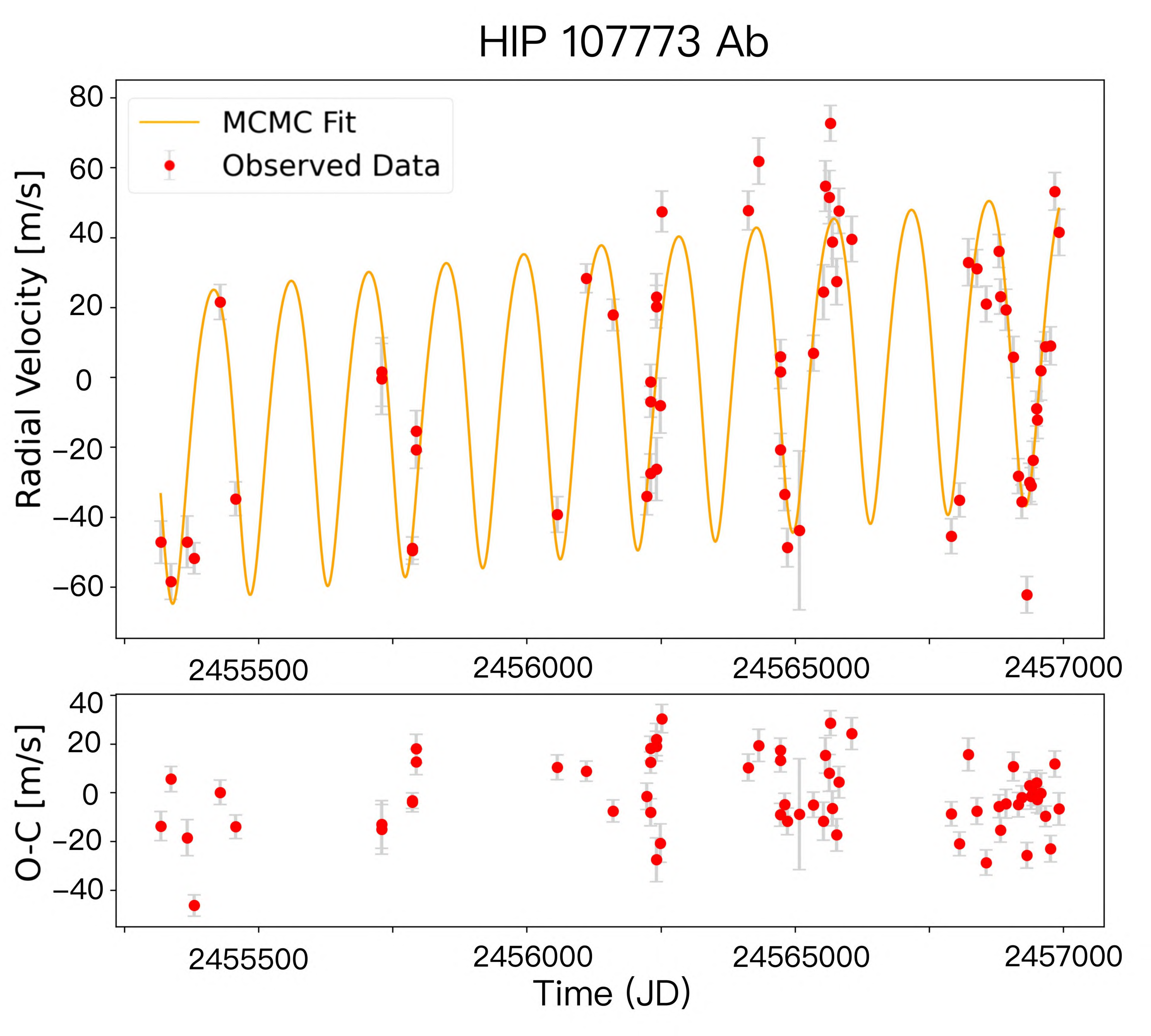}}\\
         \subfigure[]{\includegraphics[width=0.32\columnwidth,height=6cm]{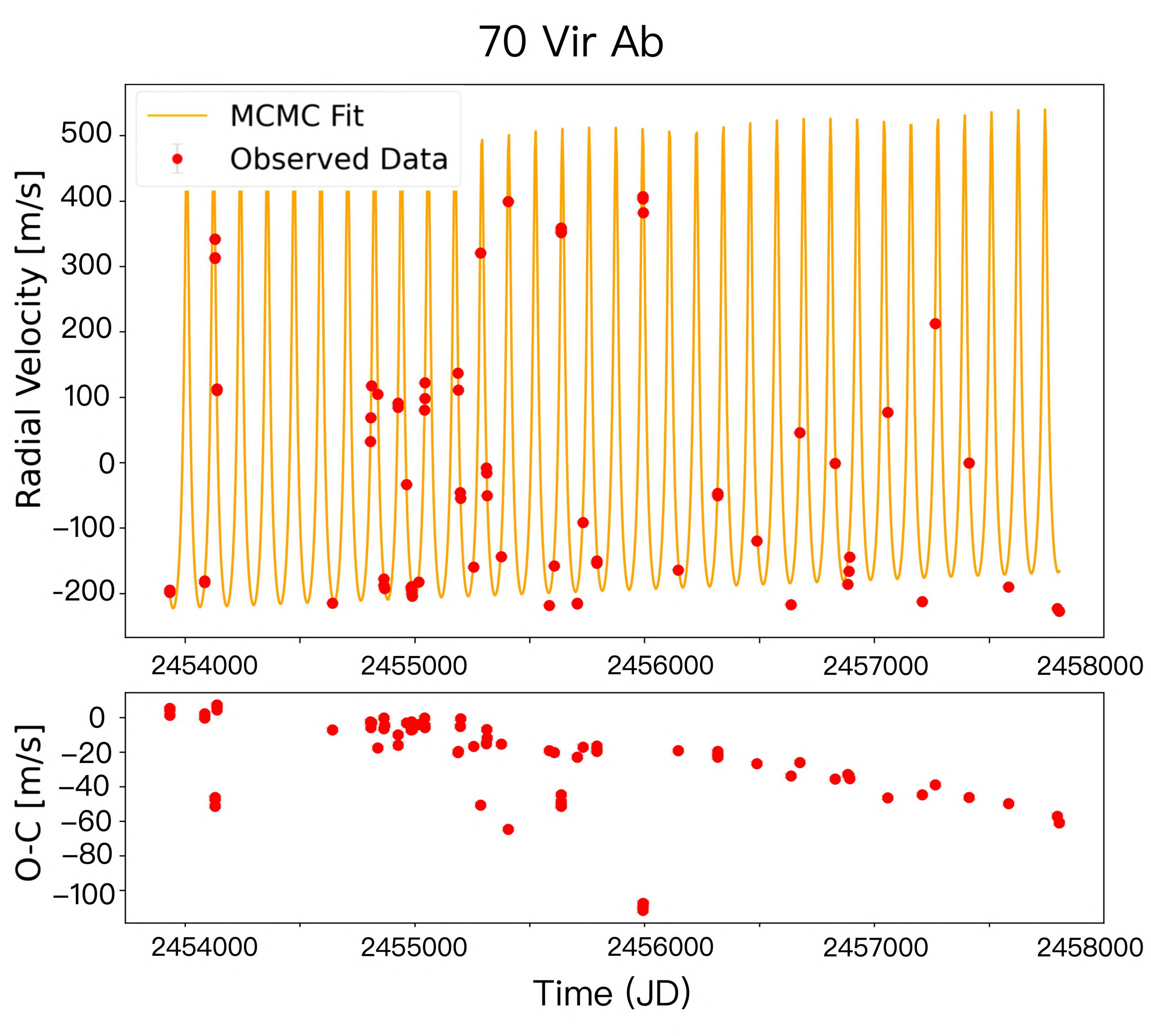}}
         \subfigure[]{\includegraphics[width=0.32\columnwidth,height=6cm]{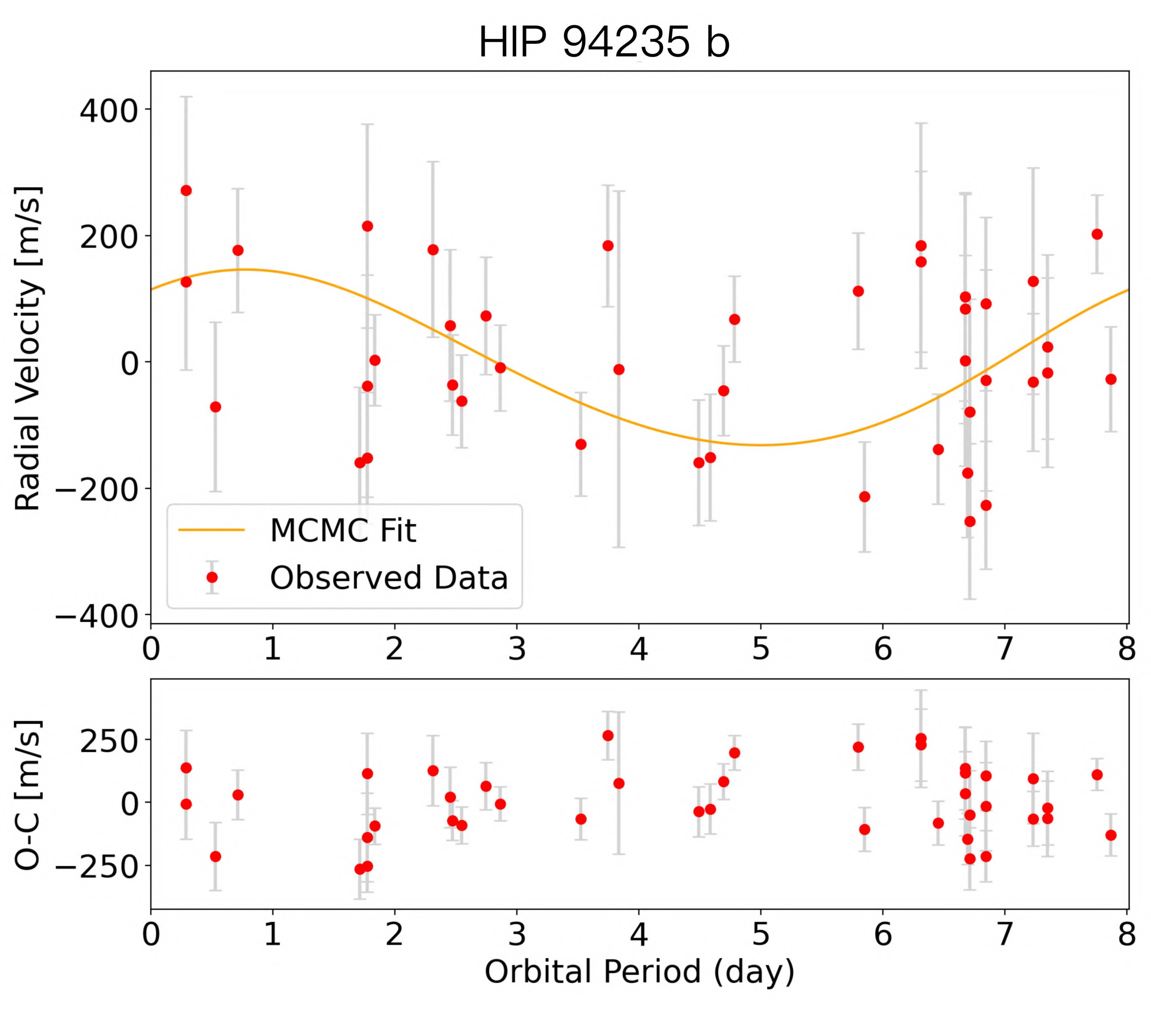}}
         \subfigure[]{\includegraphics[width=0.32\columnwidth,height=6cm]{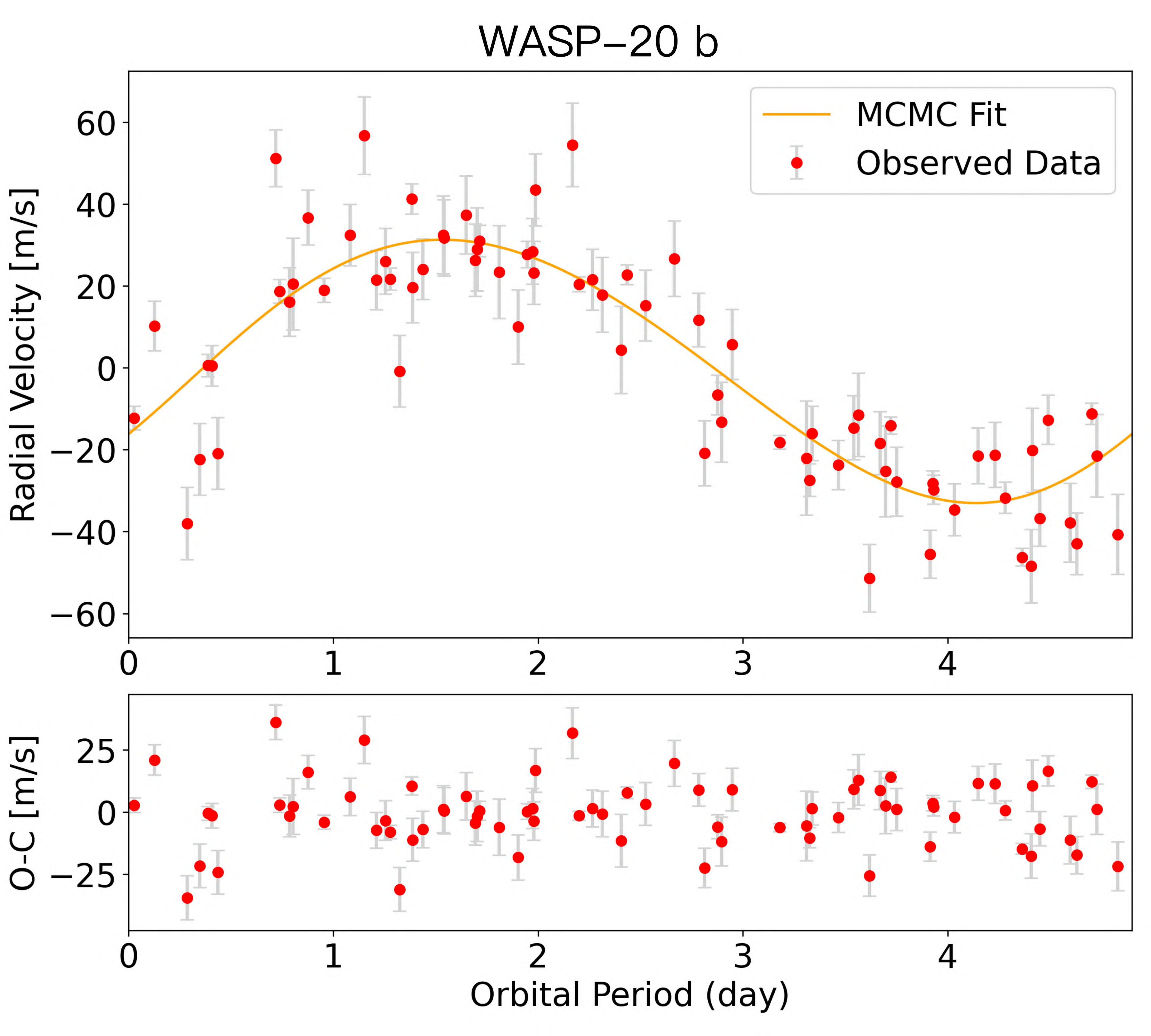}}\\
         \subfigure[]{\includegraphics[width=0.32\columnwidth,height=6cm]{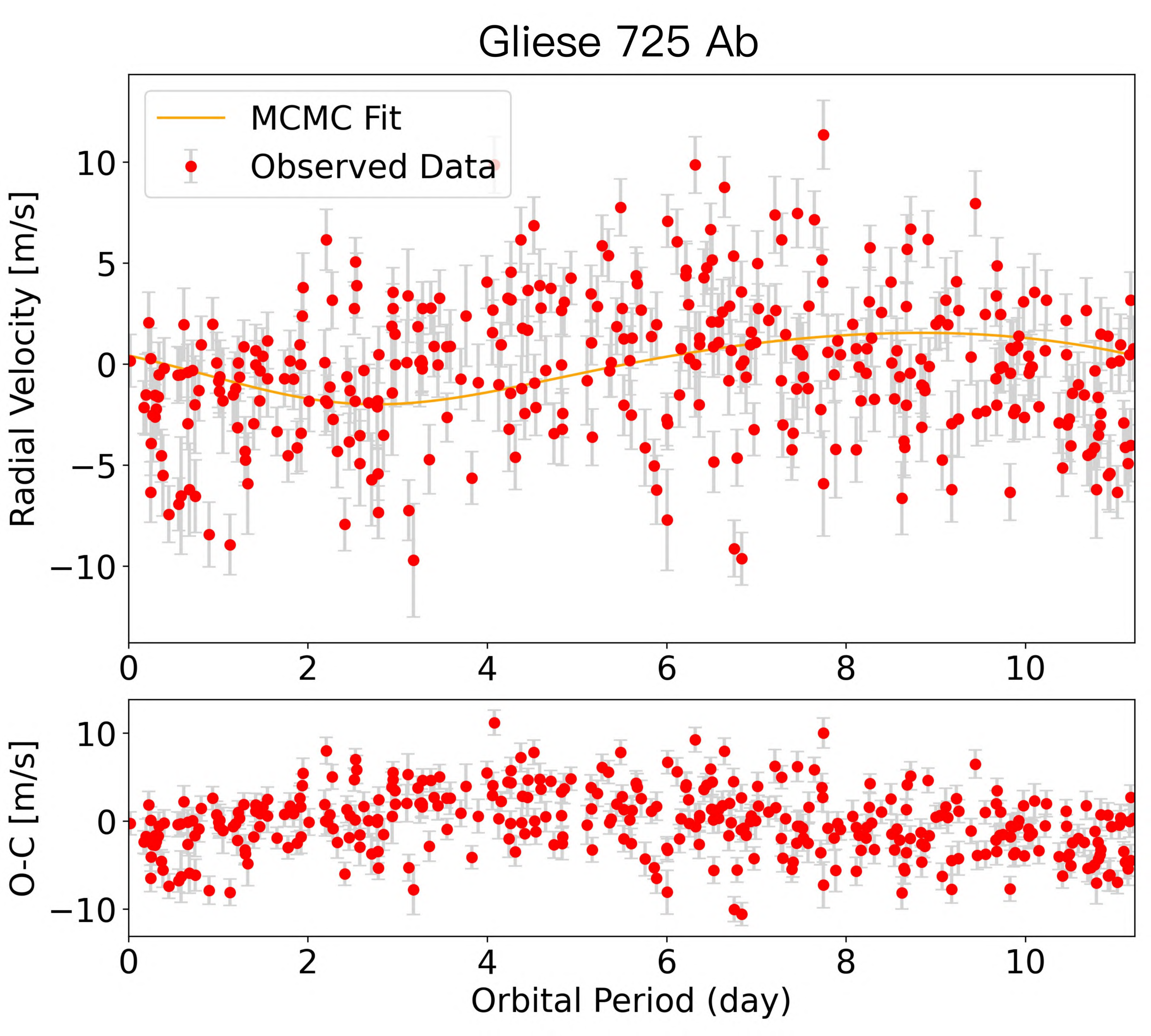}}
         \subfigure[]{\includegraphics[width=0.32\columnwidth,height=6cm]{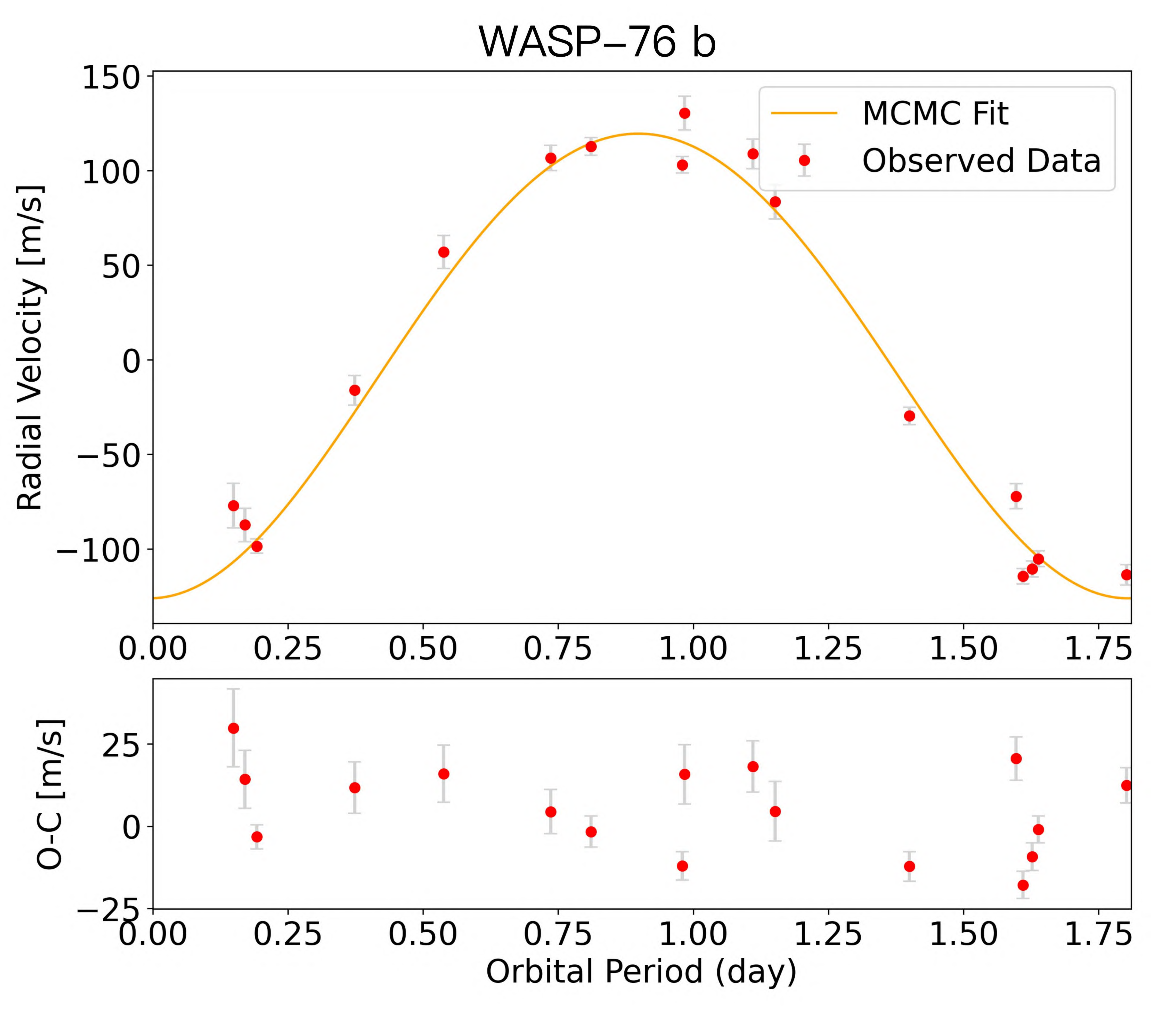}}
         \subfigure[]{\includegraphics[width=0.32\columnwidth,height=6cm]{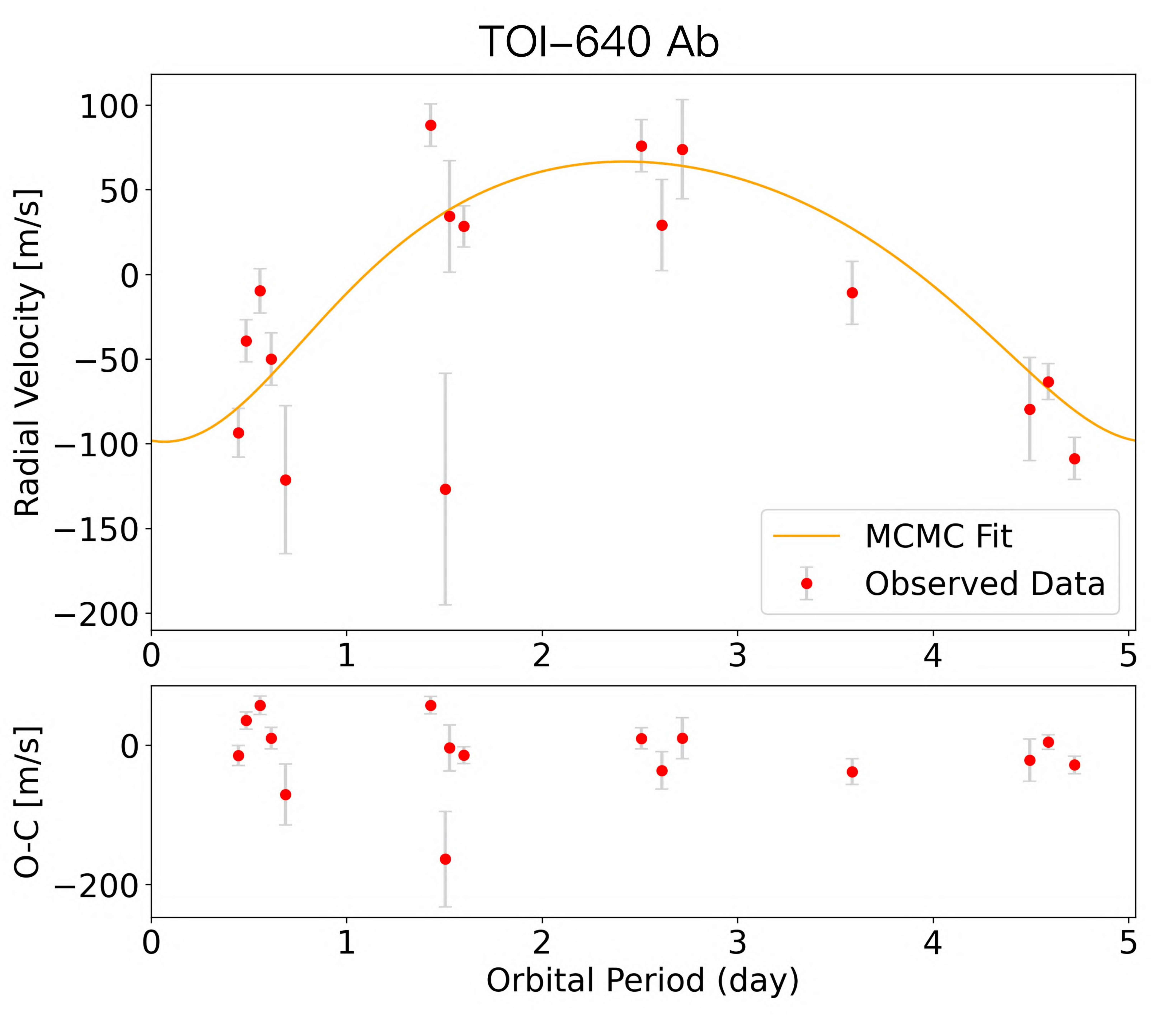}}\\
    \caption{N-body RV fitting curves of nine close-binaries with $a_{\mathrm{B}} < 85$ au, including HD 5608 Ab, HD 126614 Ab, HIP 107773 Ab, 70 Vir Ab, HIP 94235 Ab, WASP-20 Ab, Gliese 725 Ab, WASP-76 Ab and TOI-640 Ab. We derived the dynamical minimum masses of these planets with the mass deviations $|\Delta M_{\mathrm{p}}|$ = 11.7668 $M_{\oplus}$, 6.7986 $M_{\oplus}$, 10.7851 $M_{\oplus}$, 77.1565 $M_{\oplus}$, 46.8170 $M_{\oplus}$, 8.551 $M_{\oplus}$, 0.0012 $M_{\oplus}$, 9.3861 $M_{\oplus}$ and 2.9133 $M_{\oplus}$, respectively.}
    \label{fig:rv_85au}
\end{center}
\end{figure*}

\begin{figure*}
\begin{center}
       \subfigure[]{\includegraphics[width=0.32\columnwidth,height=6cm]{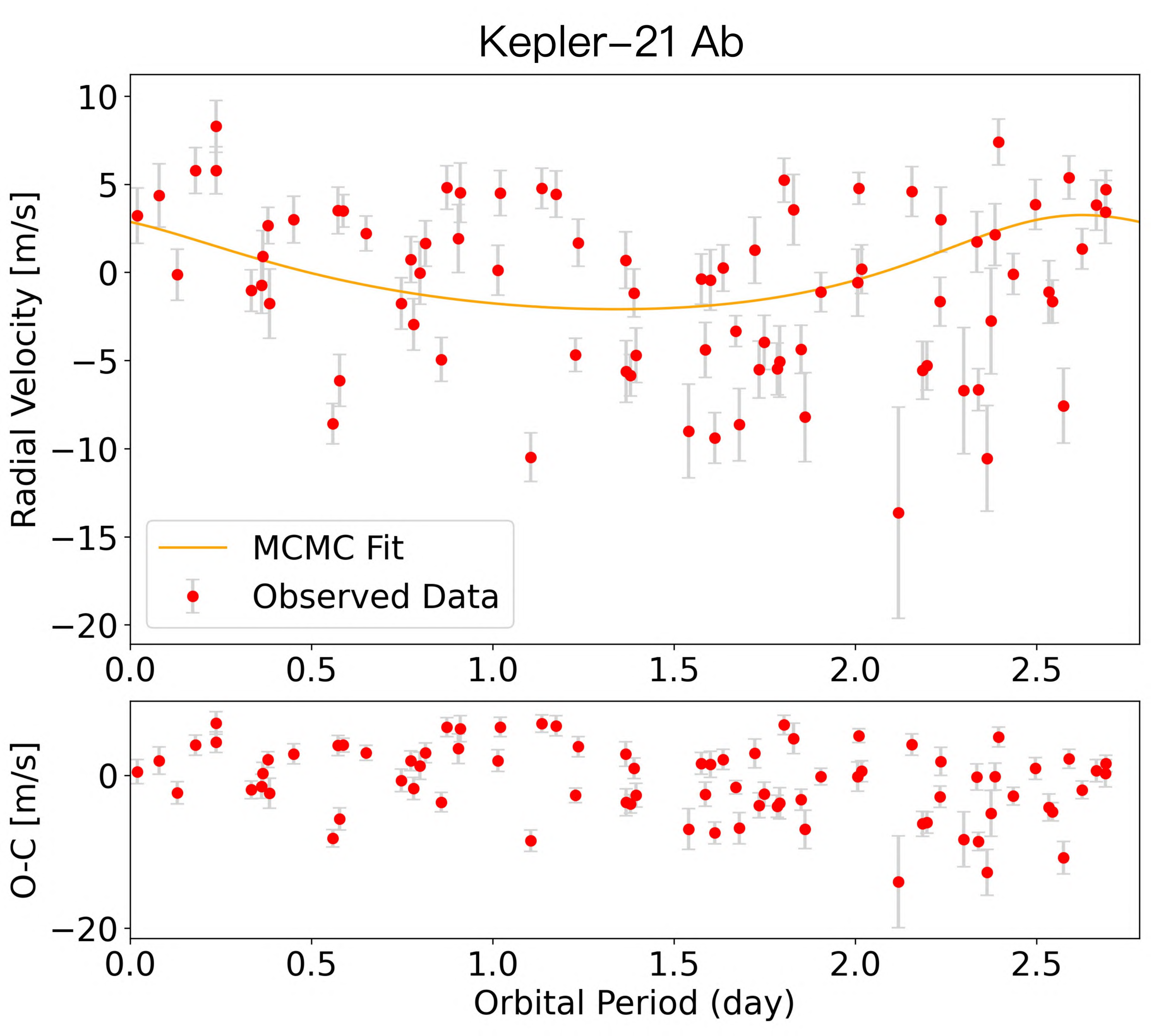}}
         \subfigure[]{\includegraphics[width=0.32\columnwidth,height=6cm]{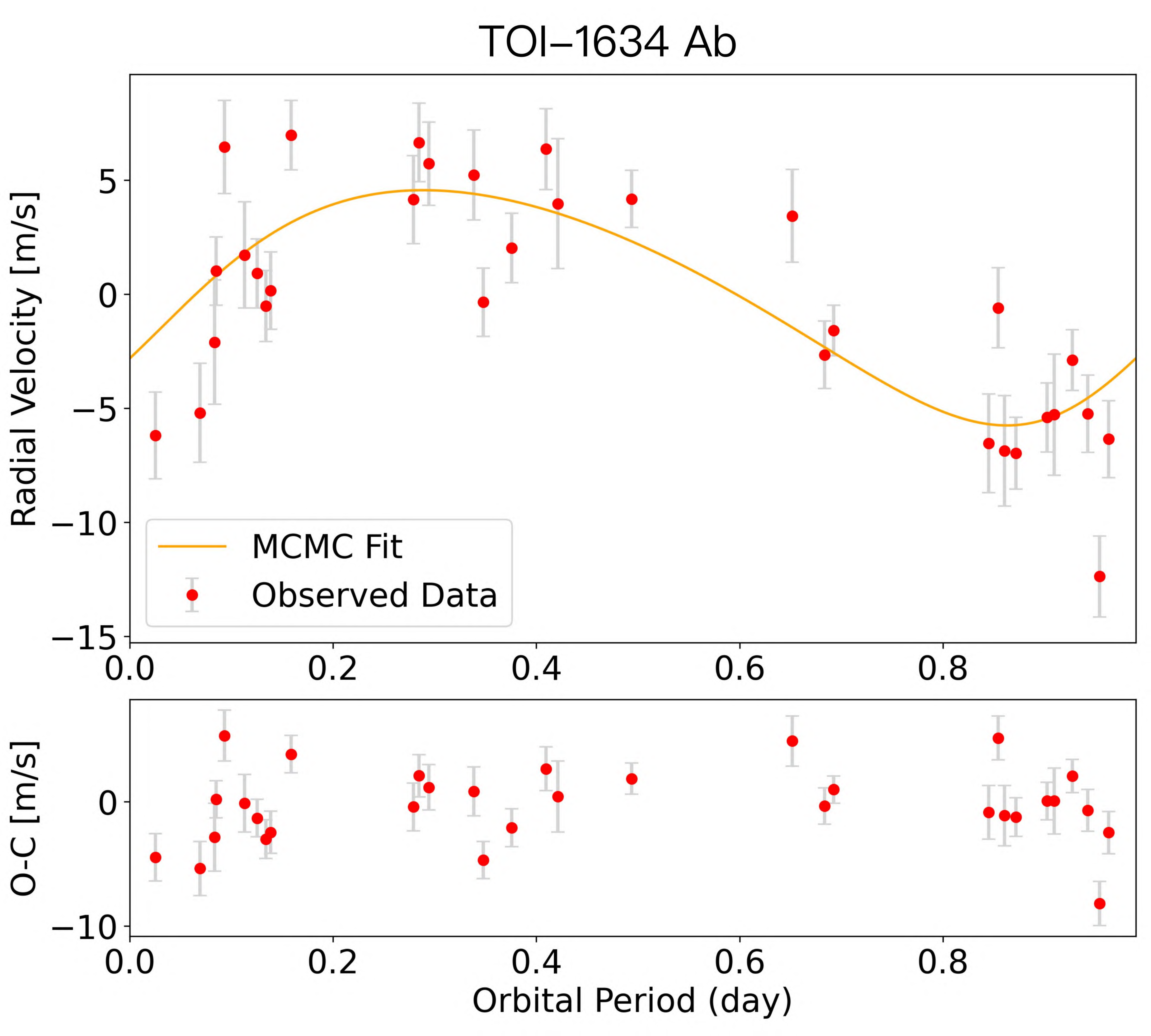}}
         \subfigure[]{\includegraphics[width=0.32\columnwidth,height=6cm]{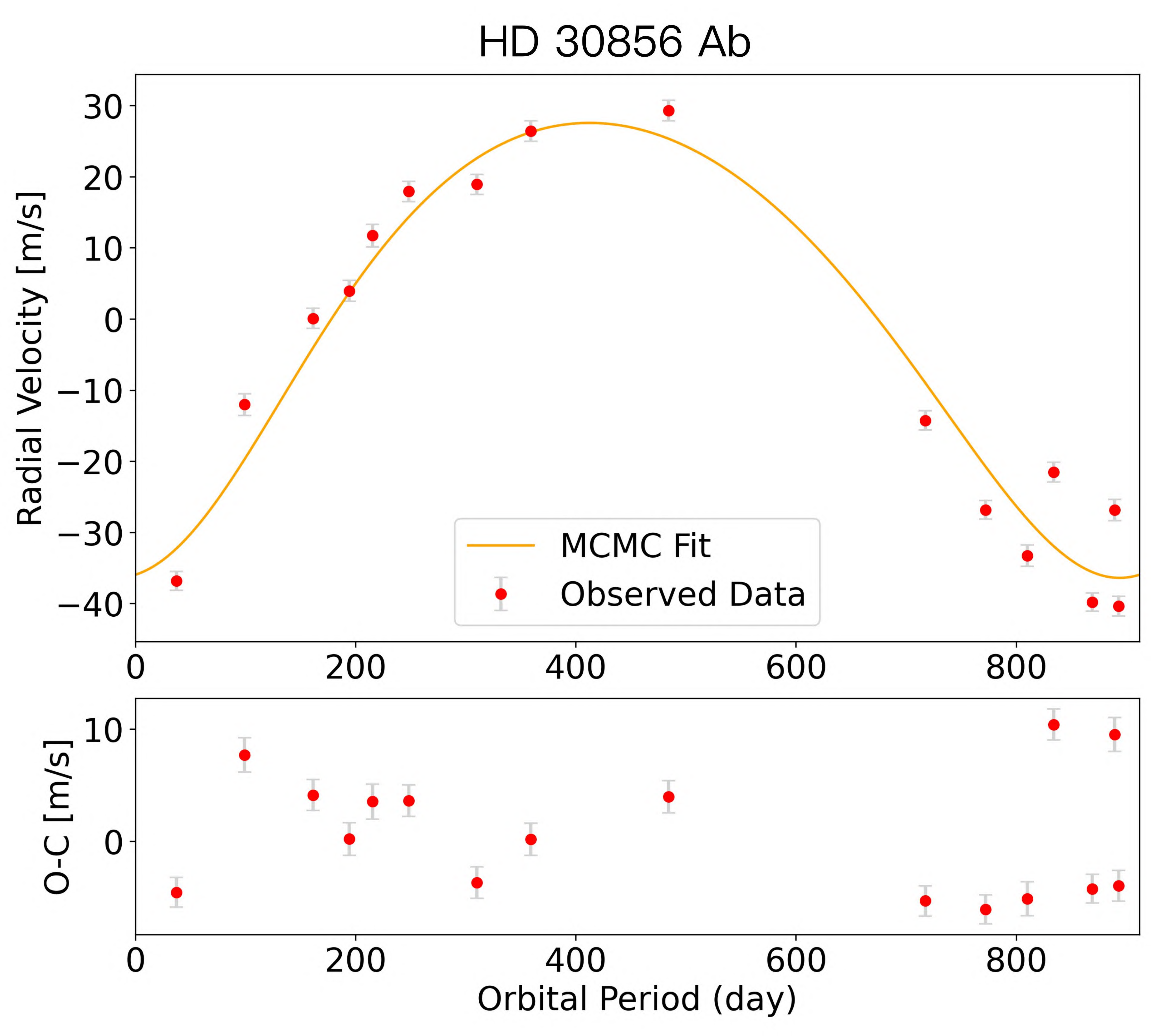}}\\
         \subfigure[]{\includegraphics[width=0.32\columnwidth,height=6cm]{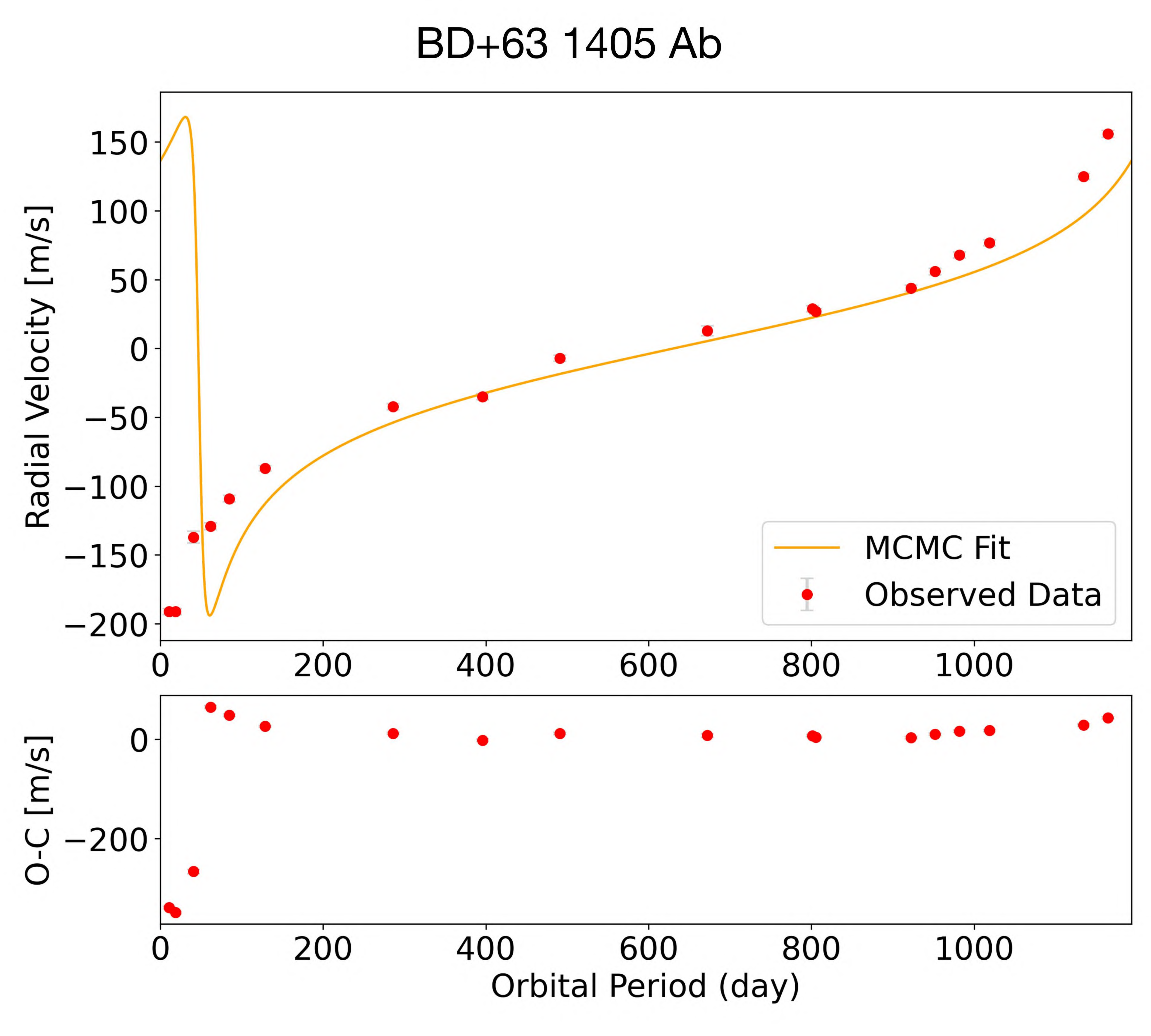}}
         \subfigure[]{\includegraphics[width=0.32\columnwidth,height=6cm]{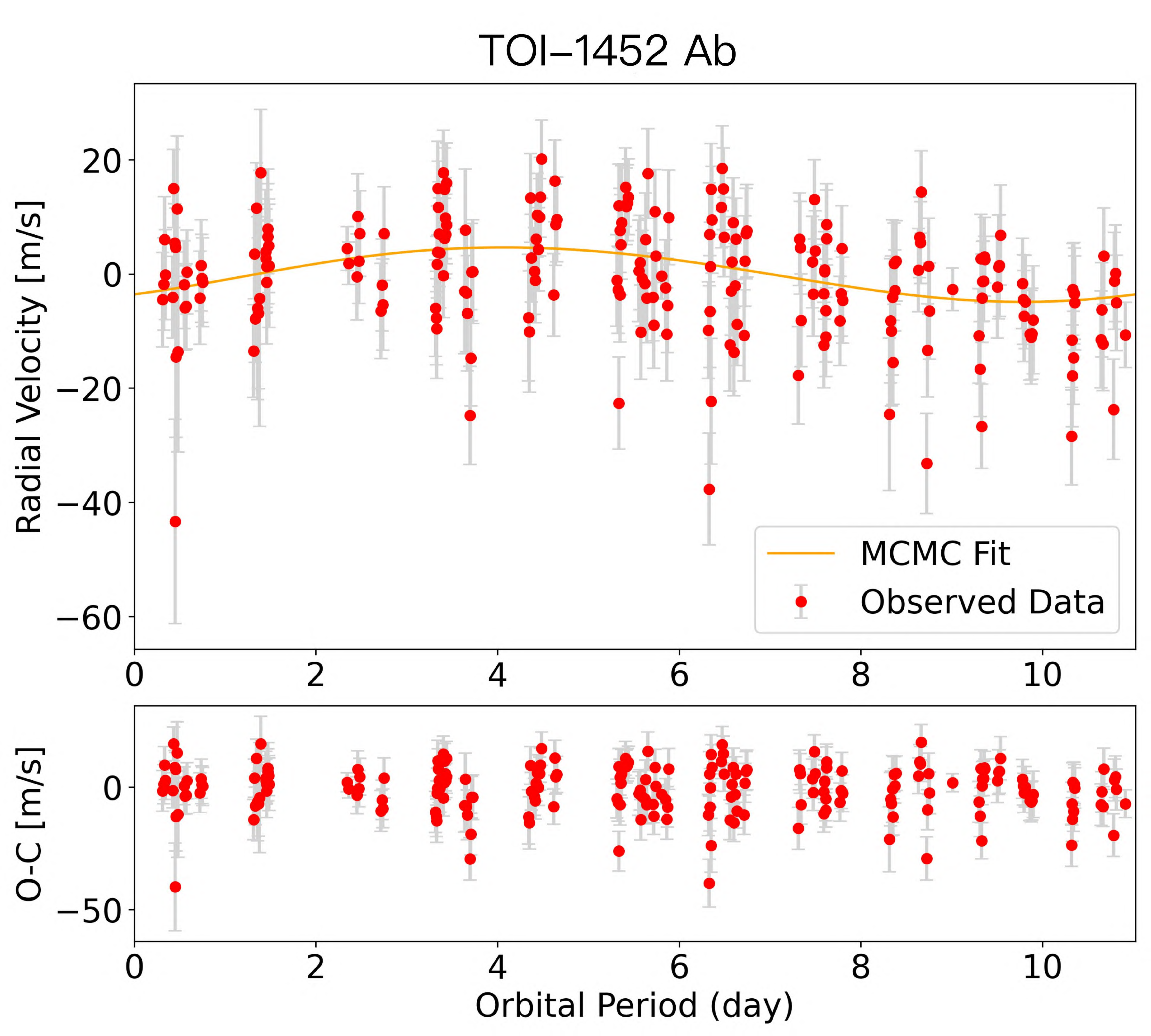}}
    \caption{N-body RV fitting curves of five more close-binaries with $a_{\mathrm{B}} < 100$ au, including Kepler-21 Ab, TOI-1634 Ab, HD 30856 Ab, TOI-1452 Ab and BD+63 1405 Ab. We derived the dynamical minimum masses of these planets with the mass deviations $|\Delta M_{\mathrm{p}}|$ = 0.1249 $M_{\oplus}$, 0.3361 $M_{\oplus}$, 2.0259 $M_{\oplus}$, 0.0581 $M_{\oplus}$ and 89.9820 $M_{\oplus}$, respectively.}
    \label{fig:rv_100au}
\end{center}
\end{figure*}

Figures \ref{fig:rv_30au} to Figure \ref{fig:rv_100au} present 23 newly fitted single S-type planets with $a_{\mathrm{B}} < 100$ au in this work. For targets where the 1-planet model is the best fitting model, we select the fast gaussian-prior N-body fitting parameters of $K_{\mathrm{1}}$, $P_{\mathrm{1}}$, $\sqrt{e_{\mathrm{1}}}$cos$\omega_1$, $\sqrt{e_{\mathrm{1}}}$sin$\omega_1$, $T_{\mathrm{p}}$ and the N-body $M_{\mathrm{p}}$. In other systems such as HIP 90988, HD 59686, HD 145934, HD 5608, HD 126614, HD 107773, and 70 Vir, we add the linear trend factor $\dot{V}$ or the parabolic trend factor $\ddot{V}$ to the original parameter space of fitting.

\subsection{N-body Masses Catalog of single S-type planets}

Table \ref{tab:catalog} list 40 transiting single S-type planets with the fast gaussian-prior N-body fitting results of $K_{\mathrm{1}}$, $P_{\mathrm{1}}$, $a_{\mathrm{1}}$, $e_{\mathrm{1}}$ and the N-body $M_{\mathrm{p}}$. In the column of planetary semi-major axis $a_{\mathrm{1}}$ , uncertainties lower than $10^{-4}$ are not listed. The N-body planetary mass solutions and the Keplerian fitting results with Gaussian prior are both derived in this paper. The rightmost column provide the corresponding references for the original transiting data, radial velocity data and the binary system parameters.

\begin{longrotatetable}\label{tab:catalog}
\movetabledown=1in
\setlength{\tabcolsep}{1pt}
\begin{deluxetable*}{lccccccccccccc}
\tabletypesize{\scriptsize}
\tablewidth{4.5cm}
\renewcommand{\arraystretch}{1.5}
\tablecaption{Revised Catalog of Single S-type planets\label{tab:catalog}}
\tablehead{
\colhead{\multirow{2}{*}{Planet}} & \colhead{$M_\mathrm{A}$} & \colhead{$M_\mathrm{B}$} & \colhead{$a_B$} & \colhead{\multirow{2}{*}{$e_B$}} & \colhead{$i_\mathrm{b}$} & \colhead{$K_b$} & \colhead{$P_b$} & \colhead{\multirow{2}{*}{$e_b$}} & \colhead{$\omega_b$} & \colhead{$T_p$} &\colhead{$M_{\mathrm{p}}$ (N-body)} & \colhead{$M_{\mathrm{p}}$ (Kepler)} & \colhead{\multirow{2}{*}{ref.}}\\
 \colhead{} & $(M_{\odot})$ & $(M_{\odot})$ & (au)  &  \colhead{} & $(^{\circ})$ & (m/s) &  (day) & \colhead{}  & (deg) & (JD) &$(M_{\oplus})$ & $(M_{\oplus})$ &  \colhead{}
}
\startdata
TOI-6963 Ab	&	$	0.70$&	0.15	&	2.3	&	--	&		--		&	$42.5420	^{+	9.7402	}_{	-9.3864	}$&	$	8.84	^{+	0.45	}_{	-0.51	}$&	$	0.1126	^{+	0.0943	}_{	-0.0945	}$&	$	45.26	^{+	23.97	}_{	-24.04	}$ &	$	2457054.27	^{+	5.28	}_{	-3.83	}$& $	101.90	^{+	25.41	}_{	-24.05	}$&	$	106.07	^{+	26.72	}_{	-25.48	}$&\citet{Barber2024}\\
HIP 90988 Ab	&	$	1.30	\pm	0.08			$&	0.097	&	7.48	&	0.289	&		--					&	$	44.2435	^{+	9.0644	}_{	-0.6138	}$&	$	455.29	^{+	12.56	}_{	-1.33	}$&	$	0.0732	^{+	0.0620	}_{	-0.0391	}$&	$	185.13	^{+	18.02	}_{	-22.62	}$&	$	2455309.70	^{+	4.81	}_{	-0.47	}$&	$	611.54	^{+	132.00	}_{	-9.07	}$&	$	617.29	^{+	85.11	}_{	-4.04	}$&	\citet{Jones2021}\\
HD 72892 b	&	$	1.02	\pm	0.05			$&	0.073	&	8.124	&	0.375	&		--					&	$	353.2604	^{+	3.1651	}_{-3.1452	}$&	$	39.47	^{+	3.01E-03	}_{	-2.95E-03	}$&	$	0.1821	\pm	0.0032			$&	$	-15.52	^{+	0.58	}_{	-0.59	}$&	$	2455701.48	\pm	0.08			$&	$	1719.43	^{+	15.45	}_{	-15.35	}$&	$	1691.21	^{+	28.19	}_{	-283.42	}$&	\citet{Jenkins2017}\\
HIP 86221 Ab	&	$	0.79					$&	0.51	&	8.5	&	0.205	&		--					&	$	137.2947	^{+	11.1648	}_{	-10.4853	}$&	$	2.22	^{+	2.85E-04	}_{	-3.17E-04	}$&	$	0.0608	^{+	0.0484	}_{	-0.0494	}$&	$	-132.03	^{+	22.72	}_{	-22.91	}$&	$	2457947.51	^{+	0.16	}_{	-0.14	}$&	$	223.64	^{+	18.20	}_{	-17.09	}$&	$	231.12	^{+	27.23	}_{	-23.65	}$&	\citet{Paredes2021}\\
HD 59686 Ab	&	$	1.90	\pm	0.20			$&	0.53	&	13.56	&	0.729	&		--					&	$	136.7892	^{+	0.5141	}_{	-0.7788	}$&	$	299.24	^{+	0.08	}_{	-0.07	}$&	$	0.0595	^{+	0.0059	}_{	-0.0077	}$&	$	132.26	^{+	2.19	}_{	-0.84	}$&	$	2451540.89	^{+	1.95	}_{	-0.31	}$&	$	2185.52	^{+	8.42	}_{	-12.61	}$&	$	2181.05	^{+	39.19	}_{	-77.76	}$&	\citet{Ortiz2016}\\
$\gamma$ Cep Ab $^{\star}$	&	$	1.40	\pm	0.12			$&	0.3991	&	18.64	&	0.3603	&	5.7	&	$	26.4	^{+	1.41	}_{	-1.18	}$&	$	901.46	^{+	2.85	}_{	-2.83	}$&	$	0.0856	^{+	0.0866	}_{	-0.0624	}$ &	$	55.37	^{+	8.81	}_{	-4.57	}$&	$	2453107.63	^{+	25.47	}_{	-30.90	}$&	$	518.20	^{+	22.18	}_{	-22.06	}$&	$	540.26	\pm	127.12	$&	\makecell[l]{\citet{Hatzes2003}\\ \citet{Torres2007} \\ \citet{Neuhauser2007}}\\
HD 41004 Ab	&	$	0.70					$&	0.4	&	20	&	0.4	&		--					&	$	108	^{+	24	}_{	-24	}$&	$	821	\pm	27			$&	$	0.8500	\pm	0.0600			$&	$	75.06	^{+	14.30	}_{	-13.75	}$&	$	2459792	^{+	244	}_{	-245	}$&	$	629.24	^{+	273.31	}_{	-203.39	}$&	$	807.21	\pm	235.17			$&\makecell[l]{\citet{Santos2002} \\ \citet{Zucker2004}}\\
HD 41004 Bb	&	$	0.40					$&	0.7	&	20	&	0.4	&		--					&	$	4290.8952	^{+	175.8025	}_{	-20.5824	}$&	$	1.33					$&	$	0.0269	^{+	0.0045	}_{	-0.0056	}$&	$	-148.65	^{+	4.88	}_{	-4.26	}$&	$	2452532.81	^{+	0.01	}_{	-0.02	}$&	$	3937.56	^{+	161.33	}_{	-18.89	}$&	$	3849.02	^{+	374.65	}_{	-1.34	}$&\makecell[l]{\citet{Santos2002} \\ \citet{Zucker2004}}\\
HD 196885 Ab$^{\star}$	&	$	1.33					$&	0.45	&	21	&	0.42	&	$	89	^{+	42	}_{	-44	}$&	$	54.84	^{+	1.87	}_{	-1.75	}$&	$	1437.90	^{+	1.90	}_{	-1.70	}$&	$	0.61	\pm	0.02			$&	$	110.01	\pm	2.29			$&	$	2451170	\pm	5			$&	$	883.48	^{+	47.67	}_{	-44.49	}$&	$	947.04	\pm	15.89			$&	\makecell[l]{\citet{Chauvin2011} \\ \citet{Chauvin2023}}\\
HD 145934 Ab	&	$	1.75	\pm	0.10			$&	0.085	&	21.774	&	0.191	&	$	81.96	^{+	40.88	}_{	-41.76	}$&	$	18.6793	^{+	0.8520	}_{	-1.4884	}$&	$	2754.58	^{+	32.90	}_{	-3.08	}$&	$	0.0176	^{+	0.0130	}_{	-0.0176}$&	$	-145.62	^{+	21.53	}_{	-51.16	}$&	$	2451429.91	^{+	0.97	}_{	-6.08	}$&	$	661.50	^{+	32.91	}_{	-52.94	}$&	$	646.20	^{+	8.93	}_{	-49.94	}$&	\makecell[l]{\citet{Feng2015} \\ \citet{Feng2022}}\\
GJ 86 Ab$^{\star}$	&	$	0.88	\pm	0.12			$&	0.5425	&	23.7	&	0.429	&		--					&	$	379.74	^{+	49.89	}_{	-44.75	}$&	$	15.53	^{+	0.10	}_{	-0.02	}$&	$	0.046^{+0.010}_{-0.020}$ &	$	268	^{+	31	}_{	-17	}$&	$	2451169	^{+	6	}_{	-23	}$&	$	1299.80	^{+	177.97	}_{	-155.72	}$&	$	1357.01	\pm	31.78			$&	\makecell[l]{\citet{Queloz2000} \\ \citet{Zeng2022}}\\
HD 2638 Ab	&	$	0.93					$&	0.48	&	25.6	&	--	&		--					&	$	69.2535	^{+	3.4859	}_{	-3.1729	}$&	$	3.44					$&	$	0.0217	\pm	0.0160			$&	$	61.43	^{+	21.85	}_{	-22.02	}$&	$	2453324.49	^{+	1.70	}_{	-1.57	}$&	$	153.73	^{+	7.76	}_{	-7.06	}$&	$	153.98	^{+	7.57	}_{	-7.46	}$&	\citet{Paredes2021}\\
HATS-37 Ab	&	$	0.84	^{+	0.02	}_{	-0.01	}$&	0.654	&	27.2	&	--	&	$	89.33	\pm	0.45			$&	$	8.9041	^{+	2.2512	}_{	-2.2081	}$&	$	3.49	^{+	0.50	}_{	-0.29	}$&	$	0.0033	^{+	0.0168	}_{	-0.0033	}$&	$	-21.41	^{+	148.34	}_{	-146.39	}$&	$	2457507.54	^{+	1.83	}_{	-2.34	}$&	$	18.77	^{+	5.81	}_{	-5.06	}$&$19.17^{+6.08	}_{	-5.51	}$&	\makecell[l]{\citet{Eisner2024} \\ \citet{Jordan2020}}\\
HD 5608 Ab	&	$	1.53	\pm	0.20			$&	0.12	&	30.5	&	0.53	&		--					&	$	28.1822	^{+	0.9077	}_{	-2.5165	}$&	$	783.29	^{+	6.06	}_{	-3.18	}$&	$	0.0236	^{+	0.0095	}_{	-0.0115	}$&	$	-54.19	^{+	12.08	}_{	-14.53	}$&	$	2452413.24	^{+	8.87	}_{	-43.09	}$&	$	472.06	^{+	16.46	}_{	-42.73	}$&	$	483.83	^{+	9.11	}_{	-8.81	}$&	\citet{Luhn2019}\\
HD 126614 Ab	&	$	1.15	\pm	0.03			$&	0.32	&	36.2	&	0.5	&		--					&	$	7.0853	^{+	0.6610	}_{	-1.6483	}$&	$	1243.99	^{+	7.42	}_{	-2.58	}$&	$	0.1991	^{+	0.1360	}_{	-0.1091	}$&	$	-122.9	^{+	14.61	}_{	-14.46	}$&	$	2453805.69	^{+	2.24	}_{	-8.14	}$&	$	116.77	^{+	11.15	}_{	-27.23	}$&	$	123.57	^{+	7.63	}_{	-6.48	}$&	\citet{Howard2010}\\
HD 164509 Ab$^{\star}$	&	$	1.13	\pm	0.02			$&	0.42	&	36.5	&	--	&		--					&	$	4.5	^{+	0.8	}_{	-0.6	}$&	$	282.60	^{+	0.70	}_{	-2.20	}$&	$	0.2600	^{+	0.0400	}_{	-0.0300	}$&	$	320.28	^{+	5.73	}_{	-8.02	}$&	$	2455726	^{+	172	}_{	-730	}$&	$	149.37	^{+	9.53	}_{	-6.36	}$&	$	152.54	\pm	28.60			$&	\citet{Giguere2012}\\
HIP 107773 Ab	&	$	2.42	\pm	0.27			$&	0.63	&	40	&	--	&		--					&	$	44.2884	^{+	1.3492	}_{	-1.5820	}$&	$	144.47	^{+	0.19	}_{	-0.17	}$&	$	0.0217	^{+	0.0098	}_{	-0.0143	}$&	$	163.29	^{+	10.68	}_{	-11.65	}$&	$	2456202.12	^{+	2.20	}_{	-0.83	}$&	$	646.29	^{+	19.98	}_{	-23.32	}$&	$657.07	^{+	15.76	}_{	-16.53	}$&	\citet{Jones2015}\\
70 Vir Ab	&	$	0.92					$&	0.08	&	52	&	--	&		--					&	$	356.4499	^{+	1.5562	}_{	-0.4135	}$&	$	116.72	^{+	0.10	}_{	-0.04	}$&	$	0.2011	^{+	0.0129	}_{	-0.0022	}$&	$	1.51	^{+	1.33	}_{	-2.03	}$&	$	2448989.57	^{+	1.25	}_{	-5.23	}$&	$	2297.48	^{+	10.69	}_{	-2.91	}$&	$	2374.63	^{+	40.83	}_{	-24.84	}$&	\citet{Luhn2019}\\
HIP 94235 Ab	&	$	1.09	^{+	0.02	}_{	-0.01	}$&	0.26	&	56	&	0.25	&	$	87.14	\pm	0.17			$&	$135.6987	^{+	9.3746	}_{	-9.9190	}$&	$	8.40	^{+	0.91	}_{-	1.62	}$&	$	0.2425^{+0.2770}_{-0.2425}	$&	$	18.04	^{+	51.94	}_{	-98.82	}$&	$	2459308.76	^{+	2.40	}_{	-1.27	}$&	398.12 $(3\sigma)$ &	351.30	$(3\sigma)$ &	\citet{zhou2022}\\
WASP-20 Ab	&	$	1.20	\pm	0.04			$&	0.88	&	61	&	--	&	$	85.57	\pm	0.02		$&	$	32.1511	^{+	0.7678	}_{	-0.7490	}$&	$	4.90					$&	$	0.0032	^{+	0.0038	}_{	-0.0032	}$&	$	-118.45	^{+	33.83	}_{	-32.87	}$&	$	2454659.35	^{+	0.33	}_{	-0.30	}$&	$	90.29	^{+	2.31	}_{	-1.96	}$&	$	98.84	\pm	5.40			$&	\citet{Anderson2015}\\
Gliese 725 Ab	&	$	0.33	\pm	0.08			$&	0.25	&	63	&	0.29	&		--					&	$	1.7686	^{+	0.1164	}_{	-0.1146	}$&	$	11.22					$&	$	0.0204	^{+	0.0252	}_{	-0.0249	}$&	$	153.38	^{+	35.87	}_{	-36.11	}$&	$	2458599.17	^{+	0.78	}_{	-0.74	}$&	$	2.92	\pm	0.19			$&	$	2.92	\pm	0.19			$&	\citet{Cortes-Zuleta2025}\\
GJ 3021 Ab$^{\star}$	&	$	0.90					$&	0.09	&	68	&	--	&		--					&	$	165	\pm	2			$&	$	133.40	\pm	0.01			$&	$	0.53	\pm	0.01			$&	$	286.7	\pm	0.03			$&	$	2451531	\pm	70			$&	$	994.71	\pm	19.07			$&	$	1070.99	\pm	28.60	$&	\makecell[l]{\citet{Naef2001} \\ \citet{Chauvin2006}}\\
WASP-76 Ab	&	$	1.46	\pm	0.07			$&	0.79	&	84.8	&	--	&	$	88	\pm	1.6			$&	$	122.8118	^{+	1.9128	}_{	-1.7442	}$&	$	1.81					$&	$	0.0007^{+0.0011}_{-0.0007}$&	$	-178.75	^{+	40.86	}_{	-39.19	}$&	$	2455961.72	^{+	0.19	}_{	-0.23	}$&	$	300.36	^{+	4.80	}_{	-4.16	}$&	$	290.97	^{+	4.50	}_{	-4.41	}$&	\citet{West2016}\\
TOI-640 Ab	&	$	1.54	^{+	0.07	}_{	-0.08	}$&	0.77	&	85	&	--	&	$	82.54	^{+	0.40	}_{	-0.60}$&	$	82.7723	^{+	7.0911	}_{	-6.1816	}$&	$	5.03	\pm	0.03			$&	$	0.0404	^{+	0.0475	}_{	-0.0404}$&	$	-165.67	^{+	26.98	}_{	-26.67	}$&	$	2458751.21	^{+	0.22	}_{	-0.34	}$&	$	291.56	^{+	25.58	}_{	-22.35	}$&	$	288.65	^{+	26.93	}_{	-20.77	}$&	\citet{Rodriguez2021}\\
Kepler-21 Ab	&	$	1.41	^{+	0.02	}_{	-0.03	}$&	0.44	&	86.7	&	--	&	$	83.2	^{+	0.28	}_{	-0.26	}$&	$	2.6718	^{+	0.2645	}_{	-0.2862	}$&	$	2.78					$&	$	0.0504	^{+	0.0604	}_{	-0.0447	}$&	$	-15.54	^{+	21.68	}_{	-23.70	}$&	$	2456797.93	^{+	0.11	}_{	-0.08	}$&	$	7.24	^{+	0.72	}_{	-0.78	}$&	$	7.11	^{+	0.97	}_{	-0.78	}$&	\citet{Bonomo2023}\\
TOI-1634 Ab	&	$	0.50	\pm	0.01			$&	0.17	&	90	&	--	&	$	88.2	\pm	1.1			$&	$	5.1577	^{+	0.5078	}_{	-0.3170	}$&	$	0.99	^{+	2.22E-04	}_{	-2.42E-04	}$&	$	0.0292	^{+	0.0518	}_{	-0.0292	}$&	$	-132.89	^{+	50.17	}_{	-46.96	}$&	$	2459068.86	^{+	0.06	}_{	-0.13	}$&	$	4.99	^{+	0.49	}_{	-0.31	}$&	$	5.33	^{+	0.74	}_{	-0.50	}$&	\citet{Cloutier2021}\\
HD 30856 Ab	&	$	1.35	\pm	0.09			$&	0.54	&	93	&	--	&		--					&	$	32.0114	^{+	0.4005	}_{	-0.5472	}$&	$	912.06	^{+	1.06	}_{	-1.00	}$&	$	0.0210	^{+	0.0058	}_{	-0.0072	}$&	$	-162.54	^{+	7.94	}_{	-10.37	}$&	$	2455274.73	^{+	17.21	}_{	-11.30	}$&	$	585.27	^{+	7.55	}_{	-10.22	}$&	$	587.29	^{+	10.24	}_{	-10.16	}$&	\citet{Johnson2011}\\
BD+63 1405 Ab	&	$	0.85	\pm	0.10			$&	0.319	&	97	&	--	&		--					&	$	181.6858	^{+	4.8903	}_{	-3.8052	}$&	$	1193.30	^{+	23.23	}_{	-21.91	}$&	$	0.7730	^{+	0.0194	}_{	-0.0180}$&	$	94.66	^{+	1.34	}_{	-1.13	}$&	$	2459436.87	^{+	23.86	}_{	-22.67	}$&	$	1254.83	^{+	42.08	}_{	-33.85	}$&	$	1344.81	^{+	75.85	}_{	-14.80	}$&	\citet{Dalal2021}\\
TOI-1452 Ab	&	$	0.25					$&	0.226	&	97	&	--	&	$	89.77	\pm	0.16			$&	$	4.7731	^{+	0.7633	}_{	-0.7682	}$&	$	11.03	\pm	0.06			$&	$	0.0018^{+0.0106}_{-0.0018}$ &	$	-136.28	^{+	169.61	}_{	-169.78	}$&	$	2459010.17	^{+	5.70	}_{	-4.26	}$&	$	6.55	^{+	1.07	}_{	-1.06	}$&	$	4.82	\pm	1.30			$&	\citet{Cadieux2022}\\
WASP-2 b	&	$	0.84	\pm	0.11			$&	0.4	&	106	&	--	&	$	84.49	\pm	0.17			$&	$	155.0174	^{+	3.9697	}_{	-3.5030	}$&	$	2.15			$&	$	0.0036	^{+	0.0022	}_{	-0.0024	}$&	$	108.25	^{+	53.49	}_{	-56.28	}$&	$	2453983.12	^{+	0.25	}_{	-0.26	}$&	$	281.47	^{+	7.41	}_{	-6.17	}$&	$	278.05	^{+	6.22	}_{	-6.24	}$&	\makecell[l]{\citet{triaud2010} \\ \citet{southworth2010}}\\
HAT-P-27 b	&	$	0.95	^{+	0.12	}_{	-0.10	}$&	0.323	&	133.8	&	--	&	$	85	\pm	0.20			$&	$	93.7609	^{+	1.4300	}_{	-1.1933	}$&	$	3.04			$&	$	0.0021	\pm	0.0014			$&	$	37.78	^{+	47.49	}_{	-47.52	}$&	$	2455191.75	^{+	0.38	}_{	-0.26	}$&	$	204.67	^{+	3.22	}_{	-2.52	}$&	$	203.16	^{+	2.72	}_{	-2.26	}$&	\makecell[l]{\citet{anderson2011a} \\ \citet{Beky2011}}\\
TOI-3540 A b	&	$	1.08	^{+	0.07	}_{	-0.07	}$&	0.15	&	189.6	&	--	&	$	81.93					$&	$	172.3243	^{+	7.9375	}_{	-7.4956	}$&	$	3.13	^{+	2.04E-03	}_{	-2.29E-03	}$&	$	0.0112	^{+	0.0084	}_{	-0.0086	}$&	$	88.49	\pm	20.77			$&	$	2459427.11	^{+	0.24	}_{	-0.19	}$&	$	405.38	^{+	32.14	}_{	-5.51	}$&	$	398.92	^{+	18.52	}_{	-19.53	}$&	\citet{yee2022}\\
KELT-24 b	&	$	1.31	^{+	0.05	}_{	-0.02	}$&	0.23	&	209.3	&	--	&	$	89.67	\pm	0.24			$&	$	459.7918	^{+	6.9646	}_{	-7.0460	}$&	$	5.55	\pm	0.0011			$&	$	0.0041	\pm	0.0033			$&	$	19.01	^{+	21.51	}_{	-21.54	}$&	$	2458567.20	^{+	0.27	}_{	-0.25	}$&	$	1462.07	^{+	22.24	}_{	-22.50	}$&	$	1458.83	\pm	57.21			$&	\citet{rodriguez2019}\\
HD 202772 A b	&	$	1.72	^{+	0.06	}_{	-0.07	}$&	1.125	&	211	&	--	&	$	84.51	\pm	1.10			$&	$	105.7082	^{+	3.4838	}_{	-4.1890	}$&	$	3.29	^{+	0.01	}_{	-0.01	}$&	$	0.0187	^{+	0.0174	}_{	-0.0187}$&	$	63.46	^{+	7.44	}_{	-10.02	}$&	$	2458368.22	^{+	0.07	}_{	-0.05	}$&	$	341.44	^{+	23.23	}_{	-2.81	}$&	$	323.72	^{+	8.13	}_{	-7.67	}$&	\citet{wang2019}\\
TrES-2 b	&	$	0.98	\pm	0.06			$&	0.509	&	232	&	--	&	$	83.57	\pm	0.14			$&	$	176.8765	^{+	3.6952	}_{	-3.2688	}$&	$	2.34	^{+	0.10	}_{	-0.07	}$&	$	0.0135	^{+	0.0097	}_{	-0.0100	}$&	$	-31.02	^{+	39.67	}_{	-40.73	}$&	$	2453949.49	^{+	0.17	}_{	-0.29	}$&	$	362.63	^{+	13.80	}_{	-9.18	}$&	$	327.75	^{+	7.85	}_{	-7.48	}$&	\makecell[l]{\citet{southworth2010} \\ \citet{odonovan2006}}\\
TOI-1201 b	&	$	0.51	\pm	0.02			$&	0.46	&	316.2	&	--	&	$	88.11	\pm	0.42			$&	$	2.9909	^{+	1.6183	}_{	-0.9571	}$&	$	1.23	^{+	0.46	}_{	-1.23	}$&	$	0.0713	^{+	0.0286	}_{	-0.0382	}$&	$	116.44	^{+	169.62	}_{	-191.22	}$&	$	2458798.24	^{+	2.21	}_{	-1.45	}$&	$	3.20	^{+	2.29	}_{	-3.01	}$&	$	3.40	^{+	1.66	}_{	-1.08	}$&	\citet{kossakowski2021}\\
TOI-3984 A b	&	$	0.49	\pm	0.02			$&	0.75	&	356	&	0.64	&	$	89.5	\pm	0.20			$&	$	26.5203	^{+	4.4916	}_{	-4.2742	}$&	$	4.36	^{+	0.02	}_{	-0.02	}$&	$	0.0366	^{+	0.0191	}_{	-0.0188	}$&	$	31.93	^{+	99.60	}_{	-97.93	}$&	$	2459451.96	^{+	1.67	}_{	-2.77	}$&	$	41.90	^{+	7.15	}_{	-6.80	}$&	$	35.92	^{+	6.84	}_{	-6.63	}$&	\citet{canas2023}\\
WASP-58 b	&	$	0.94	\pm	0.10			$&	0.3	&	379	&	--	&	$	87.4	\pm	1.50			$&	$	112.9549	^{+	4.1144	}_{	-4.0862	}$&	$	5.03	^{+	0.02	}_{	-0.02	}$&	$	0.0101	^{+	0.0069	}_{	-0.0071	}$&	$	45.9	^{+	52.56	}_{	-53.67	}$&	$	2455745.22	^{+	0.64	}_{	-0.53	}$&	$	289.39	^{+	11.55	}_{	-10.32	}$&	$	244.59	^{+	10.04	}_{	-10.12	}$&	\citet{hebrard2013}\\
WASP-145 A b	&	$	0.76	\pm	0.04			$&	0.591	&	476	&	--	&	$	83.3	\pm	1.30			$&	$	185.4799	^{+	4.6486	}_{	-4.3451	}$&	$	1.77			$&	$	0.0039	\pm	0.0015			$&	$	-86.93	^{+	96.51	}_{	-98.85	}$&	$	2456812.99	^{+	0.48	}_{	-1.31	}$&	$	293.37	^{+	7.41}_{	-6.82	}$&	$	245.36	^{+	6.08	}_{	-6.01	}$&	\citet{hellier2019}\\
Qatar-6 b	&	$	0.82	\pm	0.02			$&	0.244	&	486	&	--	&	$	86.01	\pm	0.14			$&	$	104.9685	^{+	3.6107	}_{	-3.5716	}$&	$	3.51\pm0.01$&	$	0.0061	^{+	0.0030	}_{	-0.0032	}$&	$	-35.54	^{+	106.10	}_{	-112.85	}$&	$	2457492.10	^{+	2.11	}_{	-0.80	}$&	$	218.88	^{+	7.73	}_{	-7.46	}$&	$	185.38	^{+	7.31	}_{	-7.79	}$ &	\citet{alsubai2018}\\
\enddata

\end{deluxetable*}
\end{longrotatetable}

\end{document}